\documentclass{IEEEtran}
\usepackage{cite}
\usepackage{amsmath,amssymb,amsfonts,amsthm}
\usepackage{graphicx}
\usepackage{textcomp}
\usepackage{xcolor,soul,framed} 
\usepackage{multirow}    % 一般用以设计表格，将所行合并
\usepackage{multicol}         % 合并多列
\usepackage{subcaption}
\newtheorem{rem}{Remark}
\usepackage{bm}
\usepackage{url} 
\usepackage[hidelinks]{hyperref}
\usepackage{booktabs}
\usepackage{tabularx}
\usepackage{makecell}

\newtheorem{proposition}{Proposition}
\newtheorem{problem}{Problem}
\usepackage{tikz}
\usetikzlibrary{positioning,fit,calc,arrows.meta}

\def\BibTeX{{\rm B\kern-.05em{\sc i\kern-.025em b}\kern-.08em
    T\kern-.1667em\lower.7ex\hbox{E}\kern-.125emX}}

\usepackage{algorithm}
\usepackage{algpseudocode}
\usepackage{booktabs}
\usepackage{bbding}
\ifCLASSINFOpdf
 
\else

\fi

\begin{document}

\title{Doppler-Robust Vortex Wavefront Design for Integrated Sensing and Communication}

\author{
Yuan Liu,~\IEEEmembership{Member,~IEEE}, Wen-Xuan Long,~\IEEEmembership{Member,~IEEE}, M. R. Bhavani Shankar,~\IEEEmembership{Senior Member,~IEEE}, Marco Moretti,~\IEEEmembership{Member,~IEEE}, Rui Chen,~\IEEEmembership{Member,~IEEE}, Bj\"{o}rn Ottersten,~\IEEEmembership{Fellow,~IEEE} 
\thanks{
The work of Yuan Liu, M. R. Bhavani Shankar, and Björn Ottersten was supported in whole by the Luxembourg National Research Fund (FNR) through BRIDGES Project MASTERS under grant BRIDGES2020/IS/15407066.
The work of Wen-Xuan Long and Marco Moretti was supported in part by European Union under Italian National Recovery and Resilience Plan (NRRP) of NextGenerationEU, partnership on ``Telecommunications of the Future'' (PE00000001-Program ``RESTART,'' Cascade Call SMART); and in part by Italian Ministry of Education and Research (MUR) in the Framework of the FoReLab Project (Departments of Excellence) and the Project GARDEN funded by EU in NextGenerationEU Plan through Italian ``Bando Prin 2022-D.D.1409 del 14-09-2022.''
The work of Rui Chen was supported in part by the National Natural Science Foundation of China under Grant 62271376, in part by the Guangdong Natural Science Fund for Distinguished Young Scholar under Grant 2023B1515020079. \textit{(Corresponding author: Wen-Xuan Long.)}}
\thanks{
Yuan Liu, M. R. Bhavani Shankar, and Björn Ottersten are with the Interdisciplinary Centre for Security, Reliability and Trust (SnT), University of Luxembourg, L-1855 Kirchberg, Luxembourg (e-mail: yuan.liu@uni.lu, bhavani.shankar@uni.lu, bjorn.ottersten@uni.lu).}
\thanks{Wen-Xuan Long is with the Dipartimento di Ingegneria dell’Informazione, University of Pisa, 56126 Pisa, Italy (e-mail: wenxuan.long@ing.unipi.it).}
\thanks{Marco Moretti is with the Dipartimento di Ingegneria dell’Informazione, University of Pisa, 56126 Pisa, Italy, and also with the National Inter-University Consortium for Telecommunications (CNIT), 43124 Parma, Italy (marco.moretti@unipi.it).}
\thanks{Rui Chen is with the State Key Laboratory of ISN and also with Guangzhou Institute of Technology, Xidian University, China (e-mail: rchen@xidian.edu.cn).
}
}

\markboth{  }%
{Shell \MakeLowercase{\textit{et al.}}: Bare Demo of IEEEtran cls for IEEE Journals}

\maketitle

\begin{abstract}
Integrated sensing and communication (ISAC) is a promising paradigm for future wireless systems due to spectrum reuse, hardware sharing, and joint waveform design. 
% The deployment of ISAC systems in dynamic environments introduces Doppler shifts, which can degrade both sensing accuracy and communication reliability. 
% This is particularly severe for beam-sensitive schemes, such as the emerging orbital angular momentum (OAM) vortex wavefronts. However, the Doppler-induced ambiguity problem has not been discussed in the vortex designs.
In dynamic scenes, Doppler shifts degrade both sensing and communication, which is particularly critical for beam-sensitive orbital angular momentum (OAM) wavefronts.
% Dynamic environments introduce Doppler shifts that degrade both sensing accuracy and communication reliability, especially for beam-sensitive schemes 
To address this, we propose a Doppler-robust ISAC framework, which first senses and then communicates.
% In the sensing phase, 
% We address the dynamic multi-target sensing problem and analyze the classical angle-Doppler ambiguity problem which is missing in the vortex designs. 
Specifically, in the sensing phase, multiple vortex modes are simultaneously transmitted via code-division mode-multiplexing (CDMM). 
To solve Doppler-induced inter-mode interference, we propose a velocity-consistency matching (VCM)-expectation maximization (EM) algorithm that jointly decodes the sensing matrix and estimates range, azimuth, elevation, and velocity for multiple moving targets.
In the communication phase, the joint transmitter (Tx) beamforming and receiver (Rx) beam steering are configured from the estimated channel state information (CSI). 
% Furthermore, the inherent trade-off design between sensing and communication is quantified, where allocating longer pilots for sensing improves the angle estimation accuracy and beam alignment, but reduces the communication frames, leading to a potential drop in spectral efficiency (SE).
% Furthermore, the inherent trade-off between sensing and communication is quantitatively analyzed. 
We further quantify the sensing-communication allocation trade-off by evaluating how pilot length affects estimation accuracy, beam alignment, and spectral efficiency (SE).
% Insufficient pilot length in the sensing phase degrades the estimation performance, which in turn affects beam alignment accuracy in the communication phase and reduces spectral efficiency (SE). However, allocating excessively long pilot sequences for sensing reduces the number of communication frames, also resulting in SE degradation. Therefore, the number of frames assigned to the sensing and communication phases must be carefully balanced.
% Simulations show that the proposed VCM-EM and ISAC design improve sensing accuracy and communication SE over baseline schemes in dynamic scenarios.
Simulation results show that the proposed VCM-EM and ISAC designs achieve higher sensing accuracy and communication SE than baseline schemes in dynamic scenarios.
\end{abstract}

\IEEEpeerreviewmaketitle

% \vspace{-0.2cm}
\section{Introduction}
% Introduce ISAC as a key technology for next‐generation wireless: spectrum reuse, hardware sharing, joint waveform and signal‐processing design.
% {ISAC waveform landscape for R 4-1}
The design of integrated sensing and communication (ISAC) systems is a key enabler for future wireless networks \cite{mishra2024signal,9755276,10418473,Mishra2019MmWaveJRC}. By sharing spectrum, waveforms, and hardware, ISAC has the potential to provide high-resolution sensing and reliable communication services on the joint platform \cite{9540344}.
Due to the dual merits, there are several ISAC designs built on well-known waveforms, e.g., the widely used orthogonal frequency division multiplexing (OFDM) \cite{10634583}, the radar-centric design using frequency-modulated continuous wave (FMCW) \cite{10304513,9940978}, and phase-modulated continuous wave (PMCW) \cite{8642926}. 
% \textbf{Here needs to address why use OAM for ISAC: the preliminary reason is, the OAM mode provides very good orthogonality so that different modes can support different UE without interference among users.}
For ease of incorporation into wireless communication infrastructures, OFDM-based ISAC has become popular despite known \textit{radar-related} issues, including peak-to-average power ratio (PAPR), Doppler sensitivity, sidelobe levels, and ambiguities.
In fact, OFDM is compared with PMCW for sidelobe performance in~\cite{8642926}, and FMCW under the IEEE $802.11bd$ standards in~\cite{Ehsanfar2023FMCWOFDM}.
For high-mobility ISAC scenarios, Doppler-resilient waveforms, such as orthogonal time frequency space (OTFS), have attracted increasing attention~\cite{Yuan2024DDISAC,MendezMonsanto2026OTFSISAC}.
Finally, motivated by the LoRA (long range) standard, chirp spread spectrum-based designs have been studied for outdoor Internet of Things (IoT) and LoRA communication for low-power wide-area networks (LPWANs) \cite{Azim2024LayeredCSS}.
These works indicate that ISAC waveform design should be tailored to the operating scenario, mobility condition, sensing objective, and communication requirement.

% {Why OAM for ISAC R 1-2.}
% {OAM vs UCA MIMO R 1-1, R3-1. OAM is a wavefront, which is compatable to the above-mentioned waveforms.}
{
Although existing ISAC waveforms mainly exploit the time, frequency, code, chirp, or delay-Doppler domains, the spatial structure of the transmitted electromagnetic field provides another degree of freedom.
Different from a waveform, which describes how the signal is modulated across these domains, a wavefront characterizes the spatial distribution and propagation of the electromagnetic field.
At millimeter-wave and sub-terahertz frequencies, vortex electromagnetic waves provide a structured wavefront for exploiting this spatial degree of freedom in ISAC systems~\cite{Shen2026OAMJCS}.}
Specifically, a helical wavefront carrying orbital angular momentum (OAM) has a phase varying in azimuth and a vortex electric field, where different OAM modes form orthogonal spatial modes~\cite{Allen1992Orbital,thide2007utilization}.
{It is worth noting that OAM beams can be generated by a uniform circular array (UCA), and are therefore closely related to the Fourier-mode representation of circular-aperture multiple-input multiple-output (MIMO). However, the focus here is not a generic element-domain UCA-MIMO formulation. Instead, OAM provides a physically structured mode-domain representation.
This mode-domain structure supports multiple data streams or links over orthogonal OAM modes, therefore suppressing interlink interference and potentially improving the aggregate throughput in multi-user environments~\cite{9046790}.}
Meanwhile, {compared with conventional MIMO radar systems,} vortex wavefronts can achieve high-resolution angle and rotational velocity sensing in three-dimensional (3-D) space without mechanical scanning~\cite{Liu2019Microwave,10741257}.

{These multi-function benefits make OAM a promising candidate for ISAC. The estimated mode-domain channel parameters can be used not only for target localization, but also for beam alignment and beamforming in the communication link.
Importantly, OAM should not be regarded as a replacement for baseband waveforms such as OFDM, FMCW, PMCW, or OTFS. Instead, it provides an additional spatial/mode-domain degree, which is potentially compatible with any of these to provide the aforementioned benefits.}
% Implemented as a spatially modulated superposition of plane waves from multiple antennas, vortex wavefronts enable continuous sampling in three-dimensional (3-D) space and, compared with conventional multiple-input multiple-output (MIMO) radar, offer super resolution in angle and rotational velocity estimation without mechanical beam scanning \cite{Liu2019Microwave,10741257}.
However, vortex wavefront-based communications are highly sensitive to precise beam alignment between transmitter (Tx) and receiver (Rx), which in turn requires the Tx to know the channel state information (CSI) a priori \cite{chen2023circular}. 
{While conventional communication systems typically acquire CSI through pilot-based estimation, OAM-based ISAC provides a broader sensing capability. Specifically, OAM sensing can probe both surrounding targets and communication user equipment (UE). The user-related parameters can then be used to reconstruct the communication CSI for beam alignment and beamforming.
This advantage has driven ISAC works to exploit sensing outputs to aid the communication task under an integrated design, where robust channel parameter estimation is the cornerstone.}
Most existing OAM studies {largely consider static scenes, and} adopt time-division mode-multiplexing (TDMM) to probe mode-domain response~\cite{9968273,10620284,9139401,10304513,8322276}, where the Tx transmits only one mode per symbol, thereby separating the channel impulse response (CIR) of each mode.  
Leveraging diversity across time, frequency, and mode domains, channel parameters can be estimated using classical methods, such as subspace-based multiple signal classification (MUSIC)~\cite{10620284,9968273,10304513}, the maximum-likelihood-based estimation~\cite {9139401}, and non-parametric spectral estimation~\cite{8322276}.    
Meanwhile, several enhancements have also been explored, e.g., reference~\cite{10295376} optimizes the circular array radius to improve sensing performance. 
Interestingly, simultaneous multi-mode transmission via orthogonal polyphase codes for imaging was achieved in~\cite{10311524}, {
since the spatial diversity offered by multiple OAM modes is not fully utilized in TDMM. 
}
% is proposed to achieve rapid, unambiguous imaging in a single snapshot.
%This coding-based simultaneous multi‐mode transmission is called code-division mode-multiplexing (CDMM). 
 
{Meanwhile, outdoor intelligent Internet of Things (IoT) scenarios are becoming interesting for ISAC systems, such as drone applications and autonomous vehicles \cite{9454158,11113304}.}
ISAC systems face challenges due to (i) mutual interference among multiple targets and UE; and (ii) scene dynamics that degrade sensing accuracy and communication reliability \cite{9540344,9928573}. 
{Particularly, mobility-induced Doppler introduces well-known ambiguities \cite{4602540,9524922,9309182,10446432,9591336}, which are common in practice.}
For vortex wavefronts, such dynamics can reduce mode orthogonality and lead to angle Doppler coupling under the existing and widely used TDMM scheme, which biases parameter estimates.
In OAM ISAC, the resulting CSI errors further degrade beamforming and communication spectral efficiency (SE).
% However, few works extend this analysis to vortex wavefronts. 
% The widely used TDMM scheme for sensing will suffer from angle-Doppler ambiguity and result in biased estimations. 
% In ISAC designs, it further degrades communication performance due to the imperfect CSI.
% for communication beamforming.
% , and the orthogonal polyphase codes can be ruined by time‐varying channels \cite{9591336}. 

To address these gaps, we first characterize Doppler-induced ambiguities in both TDMM and the proposed code-division mode-multiplexing (CDMM) schemes.
We then develop a Doppler-robust channel parameter estimation algorithm and a sensing-aided joint beamforming scheme for mobile UE, and integrate them into an ISAC design.
To our knowledge, this is the first study to treat Doppler-induced ambiguity in vortex wavefront sensing for dynamic multi-target scenes and to pair it with joint beamforming tailored to mobile UE.
% this is the first work to address \emph{Doppler-induced ambiguity} in vortex-wavefront sensing for \emph{dynamic multi-target} scenes as well as joint beamforming scheme tailored for mobile UE.
% , covering both the widely–used TDMM and the proposed CDMM precoding.
The main contributions of this paper are as follows:
% we analyze the Doppler ambiguity of the existing waveforms and develop a Doppler-robust estimation approach, and propose a ISAC framework design. 
\begin{enumerate}
    \item \textbf{CDMM precoding and Doppler perturbation modeling}: 
    We propose a slow‐time CDMM scheme based on orthogonal coding matrices and derive a closed‐form expression for Doppler‐induced orthogonality degradation among multiple targets under a rank-one Vandermonde structure.
    % We formulate the resulting multi-target problem, exposing the rank-one Vandermonde structure and the disturbance matrix that lead to an ambiguity-free sensing model
    % , and prove that a Doppler‐based decoding approach can restore the orthogonality for multi-mode separation in dynamic scenarios.
    \item \textbf{Doppler-robust parameter estimation via VCM-EM}:  
    We incorporate velocity-consistency matching (VCM) into the E-step of expectation maximization (EM) framework. The E-step jointly estimates Doppler, decodes, and reconstructs the sensing echo of each target; and the M-step alternatively updates the remaining parameter subsets. 
    % The proposed algorithm achieves Doppler-robust parameter estimation of multiple dynamic targets.
    % The proposed algorithm  provides Doppler-robust azimuth, elevation, and velocity estimates and enables accurate multi-target reconstruction under mobility
    \item \textbf{Sensing-aided OAM joint beamforming for mobile UE}:
    % We propose the first OAM joint beamforming scheme tailored for mobile Rxs. 
    By leveraging the estimated CSI, the beamforming and beam steering are respectively performed at the integrated Tx and communication Rx, thereby achieving beam alignment between the transceiver and enhancing communication SE.
    \item \textbf{Joint sensing and communication frame scheduling}: 
    % The sensing pilot length determines the Doppler‐resolution and CSI accuracy, i.e., a longer pilot gives higher velocity resolution and which further improves channel parameter estimation. However, it reduces the frames available for communication. We therefore introduce a joint optimization of sensing‐slot length and communication scheduling, balancing high-precision Doppler compensation against overall data throughput.    
    The sensing pilot length determines the Doppler resolution and CSI accuracy, i.e., a longer pilot yields higher velocity resolution and more accurate estimation, while reducing the frames available for communication. We theoretically characterize the {trade-off} between sensing and communication scheduling and elaborate with numerical simulations.
\end{enumerate}

The following notations are used in this paper: 
Lower/upper-case bold characters denote vectors/matrices, particularly, $\mathbf{A} \in \mathbb{R}^{N_1 \times N_2}$ and $\mathbf{A} \in \mathbb{C}^{N_1 \times N_2}$ respectively denote real and complex matrices of size $N_1 \times N_2$, and $[\mathbf{A}]_{n_1, n_2} \triangleq A_{n_1, n_2}$ %or $\mathbf{A}(n_1,n_2)$ 
denotes the $(n_1,n_2)$-th entry of the matrix, $\mathbf{A}[n_1, :] $ denotes the $n_1 $th row of the matrix, and $\mathbf{A}[:, n_2] $ denotes the $n_2 $th column of the matrix. $(\cdot)^T$ and $(\cdot)^H$ denote the transpose and the conjugate transpose of a matrix or vector, respectively. 
$\|\cdot\|_0$ and $\|\cdot\|_p$, respectively, denote the $l_0$ and $l_p$ norms, in particular, $| \cdot |$ denotes the Euclidean norm of a vector, i.e., $\|\cdot\|_2$, and the absolute value of a scalar.
% whereas $\|\cdot\|_0$ and $\|\cdot\|_p$ respectively denote the $l_0$ and $l_p$ norms. 
% The symbol $\text{mod}$ refers to a remainder operation. 
$\otimes$ denotes the Kronecker product. 
$\odot$ denotes the element-wise product of matrices.
% $\circledast$ denotes convolution.
$\text{vec}(\cdot)$ denotes the vectorization operator that turns a matrix into a vector by stacking all columns on top of one another.
% The ceiling function $\lceil A \rceil$ rounds the number $A$ up to the nearest integer, whereas the floor function $\lfloor A \rfloor$ rounds $A$ down to the nearest integer. 
% The function $\bmod{A} \text{mod}~A $ denotes reminder operation of $A$.
% the function $\modop_{A}(x)$ returns $x \bmod A$
The remainder of $A$ modulo $N$ is defined as $\langle A \rangle_{N} \triangleq A \pmod{N}$.
% $\delta(t)$ denotes the ideal impulse function in the time domain. 
% $\langle A \rangle_{N}$ denotes 

The rest of the paper is organized as follows: 
Section~\ref{Sec:signal_model_isac} introduces the system model. 
Section~\ref{Sec:radar_processing_isac} and Section~\ref{Sec:joint_design_isac} introduce the sensing and sensing-aided ISAC design. 
Section~\ref{sec:simu_isac} is the simulation part. 
% Section~\ref{Sec:ambiguity} summarizes the near- and far-field localization ambiguity.
% Section~\ref{sec:sage} elaborates on the proposed GM-SAGE and the recursive searching algorithm. 
Section~\ref{sec:conclude_isac} concludes the paper. 

% \vspace{-0.2cm}
\section{System Model}\label{Sec:signal_model_isac}
\begin{figure}[t]
\centering
    \includegraphics[width=0.45\textwidth]{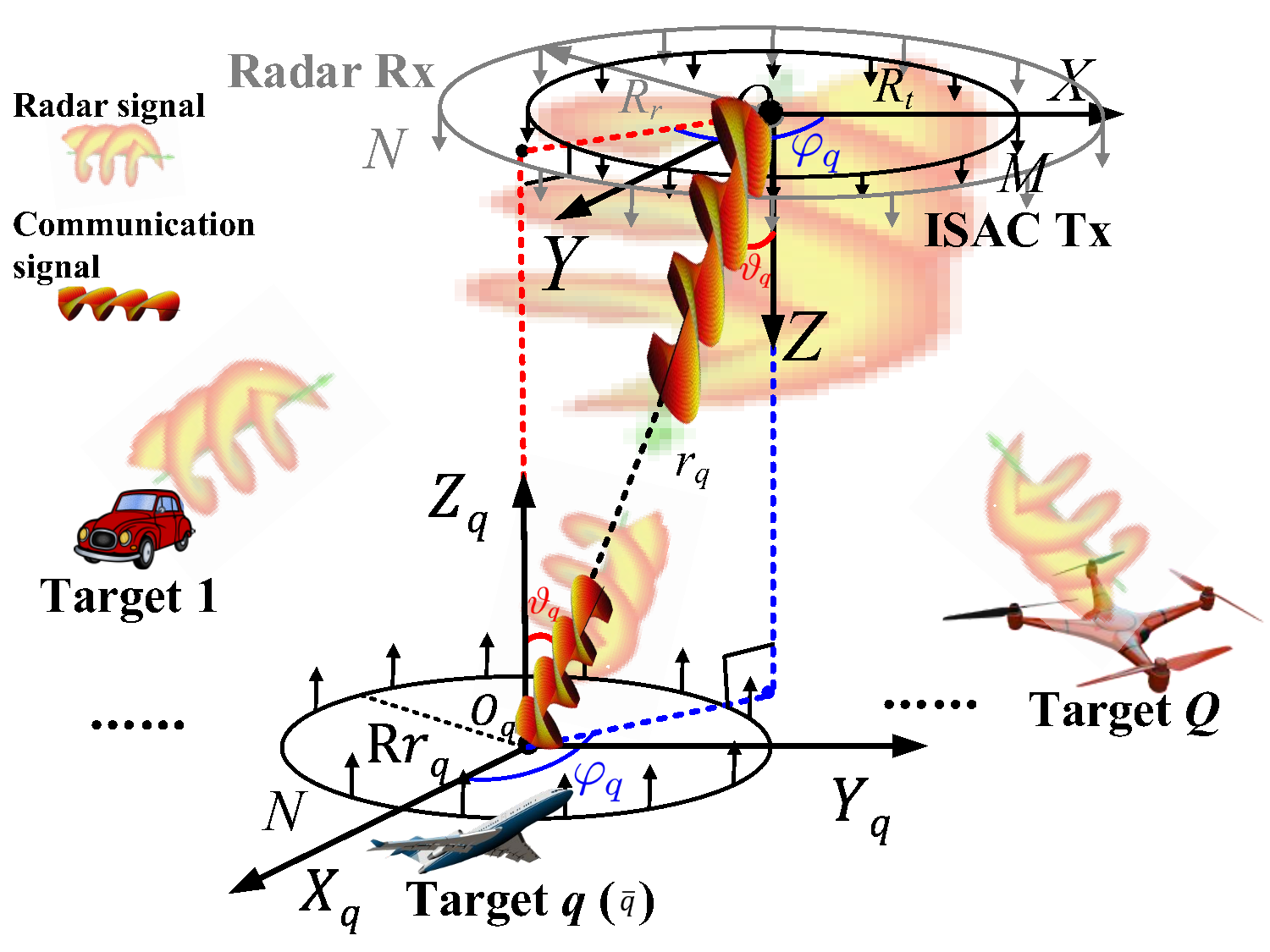}
              % \vspace{-0.2cm}
\caption{{Illustration of the ISAC system model. }  
}
\label{fig_axix_2}
 \vspace{-0.5cm}
\end{figure}
% 
%%%%%%%%%%%%%%%%%%%%%%%
\begin{figure*}[ht]
    \centering
    \begin{tikzpicture}[
        font=\footnotesize,
        >=Latex,
        block/.style={
            rectangle,
            rounded corners,
            draw,
            align=center,
            text width=2.9cm,
            minimum height=1.10cm,
            inner sep=4pt
        },
        wideblock/.style={
            rectangle,
            rounded corners,
            draw,
            align=center,
            text width=3.5cm,
            minimum height=1.20cm,
            inner sep=4pt
        },
        phase/.style={
            rectangle,
            rounded corners,
            draw,
            dashed,
            inner sep=0.12cm
        },
        arrow/.style={->, thick}
    ]

    % =======================
    % Sensing phase: first row
    % =======================
    \node[block]     (s1) at (0,0)    {CDMM-based \\ OAM sensing};
    \node[block]     (s2) at (3.8,0)  {Radar echo};
    \node[wideblock] (s3) at (8.0,0)  {VCM-EM joint decoding\\and parameter estimation\\
    $\{\hat r_q,\hat\varphi_q,\hat\theta_q,\hat v_q\}_{q=1}^Q$};
    \node[wideblock] (s4) at (12.5,0) {UE selection \\and sensing-aided\\CSI construction};

    \node[phase, fit=(s1)(s2)(s3)(s4),
          label={[font=\small\bfseries,label distance=1mm]above:Sensing Phase}] (sensephase) {};

    % =======================
    % Communication phase
    % =======================
    \node[block]     (c1) at (0,-1.9)   {Tx precoding \\design};
    \node[block]     (c2) at (3.8,-1.9) {Data transmission\\and SE evaluation};
    \node[wideblock] (c3) at (8.0,-1.9) {Rx codeword retrieval\\and post-processing};
    \node[block]     (c4) at (12.5,-1.9) {Low-overhead\\control signaling};

    \node[phase, fit=(c1)(c2)(c3),
          label={[font=\small\bfseries,label distance=1mm]below:Communication Phase}] (commphase) {};

    % =======================
    % Arrows
    % =======================
    \draw[arrow] (s1) -- (s2);
    \draw[arrow] (s2) -- (s3);
    \draw[arrow] (s3) -- (s4);

    \draw[arrow] (s4.south) -- ++(0,-0.2) -| (c4.north);
    \draw[arrow] (c4.west) -- (c3.east);
    \draw[arrow] (c1.east) -- (c2.west);
    \draw[arrow] (c3.west) -- (c2.east);
    \end{tikzpicture}
    \caption{{Operation flow of the proposed sensing-aided communication framework.}}
    \label{fig:operation_flow}
\end{figure*}
% 
% 考虑到UCA便于生成生成涡旋波以及与现有基于阵列的一体化系统的兼容性, 我们考虑一种基于UCA的OAM一体化系统, 该系统包含一个dual-function发送机和一个雷达接收机. % For the ease of generating vortex wavefronts its compatibility with array-based integrated platforms, we consider a UCA-based OAM integrated system. %一体化发送机配备有半径为R阵元数为M的UCA, 雷达接收机由半径为R, 阵元数为N的UCA构成, 且且该UCA与一体化发送机的UCA共圆心.
For the ease of generating vortex wavefronts, we consider a UCA-based integrated system.
% \footnote{{OAM transmission is closely related to UCA-MIMO, since OAM modes can be synthesized by applying a discrete Fourier transform across the elements of a uniform circular array. The difference is that conventional UCA-MIMO is usually formulated in the element or beamspace domain, whereas this work explicitly considers the OAM mode-domain channels and their mode multiplexing.}}. 
The system comprises a dual-function Tx for sensing and communication, a radar Rx, and separate communication UEs whose CSI is obtained via sensing. 
% while the communication UE is separated, whose CSI is obtained through sensing techniques. 
% 
The Tx is equipped with a UCA of $M$ elements on a circle of radius $R_t$ in the plane $z=0$, where the spherical coordinates of the $m$th element are $\mathbf{r}_m = (R_t, \phi_m, 0)$ and $\phi_m = \frac{2\pi(m-1)}{M}$, with $ m = 1, \cdots, M$.
The radar Rx employs a concentric UCA with $N$ elements, radius $R_r$, and the coordinates of the $n$th element being $\mathbf{r}_n$.
% $\mathbf{r}_n = (R_r, \phi_n, 0)$ and $\phi_n $ shares the same definition as $\phi_m$. 
% 我们假设一体化发送机和雷达接收机能够覆盖Q个动态目标, 各个目标的位置和速度可以被表示为\{()|q...\}} and \{v|q=1..\}. 其中我们假设有一个特定目标\bar{q}携带通信接收机, 能够与一体化发送机进行信息交互. 该接收机由半径为R, 阵元数为M的UCA构成.
% We consider a UCA-based integrated system with a dual-function Tx for sensing and communication and a radar Rx. The Tx is equipped with a UCA of $M$ elements on a circle of radius $R_t$, while the Rx employs a concentric UCA with $N$ elements and radius $R_r$. 
Let the $Q$ dynamic targets be indexed by $q \in \{1, \cdots, Q\}$, with spherical coordinates $\{\bm{r}_q = (r_q, \varphi_q, \vartheta_q)\}_{q=1}^{Q}$ (range, azimuth, elevation) and radial velocities $\{v_q\}_{q=1}^{Q}$.
Among them, some UEs are equipped with communication Rxs. To facilitate OAM transmission, all communication Rxs adopt the UCA architecture and are equipped with $N$ antenna elements. The dual-function Tx communicates with a designated target indexed by $\bar{q}$, whose communication Rx employs a UCA with radius $R_{r_{\bar{q}}}$.
% which receives the communication information from the integrated Tx. 
The system model is illustrated in Fig.~\ref{fig_axix_2}. 
% The communication Rx uses a UCA with $N$ elements and radius $R_{r_q}$.

% 该一体化框架被划分为感知阶段和通信阶段, 在感知阶段, 相较于现有OAM一体化方案, 我们利用基于CDMM的信号模型, 实现了Q个目标的位置及速度参数估计. 接下来, 在通信阶段, 基于第\bar{q}个目标的参数, 我们在一体化发送机和通信接收机处分别设计了预编码和后向处理矩阵, 实现了高频谱效率通信.
% In this ISAC system, the center performs sensing of $Q$ targets and communicates with some of them. 
% The sensing and communication systems share a Tx UCA with $M$ antennas. 
% while the sensing and communication Rx UCA are co-located with the Tx UCA and at the target side, respectively, with $N$ antennas.
% The Tx UCA is at the origin on the XOY plane, where the spherical coordinates of the $m$th antenna is $\mathbf{r}_m = (R_t, \phi_m, 0)$, where $R_t$, $\phi_m = \dfrac{2\pi(m-1)}{M}$, and $0$ denote the radius of the Tx UCA, the azimuth, and elevation angle of the $m$th Tx element, respectively, $ m = 1, 2, ..., M$. 
% following the description in generating vortex wavefronts in \eqref{E_m}.
The integrated framework is divided into a sensing phase and a communication phase, {where the overall joint design of the sensing-aided communication framework is illustrated in Fig.~\ref{fig:operation_flow}.}
% {as shown in Fig.~  }.
In the sensing phase,  in contrast to existing TDMM-based vortex wavefront integrated systems, we employ a CDMM scheme to estimate the position and velocity of each of the $Q$ targets. {The detailed methodology will be presented in Section~\ref{Sec:radar_processing_isac}.} {After estimating the parameters of all targets $\{\hat r_q,\hat\varphi_q,\hat\theta_q,\hat v_q\}_{q=1}^Q$, which jointly characterize the corresponding CSI, the dual-function Tx communicates with a designated target indexed by $\bar{q}$. Since the dual-function Tx and the communication Rx are physically separated, a low-overhead control signaling procedure is required prior to CSI-assisted data transmission. Through this signaling, the Tx informs the Rx of the selected OAM mode, the estimated position and velocity parameters, or equivalently, the corresponding codebook index. Upon receiving this information, the dual-function Tx and the communication Rx respectively construct the precoding and post-processing matrices based on the CSI of the $\bar{q}$-th target. This enables transmit-receive beam alignment and consequently facilitates high-rate data transmission.}
%Subsequently, in the communication phase, we focus on the communication Rx bearing target indexed by $\bar{q}$.
%Using the estimated parameters of target $\bar{q}$, i.e., the CSI associated with the $\bar{q}$th target, we design precoding and post-processing matrices at the communication Tx and Rx, respectively, to enable high SE.
%{
%Since the communication Tx and Rx are physically separated, low-overhead control signaling is required before data transmission. Through this signaling, the Tx informs the Rx of the selected OAM mode, codebook index, or associated angular parameter. The Rx then retrieves the corresponding receiving codeword from its predefined codebook and aligns its receiving beam accordingly.}
% In the sensing phase, the monostatic sensing system employs an Rx UCA with $N$ elements and radius $R_r$. The spherical coordinate of the $n$th Rx element is $\mathbf{r}_n^{(\text{rx})}$, $n = 1, 2, \ldots, N$.
% 
% Assuming that the Rx UCA receives echoes of $Q$ targets, indexed by $q = 1, 2, ..., Q$. The coordinate of the $q$th target is $(r_q, \varphi_q, \vartheta_q)$, where $r_q$, $\varphi_q$, and $\vartheta_q$ denote range, azimuth, and elevation angles, respectively. 
% 
% In the communication phase, we consider ISAC Tx communicates with the $q$th target, where the ISAC Tx transmits multi-mode vortex wavefronts embedding communication symbols.
% the $q$th target employs an Rx UCA with $N$ elements and radius ${R_r}_q$ to receive these signals and detects the symbols.

This section will first introduce the electromagnetic basis of generating vortex wavefronts, then the proposed CDMM scheme and the signal model for sensing $Q$ targets, and finally the communication model to the $\bar{q}$th target.
%%%%%%%%%%
% %
% \begin{figure}
% % \setlength{\abovecaptionskip}{-0.2cm}
% % \setlength{\belowcaptionskip}{-0.2cm}
% \begin{center}
% \includegraphics[scale=0.45]{SPAWC-2023/Fig2.pdf}%width=9cm,height=5cm scale=0.42
% \end{center}
% \caption{Coordinate of UCA Tx \textcolor{red}{(This figure has been used in SPAWC23, need to be updates)}.}
% \label{UCA_Fig2}
% \end{figure}
% %
% \vspace{-0.5cm}
\subsection{Electric Radiation Field and Vortex Wavefront}
%标题: 基于UCA的多模态涡旋波的生成
% 考虑到阵元数的限制, M阵元的UCA至多生成M个OAM模态, 即|\ell_u|<M/2 [chenTWC], 其中u = 1,..,U是OAM模态的index, and lu=一个表达式. . 然后, 由第m根发送阵元在
% \subsubsection{Electric Radiation Field of Dipole Antennas}
% We consider a unified circular array (UCA) Tx with $M$ antennas in the XOY plane of the coordinates.
% The center of the UCA is at the origin coordinates, the spherical coordinates of the $m$th antenna is $\mathbf{r}_m = (R_t, 0, \phi_m)$, where $R_t$ is the radius of the transmitting circle and $\phi_m = \dfrac{2\pi(m-1)}{M}$, with $ m = 1, 2, ..., M$ denoting the index.
An $M$-antenna Tx UCA  can generate at most $M$ vortex wavefront modes.
% , i.e., $\{ - \lfloor \frac{M}{2} \rfloor + 1, \cdots, \lfloor \frac{M}{2} \rfloor \}$.
% , $M$ is even, and $U = M$. 
% To keep the modes transmitted an even number, we neglect the maximum negative mode $-\dfrac{M}{2}$ in practice. 
Using $u$ to denote the index of mode, i.e., $u = 1, \cdots, U$, the undistorted mode ${\ell}_u$ satisfies $|{\ell}_u| < \frac{M}{2}$\footnote{The mode index $u$ is positive, while the mode $\ell_u$ takes negative and positive values symmetrically, e.g., $M = 5$, index $u = \{1, \cdots, 5\}$ and the mode $\ell_u = \{-2, \cdots, 2\}$.}~\cite{5345758}.
% the $u$th mode ${\ell}_u = (u-1)  - \frac{M}{2} $, i.e., 
% ${\ell}_u = u - \lfloor \dfrac{M}{2} \rfloor $.
In particular, to generate the $u$th vortex mode, the $m$th Tx element is modulated by $ F_{m, u} = e^{ i 2\pi \frac{(m - 1)}{M} {\ell}_u } = e^{  i \phi_m {\ell}_u } $.
% , where $F_{m,u}$ is entry of the Hermitian transpose of the (partial) discrete Fourier transform (DFT) matrix $ \mathbf{F}^{H} \in \mathbb{C}^{M \times U}$.  
% 
% To generate the vortex wavefronts, $M$ elements in the Tx UCA are modulated by the Hermitian transpose of the (partial) discrete Fourier transform (DFT) matrix $ \mathbf{F}^{H} \in \mathbb{C}^{M \times U}$, with 
% the $u$th mode of the $m$th antenna element is modulated by 
% entry $F_{m,u} F_{m, u} = e^{ i 2\pi \frac{m - 1}{M} {\ell}_u } = e^{  i \phi_m {\ell}_u } $.
% is 
% \begin{align} \label{eq:DFT_1}
%     F_{m, u} = e^{ i 2\pi \frac{m - 1}{M} {\ell}_u } = e^{  i \phi_m {\ell}_u }.
% \end{align}
% 
% In particular, 
The electric field $\bm{E}_{T,u}$ at any point in the space with coordinates $\mathbf{r}_0 = (r_0, \varphi_0, \vartheta_0)$ is given by \cite{5345758}  
% , the electric field $\bm{E}_T$ observed at point $\mathbf{r}_0 = (r_0, \vartheta_0, \varphi_0)$ is the superposition of $M$ antennas, 
%
\begin{align} \label{Et}
\bm{E}_{T,u} &= \sum_{m=1}^{M} \bm{E}_m  F_{m, u}
\approx \beta_t \frac{e^{ik_0 r_0}}{r_0} \sum_{m=1}^{M} e^{-i(\bm{k}\cdot \mathbf{r}_m-{\ell}_u \phi_m)} \nonumber\\
& \approx \beta_t \dfrac{e^{ik_0 r_0}}{r_0} M e^{i{\ell}_u \varphi_0} i^{-{\ell}_u}{J_{{\ell}_u}}(k_0 R_t \sin \vartheta_0),
\end{align}
where $\bm{E}_m \approx \beta_t \frac{e^{-ik_0 r_0}}{r_0}  e^{i \bm{k} \cdot \mathbf{r}_m }$ is the electric far-field radiated by the $m$th antenna \cite{balanis2016,6975067}, the scalar constant $\beta_t$ models all the constants relative to each Tx antenna, $i$ is the imaginary unit, $k_0$ is the wavenumber and $\bm{k} = k_0 \hat{\mathbf{r}}_0 $ with $\hat{\mathbf{r}}_0 = \frac{ \mathbf{r}_0 }{ |\mathbf{r}_0| }$ denoting the wave propagation direction.  
% Frequency $f$, can be denoted by angular frequency $\omega = 2\pi f$ and wavenumber $\bm{k} = \hat{\bm{k}} k$ with direction $\hat{\bm{k}}$, $k = 2\pi/\lambda$, and wavelength $\lambda = c/f$, where $c$ is the speed of light. 
$J_{{\ell}_u}(k_0 R_t\sin\vartheta_0)=\frac{i^{{\ell}_u}}{2\pi}\int_0^{2\pi}e^{-ik_0 R_t\sin\vartheta_0\cos\phi_0'}e^{-i{\ell}_u\phi_0'}d\phi'_0$ is the ${{\ell}_u}$th-order Bessel function and $\phi_0'=\varphi_0-\phi_m$. 
In \eqref{Et}, we assume far-field observation, and the helical phase $e^{i{\ell}_u \varphi_0}$ arises from the azimuth-dependent excitation $e^{  i \phi_m {\ell}_u }$ together with the propagation phase $e^{-i\bm{k}\cdot \mathbf{r}_m}$. 
To avoid the spatial aliasing in UCA arrays, the arc length between two neighboring antennas should be less than $\frac{\lambda}{2}$ \cite{4012435}, i.e., $R_t \frac{2 \pi}{M} \leq \frac{\lambda}{2}$, hence $ R_t \leq \frac{M\lambda}{4 \pi }$. 
\subsection{Sensing Signal Model}
We adopt an OFDM waveform for the ISAC design.
% To achieve a unified design for both sensing and communications, we employ the OFDM waveform.
% \footnote{The proposed CDMM is not limited to OFDM, which also suits conventional radar waveforms, such as FMCW.}. 
During a coherent processing interval (CPI), there are $P$ OFDM symbols; $P_\text{sen} \leq P$ of them are used for sensing, with $p = 1, 2, \cdots, P_\text{sen}$ indexing these symbols.
Each symbol has duration $T_c$ and consists of $L$ subcarriers with spacing $\Delta f$.
In the complex baseband representation, the frequency of the $l$th subcarrier is $ f_l =  (l-1) \Delta f$, $l =1, 2, \cdots, L$.
The corresponding radio frequency (RF) subcarrier frequencies are $ f_c + f_l$, where $f_c$ is the carrier frequency. 
% band is $k_l = \dfrac{2 \pi f_l}{c}$, where $ f_l =  f_c + (l-1) \Delta f$, 
% 
\subsubsection{Precoding Scheme of CDMM}
% To achieve a unified design for both radar and communications, we employ the OFDM waveform\footnote{The proposed CDMM is not limited to OFDM, which also suits conventional radar waveforms, such as FMCW.}. 
% Using $P$ to denote the number of symbols in one coherent interval (CPI) of the sensing phase, with $p = 1, 2, \cdots, P$ denoting the index, the duration is $T_c$. 
% Each symbol consists of $L$ subcarriers, with $l =1, 2, \cdots, L$ denoting the index, and the sub-carrier is $\Delta f$.
% The wave number of the $l$th band can be represented as in \eqref{E_m}, $ k_l = \dfrac{2 \pi f_l}{c} = \dfrac{2 \pi f_c + (l-1) \Delta f}{c} $, where $f_c$ is carrier frequency. 
% 
% As indicated by \eqref{Et}, 
% To generate the vortex wavefront of the $u$th mode, all $M$ antennas transmit the same baseband signals $s_{u}(p, l)$, while the $m$th antenna is modulated by $e^{i {\ell}_{u}\phi_m}$, i.e., $s_{u}(p, l) e^{i {\ell}_{u}\phi_m}$. 
% 
State-of-the-art OAM-based sensing works mainly employ TDMM to separate modes \cite{9968273,10304513,10620284}, where each symbol transmits a single mode and the symbol index is treated as the slow-time index. 
% s are transmitted in different symbols within one CPI. 
{Different from these works, we propose a slow-time precoding scheme for mode-multiplexing to make full use of the mode degree of freedom, where the UCA transmits $U$ modes simultaneously.}
The transmitted signal at the $m$th antenna on the subcarrier $l$ of the $p$th symbol is
% The equivalent baseband signal of the $m$th antenna $x_m(p,l)$ therefore is given as
\begin{equation}
\begin{aligned}  \label{eq:signal_1}
x_m(p,l)
=\mathbf{W}[\langle p\rangle_U,:]\tilde{\mathbf{s}}_U^m(p,l) \!=\!\sum_{u=1}^{U}s_u(p,l)e^{i\ell_u\phi_m}w_{p,u},
\end{aligned}
\end{equation}
where 
$\tilde{\mathbf{s}}_U^m(p,l) = \left[ s_{u}(p, l) e^{i {\ell}_{u}\phi_m} \right]_{U \times 1} $
% $\tilde{\mathbf{s}}_U^m(p,l) \in \mathbb{C}^{U \times 1}$ 
is the stacking of the symbol corresponding to the $U$ modes into a vector; 
% as  
% \begin{align}
% \tilde{\mathbf{s}}^m_U&(p,l)  = 
% % \begin{bmatrix}
% %         s_{u = 1}(p, l) e^{i \ell_{u=1}\phi_m } \notag  \\ 
% %          s_{u = 2}(p, l) e^{i \ell_{u=2}\phi_m }  \notag \\ 
% %         \vdots \\ 
% %           s_{u = U}(p, l) e^{i \ell_{u=U}\phi_m }  \notag \\
% % \end{bmatrix},
% % \\ \nonumber
% % &
% \begin{bmatrix}
%         s_{1}(p, l) e^{i \ell_{1}\phi_m }, 
%         s_{2}(p, l) e^{i \ell_{2}\phi_m }, 
%         \cdots ,
%           s_{U}(p, l) e^{i \ell_{U}\phi_m } 
% \end{bmatrix}^T, 
% \end{align}
$\mathbf{W} \in \mathbb{R}^{U \times U}$ is chosen as a scaled unitary precoding matrix, satisfying $\mathbf{W}^T \mathbf{W} = U \mathbf{I}_U$, with $w_{p,u}$ denoting the entry of the $\langle p \rangle_{U}$th row and $u$th column, $ \langle p \rangle_{U} = \left( \left( p-1 \right) \pmod{U} +1 \right)$. 
% \footnote{
% Mode-division multiplexing separates signals by OAM mode index $u$ across slow time $p$, whereas conventional MIMO radar separates signals by transmit channels.
% % Mode division multiplexing aims to separate signals of different modes, which differs from MIMO radar, whose goal is to separate signals from different Txs. Hence, mode division multiplexing is in terms of slow time $p$, and mode index $u$.
% }, i.e., the $\overline{p}$th raw and the $u$th column of coding matrix $ \mathbf{W} \in \mathbb{R}^{U \times U} $, where $\overline{p} = \left( p \bmod U \right)$. 
 
Design examples of $\mathbf{W}$:
% \begin{equation} \label{eq:ch_1_3}
% \mathbf{W} = 
% \begin{cases}
% \mathbf{I}_{U},   & \text{Slow-time TDMM precoding} \\
% \mathbf{W}^{\text{HM}}_{2^\kappa},   & \text{Slow-time CDMM precoding}
% \end{cases}  
% \end{equation}
%     \begin{equation}
%         \begin{aligned} \label{eq:bpm_wight_2}
%             \mathbf{W}^{\text{HM}}_{2}   = 
%         \begin{bmatrix}
%              1& 1\\
%               1&-1
%         \end{bmatrix}.
%             \end{aligned}
%     \end{equation}
% %%%%%%%%%%%%%%%%%%%%%%%%%%%%%%%%%%%%%%%%%%%%%%%%
% 
\begin{itemize}
    \item 
    \textbf{TDMM case}: With $\mathbf{W} = \mathbf{I}_U$ and the active mode index $u = \langle p \rangle_{U} $, the transmitted signal becomes 
    % \begin{equation} 
    %     w_{p,u} = 
    %     \begin{cases}
    %     1,   & \left( p \bmod U \right) = u \\
    %     0.   & \text{Others}
    %     \end{cases}  
    % \end{equation}
    \begin{equation}
        \begin{aligned} \label{eq:TMD_1}
        x_m(p,l) &= s_{u}(p, l) e^{i{\ell}_{u}\phi_m},   
        \end{aligned}
    \end{equation}
    which takes the form of the TDMM scheme in the literature~\cite{9968273}.
    % where the modes are separated by time-slot $p$. 
    In this sense, TDMM is a special case of CDMM. 
    \item 
    \textbf{CDMM case}: With $\kappa = \log_2 {U}, \kappa \in \mathbb{Z}^{+}$, we let $\mathbf{W} = \mathbf{W}^{\text{HM}}_{2^\kappa}$, where $\mathbf{W}^{\text{HM}}_{2^\kappa}$ are Hadamard codes obtained using the standard recursion~\cite{hedayat1978hadamard}
% %    the entries $w_{p,u} \in \{ \pm 1 \}$ are slow-time orthogonal over any $U$-length block.
%
     % ,  $ \mathbf{W}^{\text{HM}}_{2^\kappa} \in \mathbb{R}^{U \times U}$ is an orthogonal codebook. In this paper, we use the Hadamard matrix as an example \cite{hedayat1978hadamard},  
%We use the standard recursion Hadamard codes as \cite{hedayat1978hadamard}
\begin{equation}
        \begin{aligned} \label{eq:bpm_wight_1}
            \mathbf{W}^{\text{HM}}_{2^\kappa}  = \mathbf{W}^{\text{HM}}_{2} \otimes \mathbf{W}^{\text{HM}}_{2^{\kappa-1}},
        \end{aligned}
        \ \text{and} \ 
        \begin{aligned}  
         \mathbf{W}^{\text{HM}}_{2}   = 
        \begin{bmatrix}
             1& 1\\
              1&-1
        \end{bmatrix}.
            \end{aligned}
    \end{equation}
    % with $\kappa \geq 2$.
%   with $\kappa = \log_2 {U}, \kappa \in \mathbb{Z}^{+} $.  
    \end{itemize}

\subsubsection{Sensing Echo Model}
%
% Then, the RF vortex wavefront signal radiating by the $m$-Tx UCA is $x_m(p, l) e^{-i\bm{k}_l \cdot \mathbf{r}_m }$.
% \begin{align} \label{eq:cdm_0_1}
% \overline{x}_m(p, l) & = x_m(p, l) e^{-i\bm{k}_l \cdot \mathbf{r}_m } . 
% \end{align}
%
% For a monostatic radar system employing a UCA with $N$ elements and a radius of $R_r$, the coordinate of the $n$th Rx antenna element is denoted by $\mathbf{r}_n^{(\text{rx})}$, where $n = 1, 2, \ldots, N$ indexes the receiving antennas.
% % 
% Assuming that the Rx UCA receives echoes of $Q$ targets, $q = 1, 2, ..., Q$ denotes index of targets, the coordinates of the $q$th target is $(r_q, \varphi_q, \vartheta_q)$, where $r_q$, $\varphi_q$, $\vartheta_q$ denotes range, azimuth, and elevation angles, respectively. 
% 
% According to \eqref{Et}, 
The radiated RF signal by the Tx UCA for the $p$th symbol in the time domain is 
\begin{equation}
\begin{aligned}
    x_t^{\text{RF}}(p,t_{s}) \! = \! \sum_{m=1}^{M} \! \big [\mathcal{F}_L^{-1}\{ \mathbf{x}_m(p)\} \big]_{s} e^{i 2 \pi f_c t_s} \! \triangleq \! x_t^{\text{rad}}(p,t_{s})e^{i 2 \pi f_c t_s},
\end{aligned} 
\end{equation}
where $\mathbf{x}_m(p) \triangleq [x_m(p,1),\ldots,x_m(p,L)]^{T}$ collects symbols of $L$ subcarriers, $[\mathcal{F}_L^{-1}\{\cdot\}]_{s}$ denotes the $s$th time domain sample of the $L$-point inverse fast Fourier transform (IFFT), with $t_s =s T_s$ and $T_s$ being the sampling period, and $x_t^{\text{rad}}(p,t_{s})$ is the baseband signal. 

For the $q$th target, the round-trip distance is $2 r_q(v_q)$.
After downconversion to complex baseband, low-pass filtering and cyclic prefix removal, the resulting echo from $Q$ targets at the $n$th Rx antenna is 
% The corresponding baseband echo of $Q$ targets at the $n$th receive antenna is
\begin{equation}
\begin{aligned} \label{eq:yt_time_model}
y_{t,n}^{\text{rad}}(p,t_s)
\!= \!\sum_{q=1}^{Q} \dfrac{\beta \sigma_q}{r_q^2}  x_t^{\text{rad}}\left(p,t_s \!-\! \frac{2 r_q}{c}\right) e^{j 2\pi f_{D,q} t_s} \!+\! n_t(p,t_s), 
\end{aligned} 
\end{equation}
where $\beta=\beta_t\beta_r$, $\beta_r$ models all the constants related to the Rx antenna, $\sigma_q$ denotes the radar cross section (RCS) of the $q$th target, $f_{D,q} = \frac{-2v_q}{c}f_c$ is the Doppler frequency shift, $c$ is the speed of light, and $n_t(p,t_s)$ is additive complex Gaussian noise.

The received OFDM symbol in the subcarrier domain is obtained by applying
an $L$-point fast Fourier transform (FFT) over the duration $[(p-1)T_c,\, pT_c]$.
For the $l$th subcarrier, we have $y_n^{\text{rad}}(p,l)
= \frac{1}{T_c} \int_{(p-1)T_c}^{pT_c}
y_{t,n}^{\text{rad}}(p,t)\, e^{-i 2\pi f_l t}\, dt$. In practice, the continuous-time signal is sampled at $t_s$. By substituting \eqref{eq:yt_time_model} and exploiting the orthogonality of the subcarriers, we have\footnote{We do not exploit any unique features of OAM in} the derivation of sensing echoes.
Ignoring ICI, the Doppler term matches conventional MIMO radar \cite{10559769} and appears as a slow-time phase $e^{-i2 \pi {f_{D,q}} (p-1) T_c}$ that depends only on the symbol interval $T_c$. 
% Ignoring ICI, the Doppler term takes the same form as general MIMO radar echoes, where the velocities of the targets are revealed on the phase as \cite{10559769}.}  
\begin{align} \label{eq:cdm_2}
{y}_n^\text{rad}(p, l) \approx 
&  \sum_{q=1}^{Q}  \frac{ \beta \sigma_q e^{i 2k_lr_q } e^{-i2 \pi {f_{D,q}} (p-1) T_c} }{r_q^2} e^{-i\bm{k}_{l, q} \cdot \mathbf{{\overline{r}}}_n}  \nonumber \\   & \times \sum_{m=1}^{M} x_m(p, l) e^{-i\bm{k}_{l, q} \cdot \mathbf{r}_m } + n_\text{rad}(p,l) ,
\end{align}
%
% \begin{align} \label{eq:cdm_6}
% f_D = \dfrac{-2v}{c}f_s,
% \end{align}
where $\bm{k}_{l, q} = \frac{2 \pi \left( f_c + f_l \right)}{c}\hat{\bm{r}}_q$ with $\hat{\bm{r}}_q = \frac{\bm{r}_q}{|\bm{r}_q|}$, $\mathbf{\overline{r}}_n $ is the position vector of the $n$th Rx, and $n_\text{rad}(p,l)$ denotes complex Gaussian noise.
% approximation $\overset{(a)}\approx$ is defined in \eqref{E_m}, $q = 1, 2, ..., Q$ denotes the target index, 
% $\sigma_q$ denotes the radar cross section (RCS)
% and velocity of the $q$th target, respectively, 
\begin{figure*}[ht]
\begin{align} \label{eq:radar_model}
{y}_n^\text{rad}(p, l)  
= & \sum_{q=1}^{Q}  \frac{ \beta \sigma_q e^{i 2k_lr_q } e^{-i2 \pi f_{D,q} (p-1)T_c }}{r_q^2} 
   e^{-i\bm{k}_l \cdot \mathbf{\overline{r}}_n} 
   \sum_{m=1}^{M} \sum_{u=1}^{U}  
   s_{u}(p, l) e^{i{\ell}_{u}\phi_m} w_{p,u} e^{-i\bm{k}_l \cdot \mathbf{r}_m }  
   + n_\text{rad}(p,l) \nonumber\\
\overset{(a)}=& \sum_{q=1}^{Q}  
   \frac{ \beta \sigma_q e^{i 2k_lr_q } e^{-i2 \pi f_{D,q} (p-1)T_c }}{r_q^2} 
   e^{-i\bm{k}_l \cdot \mathbf{\overline{r}_n} } 
   \sum_{u=1}^{U} s_{u}(p, l)'     
   \sum_{m=1}^{M} e^{i{\ell}_{u} \phi_m} e^{-i\bm{k}_l \cdot \mathbf{r}_m } 
   + n_\text{rad}(p,l) \nonumber\\
\overset{(b)}\approx& \sum_{q=1}^{Q}  
   \frac{ \beta \sigma_q e^{i 2k_lr_q } e^{-i2 \pi f_{D,q} (p-1)T_c} }{r_q^2} 
   e^{-i\bm{k}_l \cdot \mathbf{\overline{r}}_n } 
   \sum_{u=1}^{U} s_{u}(p, l)'   
   M e^{i{\ell}_u \varphi_q} i^{-{\ell}_u} 
   J_{{\ell}_u}\!\left(k_l R_t \sin{\vartheta_q}\right) + n_\text{rad}(p,l).
\end{align}
\vspace{-0.5cm}
\begin{align} \label{eq:cdm_4}
{y}^\text{rad}  (p, l) = & \sum_{n=1}^{N} {y}_n^\text{rad}(p, l)  = \sum_{n=1}^{N} \sum_{q=1}^{Q}  \frac{ \beta \sigma_q e^{i 2k_lr_q } e^{-i2 \pi f_{D,q} (p-1)T_c} }{r_q^2} e^{-i\bm{k}_l \cdot \mathbf{\overline{r}}_n } \sum_{ u = 1 }^{ U } s_{u}(p, l)'   M e^{i{\ell}_u \varphi_q} i^{-{\ell}_u}{J_{{\ell}_u}}(k_l R_t\sin{\vartheta_q}) + n_\text{rad}(p,l)  \nonumber \\
 \approx &  \sum_{q=1}^{Q}  \frac{ \beta_1 \sigma_q e^{i 2k_lr_q } e^{-i2 \pi f_{D,q} (p-1)T_c} }{r_q^2} {J_{0}}(k_l R_r\sin{\vartheta_q})   \sum_{ u = 1 }^{ U }  e^{i{\ell}_u \varphi_q} i^{-{\ell}_u}{J_{{\ell}_u}}(k_l R_t\sin{\vartheta_q}) s_{u}(p, l)' + n_\text{rad}(p,l).
\end{align}
\vspace{-0.5cm}
\noindent\hrulefill
\end{figure*}
\begin{rem}
    {\em} 
    The Doppler term in \eqref{eq:yt_time_model} contains two effects after OFDM demodulation. 
    The first one is the intra-symbol Doppler phase rotation within one OFDM symbol, which is called inter-carrier interference (ICI)~\cite{9529026}. 
    With the configurations in Table~\ref{tab:simu_ISAC_configs_1}, the normalized Doppler ratio $\frac{|f_{D,q}|}{\Delta f}$ remains relatively small. For example, at $v_q=50~\mathrm{m/s}$ and $f_c=77~\mathrm{GHz}$, we have $\frac{|f_{D,q}|}{\Delta f} \approx 0.016$. Following the classical OFDM frequency-offset analysis in~\cite{328961}, ICI is expected to be limited in this operating regime. Thus, \eqref{eq:cdm_2} adopts the commonly used no-ICI approximation while retaining the slow-time Doppler phase.
    
    The second effect is the inter-symbol Doppler phase evolution across slow-time symbols, represented by $e^{-j2\pi f_{D,q}(p-1)T_c}$ in \eqref{eq:cdm_2}, which is retained in the model. This slow-time Doppler phase is the source of the Doppler-induced disturbance in the subsequent CDMM mode decoding analysis.
\end{rem}

% \begin{align} 
% {y}_n^\text{rad}&(p, l)  =  \sum_{q=1}^{Q}  \frac{ \beta \sigma_q e^{i 2k_lr_q } e^{-i2 \pi {f_{D,q} (p-1)T_c} }}{r_q^2} e^{-i\bm{k}_l \cdot \mathbf{r}_n } \times \nonumber\\
% & \sum_{m=1}^{M} \sum_{ u = 1 }^{ U }  s_{u}(p, l) e^{i{\ell}_{u}\phi_m} w_{p,u} e^{-i\bm{k}_l \cdot \mathbf{r}_m }  + n_\text{rad}(p,l) \nonumber\\
% \overset{(a)}= & \sum_{q=1}^{Q}  \frac{ \beta \sigma_q e^{i 2k_lr_q } e^{-i2 \pi {f_{D,q} (p-1)T_c} }}{r_q^2} e^{-i\bm{k}_l \cdot \mathbf{r}_n } \times\nonumber\\
% & \sum_{ u = 1 }^{ U } s_{u}(p, l)'     \sum_{m=1}^{M} e^{i{\ell}_{u} \phi_m  }   e^{-i\bm{k}_l \cdot \mathbf{r}_m } + n_\text{rad}(p,l) \nonumber\\
% \overset{(b)}\approx &   \sum_{q=1}^{Q}  \frac{ \beta \sigma_q e^{i 2k_lr_q } e^{-i2 \pi f_{D,q} (p-1)T_c} }{r_q^2} e^{-i\bm{k}_l \cdot \mathbf{r}_n } \times \nonumber\\
% & \sum_{ u = 1 }^{ U } s_{u}(p, l)'   M e^{i{\ell}_u \varphi_q} i^{-{\ell}_u}{J_{{\ell}_u}}(k_l R_t \sin{\vartheta_q}) + n_\text{rad}(p,l),
% \end{align}
% \end{figure*}
Substituting \eqref{eq:signal_1} into \eqref{eq:cdm_2}, we have \eqref{eq:radar_model}, 
where operation $\overset{(a)}=$ defines the product of the transmitted signal and phase code $s_{u}(p, l) w_{p,u}$ is defined as new signal as $s_{u}(p, l)'$ and interchanges the summations over $m$ and $u$.
Operation $\overset{(b)}\approx$ utilizes the same approximation as in \eqref{Et}. 
% the approximation $\overset{(a)}\approx$ is defined in \eqref{E_m}, 
% $\vartheta_q$ and $\varphi_q$ denote the azimuth and elevation angles of the $q$th target.
Subsequently, \eqref{eq:cdm_4} indicates the superposition of the $N$ Rx antennas, which is equivalent to a zero-mode operation, where $\beta_1 = \beta MN$.

\subsection{Communication Signal Model}
\subsubsection{LoS MIMO Channel}
The line-of-sight (LoS) channel between the $m$th ISAC Tx antenna and the $n$th communication Rx antenna of the target $\bar{q}$ is  
\begin{equation} \label{FreeSpaceChannel}
h^{\bar{q}}_{n,m}(p,l)=\frac{\beta}{2k_l  d^{\bar{q}}_{n,m}(p)}e^{-ik_l d^{\bar{q}}_{n,m}(p)},
\end{equation}
where
% is fully determined by the carrier wave number $k$ and 
$d^{\bar{q}}_{n,m}(p)$ denotes the physical propagation distance.
% between the $m$th Tx antenna and the $n$th Rx antenna at the $p$th symbol.
% given the array parameter $\beta$. 
% 
\begin{figure*}[ht]
\begin{align}\label{dmn}
d^{\bar{q}}_{n,m}(p)&=\big\{R_t^2+R^2_{r_{\bar{q}}}+\left[r_{\bar{q}}+(p-1)T_c v_{\bar{q}}\right]^2 +2\left[r_{\bar{q}}+(p-1)T_c v_{\bar{q}}\right]R_{r_{\bar{q}}}\sin\vartheta_{\bar{q}}\cos(\varphi_{\bar{q}}-\alpha^{\bar{q}}_n) \nonumber\\
& \qquad\qquad\quad -2\left[r_{\bar{q}}+(p-1)T_c v_{\bar{q}}\right]R_t\sin\vartheta_{\bar{q}}\cos(\varphi_{\bar{q}}-\phi_m) -2R_tR_{r_{\bar{q}}}\cos(\alpha^{\bar{q}}_n-\phi_m)\big\}^{1/2}.
\end{align}    
\vspace{-0.5cm}
\begin{align} \label{dmn appx}
d^{\bar{q}}_{n,m}(p) &\overset{(a)}{\approx} \sqrt{R_t^2+R^2_{r_{\bar{q}}}+\left[r_{\bar{q}}+(p-1)T_c v_{\bar{q}}\right]^2}\ + \frac{\left[r_{\bar{q}}+(p-1)T_c v_{\bar{q}}\right]R_{r_{\bar{q}}}\sin\vartheta_{\bar{q}}\cos(\varphi_{\bar{q}}-\alpha^{\bar{q}}_n)}{\sqrt{R_t^2+R^2_{r_{\bar{q}}}+\left[r_{\bar{q}}+(p-1)T_c v_{\bar{q}}\right]^2}}\ \nonumber\\
&\qquad\qquad\quad - \frac{\left[r_{\bar{q}}+(p-1)T_c v_{\bar{q}}\right]R_t\sin\vartheta_{\bar{q}}\cos(\varphi_{\bar{q}}-\phi_m)}{\sqrt{R_t^2+R^2_{r_{\bar{q}}}+\left[r_{\bar{q}}+(p-1)T_c v_{\bar{q}}\right]^2}}\ - \frac{R_tR_{r_{\bar{q}}}\cos(\alpha^{\bar{q}}_n-\phi_m)}{\sqrt{R_t^2+R^2_{r_{\bar{q}}}+\left[r_{\bar{q}}+(p-1)T_c v_{\bar{q}}\right]^2}}\nonumber\\
&\overset{(b)}{\approx} \left[r_{\bar{q}}+(p-1)T_c v_{\bar{q}}\right]+R_{r_{\bar{q}}}\sin\vartheta_{\bar{q}}\cos(\varphi_{\bar{q}}-\alpha^{\bar{q}}_n) -R_t\sin\vartheta_{\bar{q}}\cos(\varphi_{\bar{q}}-\phi_m) -\frac{R_tR_{r_{\bar{q}}}\cos(\alpha^{\bar{q}}_n-\phi_m)}{r_{\bar{q}}+(p-1)T_c v_{\bar{q}}}.
\end{align}
\vspace{-0.5cm}
\begin{align} \label{hmn}
h^{\bar{q}}_{n,m}(p,l) & \overset{(a)}\approx \frac{\beta}{2k_l \left[r_{\bar{q}}+(p-1)T_c v_{\bar{q}}\right]} \exp \big\{\!-ik_l\left[r_{\bar{q}}+(p-1)T_c v_{\bar{q}}\right]-i{k_l R_{r_{\bar{q}}}}\sin\vartheta_{\bar{q}}\cos(\varphi_{\bar{q}}-\alpha^{\bar{q}}_n) \nonumber\\
& \quad\quad\quad +i{k_l R_t}\sin\vartheta_{\bar{q}}\cos(\varphi_{\bar{q}}-\phi_m)+i\frac{k_l R_t R_{r_{\bar{q}}}\cos(\alpha^{\bar{q}}_n-\phi_m)}{r_{\bar{q}}+(p-1)T_c v_{\bar{q}}} \big\}.
\end{align}
\vspace{-0.5cm}
\noindent\hrulefill
\end{figure*}

The static distance in the \emph{off-axis misalignment case}, as shown in Fig.~\ref{fig_axix_2}, has been studied in \cite{9442905}, following which the time-varying $d^{\bar{q}}_{n,m}(p)$ is derived as \eqref{dmn},
%
% \begin{align}\label{dmn}
% d^q_{n,m}(p)&=\big\{R_t^2+R^2_{r_q}+\left[r_q+(p-1)T_c v_q\right]^2\nonumber\\ 
% &+2\left[r_q+(p-1)T_c v_q\right]R_{r_q}\sin\vartheta_q\cos(\varphi_q-\alpha^q_n) \nonumber\\
% &-2\left[r_q+(p-1)T_c v_q\right]R_t\sin\vartheta_q\cos(\varphi_q-\phi_m)\nonumber\\
% &-2R_tR_{r_q}\cos(\alpha^q_n-\phi_m)\big\}^{1/2},
% \end{align}
%
% where $(r_q,\vartheta_q,\varphi_q)$ is the coordinates of the target $\bar{q}$, relative to the integrated OAM Tx, $v_q$ is the corresponding velocity at the $p$th slow time, $R_t$ and 
where $\alpha^{\bar{q}}_n=[\frac{2\pi(n-1)}{N}]$.
% $m=1,2,\cdots,N$, 
% $\phi_0$ and $\alpha_0$ denote the corresponding initial azimuth angles of the reference elements in both Tx and Rx UCAs, respectively. 
% For easier analysis, we assume $\phi_0=0$ and $\alpha_0=0$ in this paper. 
% {WX: Here, except for $\alpha^p_m$, the other parameters should have already appeared in the integrated OAM transmitted signal and the OAM radar echo signal. Please ensure that the parameters that have been introduced are removed during the integration process.} 
In far-field communication scenarios, $r_{\bar{q}}\gg R_t$ and $r_{\bar{q}}\gg R_{r_{\bar{q}}}$, we can approximate $d^{\bar{q}}_{n,m}(p)$ as \eqref{dmn appx},
%
% \begin{align} \label{dmn appx}
% &d^q_{n,m}(t)\overset{(d)}{\approx} \sqrt{R_t^2+R^2_{r_q}+\left[r_q+(p-1)T_c v_q\right]^2}\ +\nonumber\\
% &\qquad\qquad\quad\frac{\left[r_q+(p-1)T_c v_q\right]R_{r_q}\sin\vartheta_q\cos(\varphi_q-\alpha^q_n)}{\sqrt{R_t^2+R^2_{r_q}+\left[r_q+(p-1)T_c v_q\right]^2}}\ -\nonumber\\
% &\qquad\qquad\quad\frac{\left[r_q+(p-1)T_c v_q\right]R_t\sin\vartheta_q\cos(\varphi_q-\phi_m)}{\sqrt{R_t^2+R^2_{r_q}+\left[r_q+(p-1)T_c v_q\right]^2}}\ -\nonumber\\
% &\qquad\qquad\quad\frac{R_tR_{r_q}\cos(\alpha^q_n-\phi_m)}{\sqrt{R_t^2+R^2_{r_q}+\left[r_q+(p-1)T_c v_q\right]^2}}\nonumber\\
% &\overset{(e)}{\approx} \left[r_q+(p-1)T_c v_q\right]+R_{r_q}\sin\vartheta_q\cos(\varphi_q-\alpha^q_n)\nonumber\\
% &\quad-R_t\sin\vartheta_q\cos(\varphi_q-\phi_m) -\frac{R_tR_{r_q}\cos(\alpha^q_n-\phi_m)}{r_q+(p-1)T_c v_q},
% \end{align}
%
where $\overset{(a)}{\approx}$ {follows by completing the square and using the condition}
% uses the method of completing a square and the condition 
$r_{\bar{q}}\gg R_t, R_{r_{\bar{q}}}$ as same as the simple case $\sqrt{a^2-2b}\approx a-\frac{b}{a}, a\gg b$; $\overset{(b)}{\approx}$ is directly obtained from the condition $r_{\bar{q}}\gg R_t, R_{r_{\bar{q}}}$. Then, substituting \eqref{dmn appx} into \eqref{FreeSpaceChannel}, we have \eqref{hmn},
%
% \begin{align} \label{hmn}
% &h^q_{n,m}(p,l) \overset{(f)}\approx \frac{\beta}{2k_l \left[r_q+(p-1)T_c v_q\right]}\times \nonumber\\
% &\exp \big\{\!-ik_l\left[r_q+(p-1)T_c v_q\right]-i{k_l R_{r_q}}\sin\vartheta_q\cos(\varphi_q-\alpha^q_n) \nonumber\\
% &+i{k_l R_t}\sin\vartheta_q\cos(\varphi_q-\phi_m)+i\frac{k_l R_t R_{r_q}\cos(\alpha^q_n-\phi_m)}{r_q+(p-1)T_c v_q} \big\},
% \end{align}
%
where $\overset{(a)}\approx$ neglects the trigonometric-function terms in the denominator and thus only $2k_l\left[r_{\bar{q}}+(p-1)T_c v_{\bar{q}}\right]$ is left. 
In this way, the free space channel matrix from the ISAC Tx to the communication Rx at the target ${\bar{q}}$ can be expressed as $\mathbf{H}_{\bar{q}}(p,l)=[h^{\bar{q}}_{n,m}(p,l)]_{N\times M}$. 
% Appendix~\ref{App:OAM_channel}
\begin{rem}
    {\em} This paper considers $M=N$, while the case $M \neq N$ is more complicated and has been discussed in \cite{10480418}.  
    When $\vartheta_{\bar{q}} =0$ and $\varphi_{\bar{q}} =0$, $\mathbf{H}_{\bar{q}}$ is a circulant matrix that can be decomposed by the $M$-dimensional Fourier matrix $\mathbf{F}_M$ as $\mathbf{H}_{\bar{q}} = \mathbf{F}_M^H \mathbf{\Lambda} \mathbf{F}_M $, where $\mathbf{\Lambda}$ is a diagonal matrix denoting eigenvalues of $\mathbf{H}_{\bar{q}}$. However, because of the non-zero $\vartheta_{\bar{q}}$ and $\varphi_{\bar{q}}$, we need to design the beamforming and beam steering at the Tx and Rx sides, respectively, to diagonalize the equivalent channel in the communication signal model.
\end{rem}

\subsubsection{Signal Model}
% \textcolor{red}{
% \textbf{change subtile: Effective OAM Channel and Signal Model}
% Yuan: would suggest, after introducing the channel model \eqref{hmn}; define effective OAM channel here , i.e., \eqref{effectivechannel} and \eqref{effectivechannel2}. Then introducing equation \eqref{y} and \eqref{xTD}.}
Similar to conventional MIMO communications, we apply the matrix $\mathbf{T}(p,l)\in\mathbb{C}^{M\times U}$ at the Tx to precode the multi-mode signals $\mathbf{s}(p,l)\in\mathbb{C}^{U \times 1}$. Thus, the received signal vector at the target $\bar{q}$ can be expressed as
\begin{align} \label{y}
\mathbf{y}_{\bar{q}}(p,l)&=\mathbf{H}_{\bar{q}}(p,l)\mathbf{T}(p,l)\mathbf{s}(p,l)+\mathbf{n}_\text{com}^{\bar{q}}(p,l),
\end{align}
where $\mathbf{s}(p,l)=[s_{u}(p,l)]_{U \times 1}$ includes the communication symbols modulated over $U$ modes,
%, $s_{u}(p,l)$ shares the same definition as in \eqref{eq:signal_1}\footnote{In the sensing period, $\mathbf{s}(p,l)$ contains deterministic waveform. In the communication period, $\mathbf{s}(p,l)$ contains communication data to be transmitted.}, 
and noise vector $\mathbf{n}_\text{com}^{\bar{q}}(p,l) \sim \mathcal{CN}(\mathbf{0},\sigma^2_n\mathbf{I}_{M})$. 
% \in\mathbb{C}^{M\times1}$ is the complex Gaussian noise vector with zero mean and covariance matrix $\sigma^2_n\mathbf{I}_{M}$. 
%
The communication Rx then applies the matrix $\mathbf{D}(p,l)\in\mathbb{C}^{U\times N}$ to post-process the received $\mathbf{y}_{\bar{q}}(p,l)$ to detect information carried by different modes, and the communication symbols at the target $\bar{q}$ can be expressed as
\begin{align} \label{xTD}
% \mathbf{x}_q(p,l)&=\mathbf{D}(p,l)\left( \mathbf{H}_q(p,l)\mathbf{T}(p,l)\mathbf{s}(p,l)+\mathbf{n}_\text{com}^q(p,l) \right),
\mathbf{x}_{\bar{q}}(p,l)&=\mathbf{D}(p,l)\mathbf{y}_{\bar{q}}(p,l),
\end{align}
where the detailed design of $\mathbf{T}(p,l)$ (in $\mathbf{y}_{\bar{q}}(p,l)$, \eqref{y}) and $\mathbf{D}(p,l)$ will be discussed in Section~\ref{Sec:joint_design_isac}.
% \textcolor{red}{Yuan: here is quite confusing. You use matrix $\mathbf{D}$ and $\mathbf{T}$, while in the later section-communication design, the matrices used are $\mathbf{B}$ and $\mathbf{P}$ in \eqref{xp}, what are the relationship and differences with them.}
% % \subsubsection{Time-varying radar echos}
% Assuming the target has a radius velocity $v$, it changes the range term $e^{i 2 k_s r}$ as
% %
% \begin{align} \label{eq:cdm_5}
% e^{i 2 k_s (r + (p -1)v )} = e^{i 2 k_s r} e^{- i 2 \pi f_D (p -1)},
% \end{align}
% The time-varying radar echo becomes
% %
% % \begin{figure*}[ht]
% \begin{align} \label{eq:cdm_7}
%  {y}^\text{rad}&(p, l)  = \sigma_t \beta_1  \frac{e^{i 2 k_s r} e^{- i 2 \pi f_D (p -1)}}{r^2}   {J_{0}}(k_s R_r\sin{\theta}) \nonumber \\ & \sum_{ u = 1 }^{ U } e^{i{\ell}_u \varphi}  i^{-{\ell}_u}{J_{{\ell}_u}}(k_s R\sin{\theta}) s_{u}(p, l)' 
%  % \nonumber \\
%  % & = \sigma_t \beta_3  \frac{e^{i 2 k_s r} e^{- i 2 \pi f_D (p -1)}}{r^2}   {J_{0}}(k_s R_r\sin{\theta}) \nonumber \\ & \sum_{ u = 1 }^{ U } e^{i{\ell}_u \varphi}  i^{-{\ell}_u}{J_{{\ell}_u}}(k_s R\sin{\theta})
%  % s_{u}(p, l) {w_{C}}_{p,u}
%  .
% \end{align}
% \end{figure*}
% We propose a sensing-aided ISAC framework work. Hence, we would first explore the radar function to estimate channel parameters. Then the channel parameters can be used for 

\section{Sensing Signal Processing}\label{Sec:radar_processing_isac}
% \subsection{Received radar echoes}
% % A UCA with $N$ elements receives the radar echoes from the arbitrary point $\bm{P}$($r$, $\theta$, $\varphi$), with a radius of $R_r$. Hence, the signal received by the $n$th antenna is 
% %
% \begin{align} 
% % \label{eq:cdm_3}
%  {y}_n^\text{rad}(p, l) & \overset{(a)}\approx d \frac{e^{ik_sr}}{r} e^{-i \bm{k}_s \cdot\bm{r}_n} \sigma_t \overline{\mathbf{x}}(p, l), \nonumber \\
% & = \sigma_t \beta_1  \frac{e^{i 2 k_s r}}{r^2}  e^{-i \bm{k}_s \cdot \bm{r}_n} M 
% % \nonumber \\
% % &
% \sum_{ u = 1 }^{ U } e^{i{\ell}_u \varphi}  i^{-{\ell}_u}{J_{{\ell}_u}}(k_s R\sin{\theta}) s_{u}(p, l)',
% \end{align}
% %
% $\sigma_t$ is RCS, $\beta_1 = d \beta $. 

\subsection{Angle Doppler Ambiguity}
%%%%%%%%%%
\begin{figure}[t]
    \centering
    \includegraphics[width=0.45\textwidth]{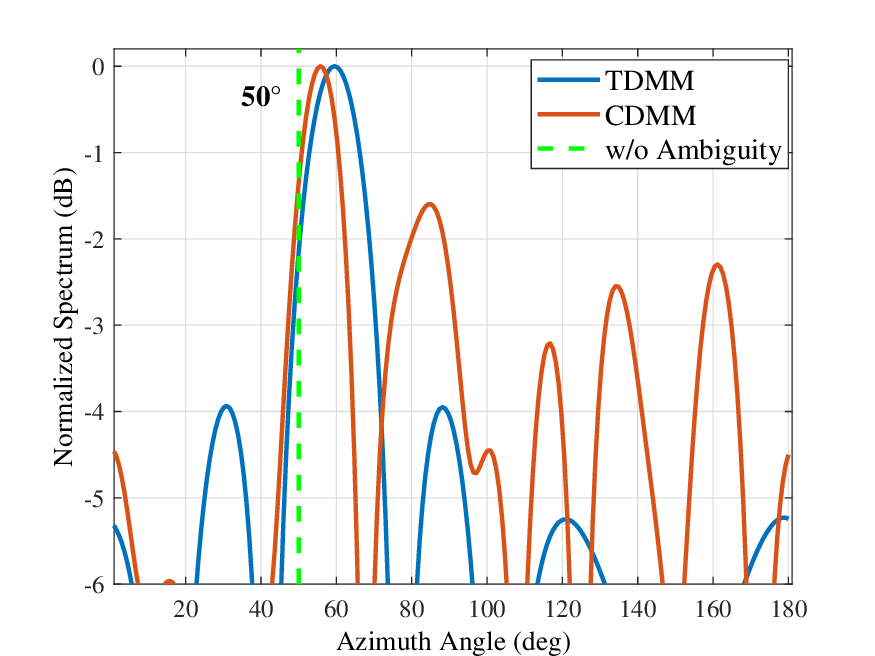}
          % \vspace{-0.3cm}
    \caption{Example of angle-Doppler ambiguity in azimuth estimation of a single target applying the FFT on the $U = 16$ modes domain: The ground truth is $50^\circ$ and target velocity is $5~\mathrm{m/s}$.  
    }\label{fig:DOp_ambiguity_eg}
     \vspace{-0.5cm}
\end{figure}
%%%%%%%%%%%%%%%
% \subsubsection{Doppler Robust}
% 
% We first present a case to illustrate the effect of the proposed Doppler-robust approach based on the proposed method. The ground truth azimuth angle is $50^\circ$ with a radial velocity of $5~\mathrm{m/s}$. Fig.~\ref{fig:DOp_ambiguity_eg} shows the FFT results in the mode domain. It can be observed that TDMM and CDMM have azimuth errors of approximately $5^\circ$. 
\subsubsection{TDMM}
In TDMM precoding of \eqref{eq:cdm_4}, the modes are separated by symbol index.  
% i.e, $\overline{y}^\text{rad}(p,:) \in \mathbb{C}^{1 \times l} $; 
Recalling \eqref{eq:TMD_1}, we can easily obtain the radar echo\footnote{{In the considered system, separate antennas are used for transmission and reception. Hence, the transmission leakage mainly arises from the direct coupling between these physically separated Tx and Rx antennas. 
The transmission leakage can be effectively suppressed by standard self-interference cancellation techniques, such as antenna isolation and adaptive interference cancellation. Therefore, in this paper, the residual transmission leakage is assumed to be negligible and is ignored in the signal model.}} of the mode $u = \langle p \rangle_{U}  $ as 
\begin{align} \label{eq:cdm_4_1_0}
  {y}_{{\ell}_u}^\text{rad}&(p, l)  
   =  \sum_{q=1}^{Q} \frac{ \beta_1 \sigma_q e^{i 2k_lr_q } e^{-i2 \pi f_{D,q} (p-1)T_c} }{r_q^2} {J_{0}}(k_l R_r\sin{\vartheta_q})  \nonumber\\
 & \times   e^{i{\ell}_u \varphi_q} i^{-{\ell}_u}{J_{{\ell}_u}}(k_l R_t\sin{\vartheta_q}) s_{u}(p, l)    + n_\text{rad}(p,l).
\end{align}
In monostatic radars, the transmitted signal $s_{u}(p, l)$, vortex mode ${\ell}_u$, and $\beta_1$ are known, hence, the radar echo model can be normalized as 
\begin{align} \label{eq:signal_4_1}
\overline{y}^\text{rad}(p, l) = & \sum_{q=1}^{Q}   \frac{\sigma_q e^{i 2 k_l r_q} e^{- i 2 \pi f_{D,q} (p -1)T_c}  }{r_q^2} {J_{0}}(k_l R_r\sin{\vartheta_q}) \nonumber \\  & \times   e^{i{\ell}_u \varphi_q} {J_{{\ell}_u}}(k_l R_t \sin{\vartheta_q}) + n'_\text{rad}(p,l).
\end{align}
%
% At the $p$th time slot, a vector containing length $L$ is received.

By collecting the $P_\text{sen}$ symbols, the sensing matrix of one CPI, $\overline{\mathbf{Y}}_\text{CPI}^\text{rad} = \left[ \overline{y}^\text{rad}(p, l) \right]_{ P_\text{sen} \times L }$, is formed.
In $\overline{\mathbf{Y}}_\text{CPI}^\text{rad}$, the azimuth angle $\varphi_q$ is associated with the mode index $u$ (and hence $p$ in TDMM), the range $r_q$ is mapped to the subcarrier index $k_l$, and the elevation angle $\vartheta_q$ appears with both $u$  and $k_l$ in the Bessel function. 
Conventional approaches, e.g., the FFT \cite{6975067} and subspace-based methods \cite{9968273}, have been used to estimate target parameters.
% \subsubsection{Angle-Doppler Ambiguity of TDMM precoding}
% In \eqref{eq:signal_4_1}, 
% different modes are transmitted in different time slots, typically as $u = p \mod{U} $. 
However, the Doppler term $f_D(p-1)T_c$ is also a function of $p$ (or $u$ in TDMM). 
% , i.e.,     
% when performing mode ${\ell}_u$ domain search, we assume the other parameters are constant. However, in TDM mode precoding, different modes are sent in different frames/slow-times $p$, i.e., the Doppler term $f_D(p-1)$ is a variation.  
%
As a result, the estimation of angular parameters by performing the mode domain search contains ambiguities due to Doppler, as shown in Fig.~\ref{fig:DOp_ambiguity_eg}. 
A similar conclusion can be drawn for Doppler estimation: ambiguity terms arise due to the $p$-dependent angular factor $e^{i{\ell}_u \varphi_q} {J_{{\ell}_u}}(k_l R_t \sin{\vartheta_q})$. 
% \textcolor{red}{We can also have a look of ambiguity function}
% %
% By performing the sub-space search in the mode domain, we can estimate the $\{ \vartheta_q, \varphi_q \}$, i.e., the azimuth-elevation angle map. 
% By performing the sub-space search in the subcarrier domain, we can estimate $\{{r_q}, \vartheta_q \}$, i.e., range-azimuth angle map. 
% Eventually, combining the $\{ \vartheta_q, \varphi_q \}$ map and the $\{ {r_q}, \vartheta_q \}$ map, we obtain 3-D image. 
% \begin{rem}
% {\em}
% % We have to note that, in TDM manner, the $U$ modes are not necessarily to be in regular orders as \eqref{eq:signal_tdm_1}, they can be random. 
% % especially using as communication symbols, they are random.
% % while changing with frame index as ${\ell}_u = f(p)$. 
% % However, in the radar Rx, we know the transmitted waveform and the OAM mode, hence we can always generate the matrix in the form of \eqref{eq:matrix_2}.
% \end{rem}
\begin{rem}
    {\em} The angle-Doppler ambiguity is not unique to vortex wavefront sensing systems such as \cite{9968273}, but is a universal problem in time division multiplexing (TDM)-based MIMO radars \cite{10559769,9272873}. Typically, the unambiguous angle-Doppler region is reduced, and work on handling multiple moving targets in the MIMO radar case is still limited.
    In the following, we focus on the proposed CDMM precoding scheme and on eliminating the angle-Doppler ambiguity.
\end{rem}    

\subsubsection{CDMM}
In CDMM precoding of \eqref{eq:cdm_4}, ${y}^\text{rad}(p, l)$ is a superposition of $U$ signals as 
\begin{align} \label{eq:cdm_7_1}
  {y}^\text{rad}(p, l)  
%  &  = \sum_{ u = 1 }^{ U }  \sum_{q=1}^{Q} \frac{ \beta_1 \sigma_q e^{i 2k_lr_q } e^{-i2 \pi f_{D,q} (p-1)T_c} }{r_q^2} {J_{0}}(k_l R_r\sin{\vartheta_q})  \nonumber\\
% & \times   e^{i{\ell}_u \varphi_q} i^{-{\ell}_u}{J_{{\ell}_u}}(k_l R_t\sin{\vartheta_q}) s_{u}(p, l) w_{p,u}  + n_\text{rad}(p,l) \nonumber \\
  & \overset{\triangle}{=}  \sum_{ u = 1 }^{ U } {y}_{{\ell}_u}^\text{rad}(p, l) {w}_{p, u} + n_\text{rad}(p,l),
\end{align}
where we define ${y}_{{\ell}_u}^\text{rad}(p, l)$ to be the equivalent representation of the radar echo of the $u$th mode, defined in \eqref{eq:cdm_4_1_0}. 
Now the goal is to separate ${y}_{{\ell}_u}^\text{rad}(p, l)$ by using the orthogonal properties of the precoding matrix $\mathbf{W}$.
% reconstruct the sensing matrix as in \eqref{eq:cdm_4_1_0}, then we can perform parameter estimations. 
% in TDM mode precoding.
% and reconstruct the sensing matrix $\mathbf{Y}_\text{CPI}^\text{rad}$ as  \eqref{eq:matrix_2_1}.    
% Recall that \eqref{eq:bpm_wight_1}, the precode matrix $\mathbf{W}_{C}^\text{CDM}$ is a Hadamard matrix of $M$ dimensions, we can separate the signals using orthogonal property $\mathbf{W}_{C}^\text{CDM} {\mathbf{W}_{C}^\text{CDM}}^T = M\mathbf{I}$ \cite{hedayat1978hadamard}.
%
% The proof of using orthogonality to separate the signal is provided in \textcolor{red}{Appendix~\ref{App:Hadamard}}.
%
% The de-coding utilizes the transpose matrix of the encoding matrix 
% \subsubsection{Time varying precoding and decoding}
In the $p$th symbol, the precoding matrix is equivalent to
$\mathbf{W}_{p} = \mathbf{P}_{\langle p \rangle_{U}} \mathbf{W}$, where $\mathbf{P}_{\langle p \rangle_{U}} \in \mathbb{R}^{U \times U} $ is a projection matrix used to move the first $\langle p \rangle_{U}$ rows of the original matrix to the last $\langle p \rangle_{U}$ rows as 
%%%%%%%%%%%%%%%%%%%%%%%%%%%%%%%%%%%%%%%%%%%%%%%%
\begin{equation}
    \begin{aligned} \label{eq:cdm_8_0}
        \mathbf{P}_{\langle p \rangle_{U}}  =
    \begin{bmatrix}
         \mathbf{0} & \mathbf{I}_{U - \langle p \rangle_{U} +1}\\
          \mathbf{I}_{\langle p \rangle_{U} -1} & \mathbf{0}
    \end{bmatrix}.
        \end{aligned}
\end{equation}
%%%%%%%%%%%%%%%%%%%%%%%%%%%%% 
% with $\langle p \rangle_{U} = \left(p \mod U \right) $.
% should be a projection of $\mathbf{W}_{C}^\text{CDM}$. 
% If $(p \mod U) = 1$, the encoding matrix is the standard $\mathbf{W}^{\text{MPM}}_{2^k}$, while in general cases, $(p \mod U) = \overline{p} $, the encoding matrix is
% %%%%%%%%%%%%%%%%%%%%%%%%%%%%%%%%%%%%%%%%%%%%%%%%
% \begin{equation}
%     \begin{aligned} \label{eq:cdm_8_0_0}
%         \mathbf{P}_{p}  =
%     \begin{bmatrix}
%          \mathbf{0} & \mathbf{I}_{U - n_u +1}\\
%           \mathbf{I}_{n_u -1} & \mathbf{0}
%     \end{bmatrix} 
%         \end{aligned}
% \end{equation}
% %%%%%%%%%%%%%%%%%%%%%%%%%%%%%%%%%%%%%%%%%%%%%%%%
%
We use every $U$ symbols (from the $p$th to the $(p + U -1)$th) and the orthogonality of the Hadamard code to decode the $p$th symbol\footnote{Similarly, $U$ symbols from the $(p-U +1)$th to the $p$th can be used.}.
Let us define $\mathbf{y}_{p,l}^\text{rad} \in \mathbb{C}^{U \times 1}$ to be the received radar signal collected over $U$ symbols starting from the $p$th symbol as 
\begin{align}
    & \mathbf{y}_{p,l}^\text{rad} = 
     \begin{bmatrix}
         {y}^\text{rad}(p, l) & \cdots & {y}^\text{rad}(p + U-1 , l)  
     \end{bmatrix}^T .
\end{align}
The decoded echo signal $\mathbf{{z}}^\text{rad}_{p,l}  = \left[ {z}_{{\ell}_u}^\text{rad}(p, l) \right]_{U \times 1}$ is obtained as 
% by applying the \textbf{Lemma}~\ref{lem:2} as 
% \begin{align}
% & \mathbf{{z}}^\text{rad}_{p,l}  \in \mathbb{C}^{(U\times 1} = 
%     \nonumber \\
%      &
%      \begin{bmatrix}
%          {z}_{{\ell}_1}^\text{rad}(p, l) & {z}_{{\ell}_2}^\text{rad}(p, l) & 
%          \cdots & 
%          {z}_{{\ell}_U }^\text{rad}(p, l)   
%      \end{bmatrix}^T,  \nonumber
% \end{align}
% %
%
\begin{align} \label{eq:cdm_9}
\mathbf{{z}}^\text{rad}_{p,l}
      &= \dfrac{{\mathbf{W}}^T_{p}}{U}
% \bm{\hat{E}}_{{\ell}_u }(p, l) = 
        \mathbf{y}_{p,l}^\text{rad}     \approx
     \overline{\mathbf{y}}^\text{rad}_{p,l}
     + 
     \mathbf{\tilde{y}}^\text{rad}_{p,l}
      + \frac{{\mathbf{W}}^T_{p} \mathbf{n}}{U},
\end{align}
where $\mathbf{n}$ is the noise vector,
% \footnote{It is worth mentioning in \eqref{eq:cdm_9}, the noise power is normalized by $U$. Hence, by introducing the CDMM precoding, the SNR is improved $U$ times.}, 
and $\overline{\mathbf{y}}^\text{rad}_{p,l} \in \mathbb{C}^{U \times 1}$ is the ideally decoded echo signals of the $U$ modes as
\begin{align} \label{eq:cdm_9_0_1}
    &\overline{\mathbf{y}}^\text{rad}_{p,l}   =      
     \begin{bmatrix}
         {y}_{{\ell}_1}^\text{rad}(p, l) & \cdots & {y}_{{\ell}_U }^\text{rad}(p,l)  
     \end{bmatrix}^T,
\end{align}
which contains $Q$ components as $\overline{\mathbf{y}}^\text{rad}_{p,l} = \sum_{q=1}^{Q} \overline{\mathbf{y}}^q_{p,l}$, with $\overline{\mathbf{y}}^q_{p,l} = \left[ {y}_{{\ell}_u}^{q}(p, l) \right]_{U \times 1}$. 
According to \eqref{eq:cdm_4_1_0}, the contribution of the $q$th target is  
\begin{align} \label{eq:cdm_13}
  {y}_{{\ell}_u}^{q}(p, l)  
 &  =   \frac{ \beta_1 \sigma_q e^{i 2k_lr_q } e^{-i2 \pi f_{D,q} (p-1)T_c} }{r_q^2} {J_{0}}(k_l R_r\sin{\vartheta_q})  \nonumber\\
 & \times   e^{i{\ell}_u \varphi_q} i^{-{\ell}_u}{J_{{\ell}_u}}(k_l R_t\sin{\vartheta_q}) s_{u}(p, l).
\end{align}
%
% \begin{rem}
% {\em}
% \end{rem}
The disturbance term $\mathbf{\tilde{y}}^\text{rad}_{p,l} \in \mathbb{C}^{U\times 1}$ in \eqref{eq:cdm_9}, takes the form  
\begin{align} \label{eq:cdm_9_0_2}
  &  \mathbf{\tilde{y}}^\text{rad}_{p,l}   = 
     \begin{bmatrix}
         \tilde{y}_{{\ell}_1}^\text{rad}(p, l) & \tilde{y}_{{\ell}_2}^\text{rad}(p, l) & \cdots & 
         \tilde{y}_{{\ell}_U}^\text{rad}(p, l)  
     \end{bmatrix}^T,
\end{align}
which is due to Doppler ambiguity. 
% Fig.~\ref{fig:DOp_ambiguity_eg} also illustrates an example of angle-Doppler ambiguity of CDMM, where the FFT is directly applied to $\mathbf{{z}}^\text{rad}_{p,l}$ after \eqref{eq:cdm_9}, regardless of the Doppler term.
The angle-Doppler ambiguity in CDMM is also illustrated in Fig.~\ref{fig:DOp_ambiguity_eg}.
The mode-domain FFT is applied to $\mathbf z^{\mathrm{rad}}_{p,l}$ after \eqref{eq:cdm_9} without considering the Doppler-induced rotations. The resulting Doppler perturbation breaks code orthogonality, producing deterministic cross-interference across angles and codes. 
Therefore, explicit Doppler modeling and compensation are required.

% The uncompensated phase ramps destroy code orthogonality and smear energy across codes and angles, which yields the observed ambiguity
% Appendix~\ref{App:Hadamard} 
\begin{proposition}
    {\em} The disturbance term $\mathbf{\tilde{y}}^\mathrm{rad}_{p,l}$ is a non-linear coupling, while we model it linearly in \eqref{eq:cdm_9}. 
\end{proposition}
\begin{proof}
% For the $u$th mode, the echo contains $Q$ targets, hence in \eqref{eq:cdm_9_0_1}, ${y}_{{\ell}_u}^\text{rad}(p, l) = \sum_{q=1}^{Q}{y}_{{\ell}_u}^{q}(p, l)$.
% defined in \eqref{eq:cdm_9_0_1}, 
% is superposition of $Q$ components.  
We define the Doppler term $e^{- i 2 \pi f_{D,q} T_c } = \tilde{\omega}_q $. The time-varying model of the $q$th target in \eqref{eq:cdm_13} can be written as 
\begin{align} \label{eq:cdm_9_1}
{y}_{{\ell}_u}^q(p +1, l) =  e^{- i 2 \pi  f_{D,q} T_c } {y}_{{\ell}_u}^q(p, l) =  \tilde{\omega}_q {y}_{{\ell}_u}^q (p, l).
\end{align}
% ${y}_{{\ell}_u}^\text{rad}(p +1, l) =  e^{- i 2 \pi  f_D } {y}_{{\ell}_u}^\text{rad}(p, l) =  \tilde{\omega}_q {y}_{{\ell}_u}^\text{rad}(p, l)$.
%
Hence, due to the disturbance term of Doppler, at the $p$th symbol, the precoding matrix is equivalent to $\mathbf{W}_p \odot \tilde{\mathbf{\Omega}}$, with $\widetilde{\mathbf{\Omega}}=\sum_{q=1}^{Q} \tilde{\mathbf{\Omega}}_q$, $\widetilde{\mathbf{\Omega}}_q$ is a rank one Vandermonde matrix as
%%%%%%
\begin{equation}
    \begin{aligned} \label{eq:cdm_9_2}
        \mathbf{\tilde{\Omega}}_q^T  =
    \begin{bmatrix}
1 & \tilde{\omega}_q & \tilde{\omega}_q^2 & \cdots & \tilde{\omega}_q^{U-1} \\
1 & \tilde{\omega}_q & \tilde{\omega}_q^2 & \cdots & \tilde{\omega}_q^{U-1} \\
% 1 & \tilde{\omega}_q & \tilde{\omega}_q^2 & \cdots & \tilde{\omega}_q^{U-1} \\
\vdots & \vdots & \vdots & \ddots & \vdots \\
1 & \tilde{\omega}_q & \tilde{\omega}_q^2 & \cdots & \tilde{\omega}_q^{U-1}
\end{bmatrix}.
        \end{aligned}
\end{equation}
Now, $\left(\mathbf{W}_p  \odot \mathbf{\widetilde{\Omega}}\right)$ is not orthogonal to ${\mathbf{W}_p}$. We define the matrix $\widetilde{\mathbf{H}}_O \in \mathbb{C}^{U \times U} = \frac{{{\mathbf{W}^T_{p}}}}{U} \left( {\mathbf{W}}_{p} \odot \mathbf{\widetilde{\Omega}} \right) $ to quantify interference due to the Doppler perturbation $\tilde{\mathbf{\Omega}}$.
The decoded signal of \eqref{eq:cdm_9}
can be written as %
\begin{align} \label{eq:cdm_11_0}
    \mathbf{{z}}^\text{rad}_{p,l}
      &= \dfrac{{\mathbf{W}}^T_{p}}{U} \sum_{q=1}^{Q} 
% \bm{\hat{E}}_{{\ell}_u }(p, l) = 
     \left(  \mathbf{y}_{p,l}^\text{q}  + \mathbf{n}_q \right)   
     % \nonumber \\ &
     = \sum_{q=1}^{Q} \dfrac{{\mathbf{W}}^T_{p}}{U}
     \left(   {\mathbf{W}}_{p} \odot \mathbf{\widetilde{\Omega}}_q  \overline{\mathbf{y}}^q_{p,l} + \mathbf{n}_q \right)   \nonumber \\
     % &= \dfrac{{\mathbf{W}}^T_{p}}{M}
     % \left( {\mathbf{W}}_{p} \odot \mathbf{\tilde{\Omega}} \right) \overline{\mathbf{y}}^\text{rad}_{p,l} + \dfrac{{\mathbf{W}}^T_{p}}{M} \mathbf{n}    \nonumber \\
    & = \sum_{q=1}^{Q} \left( \mathbf{\widetilde{H}}_O \overline{\mathbf{y}}^q_{p,l} + \frac{ {\mathbf{W}}^T_{p} \mathbf{n}_q}{U} \right).
\end{align}
% where $\mathbf{y}_{p,l}^\text{rad} = \sum_{q=1}^{Q}\mathbf{y}_{p,l}^\text{q}$.
%
\begin{itemize}
    \item The diagonal elements of $\widetilde{\mathbf{H}}_O$, 
\begin{align}\label{eq:cdm_11_0_1}
 \widetilde{\mathbf{H}}_O(u,u) 
 & = \dfrac{1}{U} \frac{1 - \tilde{\omega}_q^{U} }{1 - \tilde{\omega}_q} \approx 1,
\end{align}
where the derivation is in Appendix~\ref{App:CDM_disturbance_dia}.
The $\approx$ holds, when the Doppler turbulence is small, i.e., $\tilde{\omega}_q$\footnote{A simulation example: $f_c = 77$ GHz, $v = 5$ m$/$s, $\tilde{\omega}_q = e^{-i0.108}$.} $\to 1$, the proof is in Appendix~\ref{App:proof_1}.
% 
% \begin{proof}
%     $\lim_{|\tilde{\omega}_q| \to 1^-}, 
% % |\tilde{\omega}_q| < 1$, 
% \frac{1 - \tilde{\omega}_q^{U} }{U (1 - \tilde{\omega}_q )} = \dfrac{0}{0}$, by applying L'Hospital's rule $\frac{1 - \tilde{\omega}_q^{U} }{U (1 - \tilde{\omega}_q )} = \dfrac{-U \tilde{\omega}_q^{U-1}}{-U} \to 1$.
% \qedhere % 若最后一行是文字，用 \qedhere
% \end{proof}
% 
\item  The off-diagonal elements of $\widetilde{\mathbf{H}}_O$, $u \neq u'$,
\begin{align} 
 \widetilde{\mathbf{H}}_O(u,u') 
 & = \dfrac{1}{U} \sum_{u'' =1 }^{U} {\mathbf{W} }_{u'',u}   {\mathbf{W} }_{u'',u'} \tilde{\omega}_q^{u'' -1} , 
\end{align}
where $u'' \in [1, U]$, the derivation is in Appendix~\ref{App:CDM_disturbance_off}.
\end{itemize}

Substituting $\widetilde{\mathbf{H}}_O$ into \eqref{eq:cdm_11_0}, the decoded echo of the $u$th mode ${z}_{{\ell}_u}^\text{rad}(p, l) \in \mathbf{{z}}^\text{rad}_{p,l}$ can be obtained. Specifically, 
% ${z}_{{\ell}_u}^\text{rad}(p, l) = \sum_{q=1}^Q {z}_{{\ell}_u}^q(p, l)$, 
\begin{align} \label{eq:cdm_12}
{z}_{{\ell}_u}^\text{rad}(p, l)  
%  \left( {y}_{{\ell}_u}^\text{rad}(p, l) +  {\tilde{y}_{{\ell}_u}^\text{rad}(p, l) }  \right) + \frac{ n_\text{rad}(p,l) }{U}
% \nonumber \\
    &= \sum_{q=1}^{Q} \left( {y}_{{\ell}_u}^{q}(p, l) +  \tilde{y}_{{\ell}_u}^{q}(p, l) \right) + \frac{ n_\text{rad}(p,l) }{U},
\end{align}
where the $q$th decoded echo 
\begin{align} \label{eq:cdm_11_1}
    & {z}_{{\ell}_u}^q(p, l) = \frac{1 - \tilde{\omega}_q^{U} }{U (1 - \tilde{\omega}_q )}  {y}_{{\ell}_u}^q(p, l) +  \sum_{ u'   }^{ }  
  \widetilde{\mathbf{H}}_O(u,u') {y}_{{\ell}_{u'}}^q(p, l)   + \frac{ {n_q}'}{U}   \nonumber \\
     &\approx {y}_{{\ell}_u}^q(p, l)    +   \sum_{ u'   }^{ }  
   \sum_{u'' =1 }^{U} \dfrac{ {\mathbf{W} }_{u'',u}   {\mathbf{W} }_{u',u''} \tilde{\omega}_q^{u'' -1} {y}_{{\ell}_{u'}}^q(p, l) }{U}  + \frac{ {n_q}'}{U}   \nonumber \\
   %   & \qquad  \dfrac{1}{M}  \sum_{{\ell}_u' \neq  {\ell}_u }^{ }  
   % \sum_{\ell \neq  {\ell}_u }^{} {\mathbf{W}^{\text{CDM}}_{C}}_{p,\ell,{\ell}_u}   {\mathbf{W}^{\text{CDM}}_{C}}_{p,{\ell}_u',\ell} \tilde{\omega}_q^{\ell -1}  {y}_{{\ell}_u'}^\text{rad}(p, l)   ,\nonumber \\
     & \overset{\triangle}{=}   {y}_{{\ell}_u}^q(p, l) +  \tilde{y}_{{\ell}_u}^q(p, l) + \frac{ {n_q}'}{U}.
     % & = {y}_{{\ell}_u}^\text{rad}(p, l) +  {\tilde{y}_{{\ell}_u}^\text{rad}(p, l) } + \frac{ \mathbf{n}'}{M}.
\end{align}
 % 
% This shows the disturbance term can be approximately linear as $\tilde{y}_{{\ell}_u}^q(p, l)$, which is an alternating geometry series containing the Doppler  
% 
% For the $u$th mode, the echo contains $Q$ targets, we can write ${z}_{{\ell}_u}^\text{rad}(p, l)$ as 
%
% \begin{align} 
%   {y}_{{\ell}_u}^{q}(p, l)  
%  &  =   \frac{ \beta_1 \sigma_q e^{i 2k_lr_q } e^{-i2 \pi f_{D,q} (p-1)T_c} }{r_q^2} {J_{0}}(k_l R_r\sin{\vartheta_q})  \nonumber\\
%  & \times   e^{i{\ell}_u \varphi_q} i^{-{\ell}_u}{J_{{\ell}_u}}(k_l R_t\sin{\vartheta_q}) s_{u}(p, l).
% \end{align}
% The disturbance term $\tilde{y}_{{\ell}_u}^{q}(p, l)$ will be analyzed.  
% \qedhere % 若最后一行是文字，用 
The disturbance term $\tilde{y}_{\ell_u}^q(p,l)$ is an alternating geometric-series structure that contains Doppler-induced cross-mode interference among multiple targets.
\qedhere
\end{proof}
{ \begin{rem} {\em}
In \eqref{eq:cdm_11_1}, the disturbance term $\tilde{y}^q_{\ell_u}(p,l)$ is deterministic once the target parameters are specified. Each term corresponds to the interference contribution from other targets due to Doppler-induced cross-mode coupling. Directly estimating the parameters from this coupled disturbance is computationally intensive. 
To address this challenge, we design a new approach in Section~\ref{Sec:VCM-EM}, which decouples the targets through velocity-consistent matching and mode projection, enabling robust Doppler and multi-parameter estimation in a tractable iterative framework.
\end{rem}
}
% \begin{rem}
%     {\em} The disturbance term $\tilde{y}_{\ell_u}^q(p,l)$ is an alternating geometric-series structure that contains Doppler-induced cross-mode interference among multiple targets.
% \end{rem}
% The decoding strategy in \eqref{eq:cdm_9} is
% based on the assumption that the channel is time-invariant, i.e., ${y}_{{\ell}_u}^\text{rad}(p, l) = {y}_{{\ell}_u}^\text{rad}(p+1, l) = ... = {y}_{{\ell}_u}^\text{rad}(p+ U -1, l) $. However, in a scenario of time-varying, ${y}_{{\ell}_u}^\text{rad}$ in a different time interval $p$ is slightly different, which results in Doppler ambiguity. 
% \begin{rem}
%     {\em} It is worth mentioning in \eqref{eq:cdm_9}, the noise power is normalized by $U$. Hence, by introducing the CDMM precoding, the SNR is improved $U$ times. 
% \end{rem}

\subsection{Joint Ambiguity Elimination and Parameter Estimation}\label{Sec:VCM-EM}
\subsubsection{Sensing Problem}
Obtaining $\mathbf{z}_{p}^\text{rad} \in \mathbb{C}^{U\times 1}$ in \eqref{eq:cdm_9}, and collecting $P_\text{sen}$ symbols, each with $L$ subcarriers, we have the decoded 3-D data cube\footnote{In references, e.g., \cite{11075598}, the 3-D data processing is referred to as a tensor to distinguish it from a two-dimensional matrix.} $\mathbf{{Z}}^\text{rad} \in \mathbb{C}^{P_\text{sen} \times L \times U}$. 
We can also write the 3-D matrix form of \eqref{eq:cdm_9} as
\begin{align} \label{eq:cdm_11}
\mathbf{Z}^\text{rad}
     =
     \mathbf{\overline{{Y}}}^\text{rad}
     + 
     \mathbf{\tilde{Y}}^\text{rad}
      + \mathbf{N}^\text{rad},
\end{align}
following the same definition as $\mathbf{{Z}}^\text{rad}$, we can define the 3-D data cube of echo signal $\mathbf{\overline{{Y}}}^\text{rad} \in \mathbb{C}^{P_\text{sen} \times L \times U}$, the disturbance term $\mathbf{\tilde{Y}}^\text{rad} \in \mathbb{C}^{P_\text{sen} \times L \times U}$, and noise term $\mathbf{\tilde{N}}^\text{rad} \in \mathbb{C}^{P_\text{sen} \times L \times U}$.
Now, the problem becomes twofold:
\begin{itemize}
    \item[i)] Eliminate the disturbance matrix $\mathbf{\tilde{Y}}^\text{rad}$, and obtain the ambiguity-free sensing matrix ${\overline{\mathbf{Z}}^\text{rad}} = \mathbf{\overline{{Y}}}^\text{rad} + \mathbf{N}^\text{rad}$;
    \item[ii)] Estimate the range $r_q$, azimuth angle $\varphi_q$, elevation angle $\vartheta_q$, and Doppler ${f_{D,q}}$ of $Q$ targets from ${\overline{\mathbf{Z}}^\text{rad}}$
    % decoded data cube $\mathbf{{Z}}^\text{rad}$ is corrupted by . 
\end{itemize}
\begin{rem}
    {\em} Eliminating the disturbance matrix $\mathbf{\tilde{Y}}^\text{rad}$ is easy in a single-target system. Applying the classic conjugate‐matched principle \cite{4608989,kim2022mimo}, i.e., multiplying the decoding matrix $\mathbf{W}_p^T$ element-wise by the complex conjugate of the transmit Vandermonde phase $\mathbf{\tilde{\Omega}}_q^H$, each unit module Doppler perturbation is exactly canceled as $\left(\mathbf{W}_p^T \odot \mathbf{\widetilde{\Omega}}_q^H  \right) \left(\mathbf{W}_p \odot \mathbf{\widetilde{\Omega}}_q  \right) =  \mathbf{W}_p^T  \mathbf{W}_p  = \mathbf{I} $. 
% 
% \textcolor{red}{(This is the a bunch of mismatched filters, needs to be explained by simulations)}
However, in this multi-target problem, the precoding matrix is equivalent to $\mathbf{W}_p \odot \left( \sum_{q=1}^{Q} \widetilde{\mathbf{\Omega}}_q \right)$, where the simple conjugate‐matched principle is not applicable. 
% is applied to the echo $\mathbf{y}_p^\text{rad}$ consisting of $Q$ targets, as illustrated in \eqref{eq:cdm_11_1}, where the Dopplers of the $Q$ targets are different. Hence, the conventional conjugate‐matched principle is not applicable. 
\end{rem}

\subsubsection{Problem Formulation}
For a $Q$-target multi-parameter problem, we follow the iterative EM framework for joint ambiguity-free decoding and alternating parameter estimation \cite{feder1988parameter,fleury1999channel}.
% 
% The exact sensing matrix ${\overline{\mathbf{Z}}^\text{rad}}$ contains echoes of $Q$ targets. 
% Following the concept of the conventional SAGE algorithm \cite{feder1988parameter,fleury1999channel}, 
Let us define the $Q$ hidden (complete) data in the ambiguity-free sensing matrix as 
\begin{align}\label{eq:pro_2_1}
{\overline{\mathbf{Z}}^\text{rad}} \overset{\triangle}{=} \sum_{q=1}^{Q} \overline{\mathbf{Z}}^q = \sum_{q=1}^{Q} \left( {\overline{\mathbf{Y}}^{q} } + \beta_q \mathbf{N}^\text{rad} \right),
\end{align}
where $\overline{\mathbf{Y}}^\text{rad} = \sum_{q} \overline{\mathbf{Y}}^q$, 
% ${\overline{\mathbf{Y}}^{q}} \in \mathbb{C}^{P \times L \times U }$ represents the exact echo of the $q$th target 
and $\sum_q \beta_q^2 =1$ to constrain the noise power. 
The summation of $Q$ echoes characterized by the $p$th symbol, $l$th sampling, and $u$th mode is given in \eqref{eq:signal_4_1}, hence the vectorized matrix, $\text{vec}\left({\overline{\mathbf{Y}}^{q} } \right) \in \mathbb{C}^{LP_\text{sen}U \times 1}$, is    
\begin{align}\label{eq:pro_2_2}
\text{vec}\left({\overline{\mathbf{Y}}^{q} } \right)   
  = & \sigma_q   \text{vec}\left( \mathbf{A}_{r_q, \varphi_q, \vartheta_q} \right) \otimes \mathbf{a}_{f_{D,q}} ,
\end{align}
where $\text{vec}\left( \mathbf{A}_{r_q, \varphi_q, \vartheta_q} \right) \in \mathbb{C}^{LU \times 1 } $, with  $\mathbf{A}_{r_q, \varphi_q, \vartheta_q} \in \mathbb{C}^{L \times U }$, the element at the $l$th row and $u$th column denoted as $\frac{e^{i 2 {k}_l r_q } e^{i{\ell}_u \varphi_q} {J_{0}}( {k}_l R_r\sin{\vartheta_q} )  {J_{{\ell}_u}}(k_l R_t \sin{\vartheta_q})  }{r_q^2} $, and 
% $\mathbf{a}_{f_{D,q}}  \in \mathbb{C}^{P_\text{sen} \times 1}$,
\begin{align}
    \mathbf{a}_{f_{D,q}}  = [1, e^{- i 2 \pi {f_{D,q}} T_c}, ..., e^{- i 2 \pi {f_{D,q}} (P -1)T_c} ]^T.
\end{align}

Therefore, we formulate the joint sensing matrix decoding and parameter sensing Problem~\ref{prob:1} as   
\begin{problem}\label{prob:1}
\begin{subequations}\label{eq:pro_1_1}
\begin{align}
    \min_{\bm{\Theta}} \quad  &  \| {\mathbf{Z}}^\mathrm{rad} - \mathbf{\overline{Y}}^\mathrm{rad}(\mathbf{ {\Theta}}) - \mathbf{\tilde{Y}}^\mathrm{rad}(\mathbf{ {\Theta}}) \|_F , \label{eq:pro_1_1_1} \\ 
      \text{s.t.} \quad &    
      \mathbf{\Theta} = \left[\boldsymbol{\theta}_q\right]_{Q\times 1},  \quad \boldsymbol{\theta}_q = \left[\sigma_q, r_q, f_{D,q}, \varphi_q , \vartheta_q \right],              \label{eq:pro_1_1_2} \\  
      \quad &
    \boldsymbol{{\theta}}_q = \arg \underset{\boldsymbol{\theta}_q}{ \min }  \| \overline{\mathbf{Z}}^q(\hat{f}_{D,q}) - \mathbf{\overline{Y}}^q(\boldsymbol{ \theta}_q) \|_F , \label{eq:pro_1_1_3} 
        % &   \overline{\mathbf{Z}}^q = {\mathbf{Z}}^q - \mathbf{\tilde{Y}}^q(  \mathbf{ \theta }_q) , \label{eq:pro_1_1_4} \\
        % & \overline{\mathbf{Z}}^\text{rad} = \sum_{q=1}^{Q} \overline{\mathbf{Z}}^q,   \overline{\mathbf{Y}}^\text{rad} = \sum_{q=1}^{Q} \overline{\mathbf{Y}}^q ,
       % &   \tilde{\mathbf{Y}}^\text{rad} = \sum_{q=1}^{Q} \tilde{\mathbf{Y}}^q, \label{eq:pro_1_1_5}
    \end{align} 
\end{subequations}
\end{problem}
\noindent where the Frobenius norm $\|\mathbf{A} \|_F$ is equivalent to the vector $\ell_2$ norm of its vectorized form, i.e.,
$\|\mathbf{A}\|_F = \|\mathrm{vec}(\mathbf{A})\|_2$.
% where $\mathbf{\Theta} = \left[\bm{\theta}_1; \bm{\theta}_2; ...; \bm{\theta}_Q  \right]$, with $\bm{\theta}_q = [r_q, {f_{D,q}}, \psi_q , \phi_q ]$, and $\Delta \tilde{\omega}_q = e^{-i 2 \pi f_D }, \Delta \tilde{\omega}_q \in   \Delta \mathbf{\tilde{\Omega}}$.
The Doppler parameter $f_{D,q}$ is estimated together with the parameter set in \eqref{eq:pro_1_1_3}. However, estimating the parameter set $\boldsymbol{\theta}_q$ requires a priori Doppler information to eliminate the disturbance and obtain $\overline{\mathbf{Z}}^q(\hat{f}_{D,q})$. 
Therefore, it is natural to consider an iterative and adaptive scheme. 
    % \item The $q$th measured echo $\mathbf{Z}^q$ cannot be directly obtained from $\mathbf{Z}^\text{rad}$, hence we need to further consider it in the algorithm design. 
 
% helps to eliminate the effects of Doppler perturbation of the decoded sensing matrix $\overline{\mathbf{Z}}^\text{rad}$.
% % i.e., decoupling of angle-Doppler ambiguity.
% Parameter estimations are based on a decoded sensing matrix $\overline{\mathbf{Z}}^\text{rad}$. 
% Then,   
% adaptive 
% space-alternating generalized expectation-maximization (Ad-SAGE) to solve the joint sensing matrix decoding and parameter estimation problem in an iterative style.  \eqref{eq:pro_1_1}

\subsection{Solution to Problem~\ref{prob:1}: Doppler-Robust Parameter Estimation}
 %
 % \subsubsection{Road Map of Ad-SAGE}
% %
% \begin{figure}[t]
% % \setlength{\abovecaptionskip}{-0.2cm}
% % \setlength{\belowcaptionskip}{-0.2cm}
% \begin{center}
% \includegraphics[scale=0.35]{figures/Joint_flow.pdf}%width=9cm,height=5cm scale=0.42
% \end{center}
% \caption{Road map of adaptive approach for joint decoupling of sensing matrix and target parameters estimation. \textcolor{red}{WX: Please enhance the image by unifying the font size within it, so that the text in the image matches the font size used in the main body text.}  }
% \label{fig_RaodMap}
% \end{figure}
%
% \subsubsection{Ad-SAGE}
% The Adaptive SAGE includes two loops: the adaptive loop and the SAGE loop.
% \textcolor{red}{Hence, we utilize an adaptive framework, where for each target, the estimated $f_{D,q}$ is used to refine and obtain Doppler-decoupled radar echo $\overline{\mathbf{Z}}^q$.}
Unlike the conventional EM algorithm \cite{fleury1999channel}, we propose a VCM approach for joint Doppler selection, signal decoding, and estimation in the E-step of the EM-based framework, and the parameter subsets are estimated alternately in the M-step. 
% In the two-loop framework of adaptive EM, for the $\mu$th iteration, the problems \eqref{eq:pro_1_1_3} and \eqref{eq:pro_1_1_4} are transferred into 
% \begin{align}
%     % \mathcal{P}.~1.1 \quad  
%     \begin{cases}\label{eq:pro_1_2}
%     \textbf{Adaptive loop:} \quad & {\overline{\mathbf{Z}}^q}^{(\mu)} = {\mathbf{Z}}^q - \mathbf{\tilde{Y}}^q( \boldsymbol{\theta}_q^{(\mu-1)} ) , \\
%     \textbf{EM loop:} \quad 
%     & \mathbf{\hat{\theta}}_q^{(\mu)} = \arg \underset{\boldsymbol{\theta}_q}{ \min }  \| {\overline{\mathbf{Z}}^q}^{(\mu)} - \mathbf{\overline{Y}}^q(\boldsymbol{\theta}_q) \|_F. 
%     % \\
%     % {\Delta \mathbf{\tilde{\Omega}}}^{(i+1)}
%     \end{cases} 
% \end{align}
% % (In radar functions, we need to define parameters, $Q$ targets, with the $q$th one denoted $\mathbf{\Theta} = \left[\bm{\theta}_1; \bm{\theta}_2; ...; \bm{\theta}_Q  \right]$, with $\bm{\theta}_q = [r_q, {f_{D,q}}, \vartheta_q , \varphi_q ]$ ).
% % where the $\mu$th decoupled sensing matrix is based on the $u$th parameter estimations; again the the $(\mu)$th parameter estimations is based on the $(\mu)$th decoupled sensing matrix.
% % \subsubsection{EM Loop}
% Since the $Q$ hidden data cannot be separated directly from the sensing matrix, we apply the conditional expectations to estimate the $q$th hidden data $\overline{\mathbf{Y}}^q $ in the E-step and estimate the parameter set $\boldsymbol{\theta}_q$ in the M-step.

\subsubsection{VCM-EM for Joint Doppler Selection, decoding, and estimation}
Assume the velocity space $\mathcal{V} \in [v_{\text{min}}, v_{\text{max}}]$, we apply the VCM to jointly estimate Doppler and decode the sensing matrix of the $q$th target at the $\mu$th EM loop. 

\textbf{VCM-step}: For $v_{\ast} \in \mathcal{V}$, we can decode the echo of the $q$th target by \eqref{eq:estep_0_1} and \eqref{eq:estep_0_2} as\footnote{With $v_{\ast}$, we can calculate ${f_{D,q}}_{\ast}$ and ${\tilde{\mathbf{\Omega}}}_{q\ast}$ accordingly.} 
\begin{equation} \label{eq:estep_0_1}
\begin{aligned}
    {\mathbf{y}_{p,l}^{q}}^{(\mu)}  =  \mathbf{y}_{p,l}^\text{rad}  - \sum_{q' \neq q, q'= 1}^{Q} \text{vec} \left( {\overline{\mathbf{Y}}^{q} } (\hat{\boldsymbol{\theta}}_{q'}^{(u-1 )}) \right),  
    \end{aligned} 
\end{equation}
where $q' = 1, 2, \dots, Q$. 
\begin{align} \label{eq:estep_0_2}
{\mathbf{{z}}^q}^{(\mu)}(v_{\ast})
      &= \dfrac{ \left( {\mathbf{W}}^T_{p} \odot {\tilde{\mathbf{\Omega}}}_{q\ast}^H  \right) }{U}
        {\mathbf{y}_{p,l}^{q}}^{(\mu)}   .
\end{align}
As in \eqref{eq:cdm_11}, by collecting all the $L$ samples and $P_\text{sen}$ symbols can obtain ${\mathbf{{Z}}^q}^{(\mu)}(v_{\ast})$. 
Then, the velocity is estimated by consistency matching as
\begin{subequations}\label{eq:estep_0_3}
\begin{align}
  &  \hat{v}_q^{(\mu )} = \arg \min_{v_{\ast} \in \mathcal{V} }    \| v_{\ast} - \hat{v}_q^{(\mu )}(v_{\ast})  \|  , \label{eq:estep_0_3_1} \\ 
      \text{s.t.} \quad &   
      \hat{v}_q^{(\mu )}(v_{\ast})  = \arg\mathop{\min}_{ v_q } \| {\mathbf{{Z}}^q}^{(\mu)}(v_{\ast}) -  \nonumber \\
  &  \quad \quad\quad   {\overline{\mathbf{Y}}^{q}}(  \{ \hat{\sigma}_q, r_q, \vartheta_q , \hat{\varphi}_q \}^{(\mu-1)}, v_q )  \|_F    .          \label{eq:estep_0_3_2} 
    \end{align} 
\end{subequations}

\textbf{E-step}: Estimate the $q$th hidden data of the $\mu$th iteration based on $\hat{v}_q^{(\mu )}$ as
\begin{equation} \label{eq:estep_1}
\begin{aligned}
    % &{\overline{\mathbf{Z}}^q}^{(\mu)}  =  {\overline{\mathbf{Y}}^{q}}(\hat{\bm{\theta}}_q^{(\mu-1)})   \\
    % &+ \beta_q \left( {\overline{\mathbf{Z}}^\text{rad}}  - \sum_{q' = 1}^{Q}{\overline{\mathbf{Y}}^{q} }(\hat{\boldsymbol{\theta}}_{q'}^{(u-1 )})     - \sum_{q'=1}^{Q}{\tilde{\mathbf{Y}}^{q'}}(\hat{\boldsymbol{\theta}}_{q'}^{(u )}) \right)  ,  
    {\overline{\mathbf{Z}}^q}^{(\mu)}  =  \dfrac{ \left( {\mathbf{W}}^T_{p} \odot {\tilde{\mathbf{\Omega}}}_{q}(\hat{v}_q^{(\mu )})^H  \right) }{U} {\mathbf{y}_{p,l}^{q}}^{(\mu)} .
    \end{aligned} 
\end{equation}
% where $q' = 1, 2, \dots, Q$. 
% \textcolor{red}{Here give more details on how to apply the initial Doppler in the E-step}

\textbf{M-step}: Estimate the $q$th parameter set of the $\mu$th iteration
\begin{align} \label{eq:mstep_1}
    \hat{\boldsymbol{\theta}}_q^{(\mu )} =  \arg\mathop{\min}_{\boldsymbol{\theta}_{q}} \| {\overline{\mathbf{Z}}^{q}}^{(\mu )} - {\overline{\mathbf{Y}}^{q}}(\boldsymbol{\theta}_q)  \|_F.
    %~ {\rm i.e.,} \nonumber
\end{align}
% where each \textbf{M-step} is a one-target MLE. 
% (${\mathbf{Y}}^\text{rad}$  $\hat{\mathbf{\Theta}}^{(\mu)} = \left[\hat{\bm{\theta}}_1^{(\mu)}; \hat{\bm{\theta}}_2^{(\mu)}; ...; \hat{\bm{\theta}}_Q^{(\mu)}  \right]$)
% \textcolor{red}{What other changes, in terms of vortex wavefront here?}
The implementation of the M-step follows the space-alternating framework.
In particular, the elevation angle ${\vartheta}_q$ is coupled with the mode index $u$ and subcarrier index $l$ in $\mathbf{A}_{r_q, \varphi_q, \vartheta_q}$.  
Hence, it is obtained using two-dimensional spectra as
\begin{align} \label{eq:mstep_2_1}
         & \hat{\vartheta}_q^{(\mu)}  = \arg \mathop{\min}_{ \vartheta_q }  \dfrac{1}{\beta_{q} \sigma_0^2} ( \text{vec}\left({\overline{\mathbf{Y}}^{q}}^{(\mu)} \right) - \nonumber \\
        &   \text{vec}\left( {\overline{\mathbf{Y}}^{q}}(   \hat{\sigma}_q^{(\mu-1)}, \hat{r}_q^{(\mu)}, \vartheta_q, \hat{\varphi}_q^{(\mu-1)}, {\hat{f}_{D,q}}^{(\mu)} )    \right)  )^{H} (  \text{vec}\left({\overline{\mathbf{Y}}^{q}}^{(\mu)} \right) \nonumber  \\ 
        &    - \text{vec}\left(  {\overline{\mathbf{Y}}^{q}}(   \hat{\sigma}_q^{(\mu-1)}, \hat{r}_q^{(\mu)}, \vartheta_q, \hat{\varphi}_q^{(\mu-1)}, {\hat{f}_{D,q}}^{(\mu)}  )  \right) ) .
\end{align}
The other parameters are estimated sequentially, as in other multi-parameter estimation approaches \cite{fleury1999channel,10878412}.
\begin{rem}
    {\em} This joint estimation strategy is different from the widely used recursive subspace-based method \cite{9968273,10620284}, where parameters are searched separately and require association.   
After the \textbf{M-step} of the $q$th target, we continue with the \textbf{E-step} of the $(q+1)$th target. After estimating all $Q$ targets in the $\mu$th EM loop, we perform the $(\mu+1)$th loop, until the cost function converges. 
\end{rem}
The Doppler-Robust Parameter Estimation is summarized in Algorithm~\ref{algo:DM_SAGE}.  
% {\em Remark :}   
% \textcolor{red}{Particularly, what other changes, in terms of vortex wavefront here}
% 
\begin{algorithm}[t]
\caption{Doppler-Robust Parameter Estimation}\label{algo:DM_SAGE}
\begin{algorithmic}[1] % This number denotes the frequency of line numbering
    % \State \textbf{Input:} $\mathbf{Y}^\text{rad}=\left[ \mathbf{y}_p^\text{rad} \right]_{P \times 1}$, $Q$, 
    \State \textbf{Input:} $\mathbf{Y}^\text{rad}$, $Q$, $\mu =0$  
    \State \textbf{Output:} $\hat{\mathbf{\Theta}}$
    % \State \textbf{Initialization:} Obtain $\mathbf{z}_{p}^\text{rad} $ in \eqref{eq:cdm_9}, $\mathbf{{Z}}^\text{rad} = \left[ \mathbf{z}_{p}^\text{rad} \right]_{P \times L}$, $\mathbf{\Theta}^{(0)}$
    % 
    \For{ the $\mu$th iteration, $\mu \geq 1$}
        \For{$q = 1, 2, \cdots, Q$}
            \State Obtain ${\mathbf{y}_{p,l}^{q}}^{(\mu)}$ in \eqref{eq:estep_0_1}
            \State \textbf{VCM} to estimate $\hat{v}_q^{(\mu )}$: 
            \For{$v_{\ast} \in \mathcal{V}$}
                \State Obtain ${\mathbf{{z}}^q}^{(\mu)}(v_{\ast})$ in \eqref{eq:estep_0_2}
                \State Estimate $\hat{v}_q^{(\mu )}$ in \eqref{eq:estep_0_3}
            \EndFor
            \State \textbf{E-step} based on $\hat{v}_q^{(\mu )}$: 
            \State Obtain ${\overline{\mathbf{Z}}^q}^{(\mu)}$ in \eqref{eq:estep_1}  
            \State \textbf{M-step}:
            \State Calculate $\hat{\boldsymbol{\theta}}_q^{(\mu )}$ in \eqref{eq:mstep_1}
        \EndFor
        \State Update $\hat{\mathbf{\Theta}}^{(\mu)} = \left[ \hat{\boldsymbol{\theta}}_q^{(\mu )} \right]_{Q \times 1 }$    
        % \If{cost function converges }
        %     \State $\hat{\mathbf{\Theta}} = \hat{\mathbf{\Theta}}^{(\mu)} $
        %     \State End the algorithm
        % \Else
        %     \State $\mu = \mu +1$
        % \EndIf
    \EndFor
\end{algorithmic}
\end{algorithm}
% \vspace{-0.5cm}
%Analysis 
% 
\begin{figure*}[hbt]
\setcounter{equation}{43}
\begin{align} \label{effectivechannel2}
h_{\bar{q}}^{\textrm{OAM}}(u,u') & =\sum\limits_{m = 1}^M \sum\limits_{n = 1}^M h^{\bar{q}}_{n,m}(p,l)e^{\left(-i{\ell_u}\alpha^{\bar{q}}_n+i{\ell_{u'}}\phi_m \right)}  
  =\frac{\beta}{2k_l \left[r_{\bar{q}}+(p-1)T_c v_{\bar{q}}\right]}e^{\big\{-ik_l\left[r_{\bar{q}}+(p-1)T_c v_{\bar{q}}\right]\big\}} \nonumber\\ 
& \quad \times \sum\limits_{m = 1}^M \sum\limits_{n = 1}^M e^{\bigg\{ \!-i{\ell_u}\alpha^{\bar{q}}_n+i{\ell_{u'}}\phi_{m} +i\frac{k_lR_tR_{r_{\bar{q}}}\cos\left(\alpha^{\bar{q}}_n-\phi_{m}\right)}{r_{\bar{q}}+(p-1)T_c v_{\bar{q}}} +i{k_l R_t}\sin\vartheta_{\bar{q}}\cos\left(\varphi_{\bar{q}}-\phi_{m}\right) -i{k_l R_{r_{\bar{q}}}}\sin\vartheta_{\bar{q}}\cos(\varphi_{\bar{q}}-\alpha^{\bar{q}}_n)
\bigg\}}.
\end{align}
\vspace{-0.5cm}
\setcounter{equation}{46}
\begin{align} \label{diagonal}
&\bar{h}^{\text{OAM}}_{\bar{q}}(u,u)= \frac{N\beta}{2k_l \left[r_{\bar{q}}+(p-1)T_c v_{\bar{q}}\right]} e^{\left\{-ik_l\left[r_{\bar{q}}+(p-1)T_c v_{\bar{q}}\right]\right\} } \sum\limits_{w = 1}^N e^{\bigg(i\frac{2\pi w}{N}\ell_u+i \frac{kR_tR_{r_{\bar{q}}}}{r_{\bar{q}}+(p-1)T_c v_{\bar{q}}}\cos\frac{2\pi w}{N}\bigg)}.
\end{align}
\vspace{-0.5cm}
\noindent\hrulefill
\end{figure*}
\subsubsection{Complexity Analysis}
We first investigate the complexity per iteration of the conventional E-M frame, assuming the ambiguity-free OAM sensing model in \eqref{eq:signal_4_1}.
For the $q$-th target, the range is estimated across $L$ subcarrier indices with complexity $\mathcal{O}(L\log{L})$, azimuth across $U$ modes with $\mathcal{O}(U\log{U})$, and Doppler across $P_\text{sen}$ chirps with $\mathcal{O}(P_\text{sen}\log{P_\text{sen}})$. Elevation is obtained using the Bessel template $J_{\ell_u}(k_l R_t \sin\vartheta)$, which couples the subcarrier index $l$ and the mode index $u$; thus, one elevation estimation sums over $LU$, and scanning an elevation grid of size $|\boldsymbol{\theta}|$ costs $\mathcal{O}(LU |\boldsymbol{\theta}|)$. The amplitude update is a scalar least-squares inner product and is negligible. In total, the cost per target is $\mathcal{O}\!\big(L\log{L}+U\log{U}+P_\text{sen}\log{P_\text{sen}}+LU|\boldsymbol{\theta}|\big)$.
% and for the targets $Q$, it is $\mathcal{O}!\big(Q,(L\log{L}+U\log{U}+P\log{P}+LU|\boldsymbol{\theta}|)\big)$.
% 
In the proposed approach, we consider the realistic ambiguity due to the Doppler perturbation of multiple targets. By applying the VCM-step, the velocity matching across $|\mathcal{V}|$ space, with each $\mathcal{O}(P_\text{sen}\log{P_\text{sen}})$, hence the complexity becomes $\mathcal{O}( |\mathcal{V}| P_\text{sen}\log{P_\text{sen}})$. Therefore, the cost of the Doppler-robust parameter estimation for the targets $Q$ is $\mathcal{O}\!\big( Q \left(L\log{L}+U\log{U}+|\mathcal{V}| P_\text{sen}\log{P_\text{sen}}+LU|\boldsymbol{\theta}| \right)\big)$.
\section{Sensing Aided Communication Design}\label{Sec:joint_design_isac}

\subsection{Communication Beamforming Design}
The communication to the target $\bar{q}$ requires designing the beamforming matrix $\mathbf{T}(p,l)$ and beam steering matrix $\mathbf{D}(p,l)$ defined in \eqref{xTD}.
% using the estimated angular information in the sensing phase.  
We have discussed the LoS MIMO channel of UCA transceivers in \eqref{hmn}. 
In the communication phase, each Tx antenna transmits $U$ modes simultaneously using the (partial) DFT matrix $\mathbf{F}^H$ and each Rx antenna separates the $U$ modes using the conjugate transpose DFT matrix $\mathbf{F}$. 
% the integrated transmitted UCA can generate the multi-mode OAM beam with  baseband (partial) discrete Fourier transform (DFT) matrix $\mathbf{F}$. The communication Rx at the target $\bar{q}$ target then utilizes the conjugate transpose matrix of $\mathbf{F}$ to separate and decouple the different modes of the OAM beams. 
After that, the effective OAM channel matrix from the Tx to the target $\bar{q}$ can be expressed as
\setcounter{equation}{42}
\begin{align} \label{effectivechannel}
\mathbf{H}^{\rm OAM}_{\bar{q}}(p,l)=\mathbf{F}\mathbf{H}
_{\bar{q}}(p,l)\mathbf{F}^H,
\end{align}
where the element in $\mathbf{H}^{\rm OAM}_{\bar{q}}(p,l)$ can be written as \eqref{effectivechannel2}.
% \begin{align}\label{effectivechannel2}
% &h_q^{\textrm{OAM}}(u,u')=\sum\limits_{m = 1}^M \sum\limits_{n = 1}^M h^q_{n,m}(p,l)\exp \left(-i{\ell_u}\alpha^q_n+i{\ell_{u'}}\phi_m \right) \nonumber\\
% &=\frac{\beta}{2k_l \left[r_q+(p-1)T_c v_q\right]}\exp\big\{-ik_l\left[r_q+(p-1)T_c v_q\right]\big\}\nonumber\\ 
% &\sum\limits_{m = 1}^M \sum\limits_{n = 1}^M \exp^\bigg\{ \!-i{\ell_u}\alpha^q_n+i{\ell_{u'}}\phi_{m} +i\frac{k_lR_tR_{r_q}\cos\left(\alpha^q_n-\phi_{m}\right)}{r_q+(p-1)T_c v_q}\nonumber \\
% &+i{k_l R_t}\sin\vartheta_q\cos\left(\varphi_q-\phi_{m}\right) -i{k_l R_{r_q}}\sin\vartheta_q\cos(\varphi_q-\alpha^q_n)
% \bigg\}.
% \end{align}
%
Upon obtaining the estimated position $(\hat{r}_{\bar{q}},\hat{\theta}_{\bar{q}},\hat{\varphi}_{\bar{q}})$ and the velocity $v_{\bar{q}}$ feedback from the integrated Tx, we can derive the complete CSI of the effective OAM channel based on \eqref{effectivechannel2}. 
% From \cite{9139401}, 
$\varphi_{\bar{q}}$ and $\vartheta_{\bar{q}}$ can result in inter-mode interference among different OAM modes, even if $\varphi_{\bar{q}}$ or $\vartheta_{\bar{q}}$ is small.
% \textcolor{red}{Yuan: Here, can we say more directly, use estimated angles to eliminate which term in \eqref{effectivechannel2}.}
 
To alleviate the inter-mode interference induced by the misalignment, we apply beamforming and beam steering at the Tx and communication Rx, respectively. First, the beamforming at the Tx side adjusts the generated multi-mode OAM beams towards the direction of the communication Rx, thereby compensating for the phase variations caused by $\varphi_{\bar{q}}$ and $\vartheta_{\bar{q}}$ at the transmitted UCA. Based on \eqref{effectivechannel2}, the transmit beamforming matrix $\mathbf{P}(k_l)$ can be designed as $\mathbf{P}(k_l)=\mathbf{1}\otimes\mathbf{\mathfrak{p}}(k_l)$, where $\mathbf{\mathfrak{p}}(k)=[e^{iW_1(k_l)}, e^{iW_2(k_l)}, \cdots, e^{iW_M(k_l)}]$, and
\setcounter{equation}{44}
\begin{equation} \label{W_m}
W_m(k_l)=-k_l R_t\sin\hat{\theta}_{\bar{q}}\cos\left(\hat{\varphi}_{\bar{q}}-\phi_m\right).
\end{equation}
%
% $m=1,\cdots,M$. 
Similarly, the beam steering matrix $\mathbf{B}(k_l)$ at the Rx steers the beam pattern towards the direction of the incident OAM beams,  which can be designed as $\mathbf{B}(k_l)=\mathbf{1}\otimes\mathbf{\mathfrak{b}}(k_l)$, where $\mathbf{\mathfrak{b}}(k_l)=[e^{iW_1(k_l)}, e^{iW_2(k_l)}, \cdots, e^{iW_N(k_l)}]$, and
\begin{equation} \label{W_m_}
W_n(k_l)=k_l R_{r_{\bar{q}}}\sin\hat{\theta}_{\bar{q}}\cos\left(\hat{\varphi}_{\bar{q}}-\alpha^{\bar{q}}_n\right).
\end{equation}
%
% $n=1,\cdots,N$. 
After performing the beamforming $\mathbf{P}(k_l)$ and beam steering $\mathbf{B}(k_l)$, the effective OAM channel matrix $\mathbf{\bar{H}}^{\text{OAM}}_{\bar{q}}(p,l)=\left(\mathbf{F}\odot\mathbf{B}(k_l)\right)\mathbf{H}_{\bar{q}}(p,l)\left(\mathbf{F}^H_U\odot\mathbf{P}(k_l)\right)$ becomes a diagonal matrix, {i.e.,  the inter-mode interference caused by $\varphi_{\bar{q}}$ and $\vartheta_{\bar{q}}$ has been effectively eliminated}\footnote{{The purpose of this phase-only steering is to align the effective LoS OAM channel rather than to reconstruct the full spatial amplitude profile of a tilted vortex beam. The detailed proof process is similar in \cite[Theorem 1]{9139401}.}}. 
% , which implies that 
% 
Then, the $u$th diagonal element $\bar{h}^{\text{OAM}}_{\bar{q}}(u,u)$ in $\mathbf{\bar{H}}^{\text{OAM}}_{\bar{q}}(p,l)$ is given by \eqref{diagonal},
%
% \begin{align} \label{diagonal}
% &\bar{h}^{\text{OAM}}_q(u,u)= \frac{N\beta}{2k_l \left[r_q+(p-1)T_c v_q\right]}\times\nonumber\\
% &\exp\left\{-ik_l\left[r_q+(p-1)T_c v_q\right]\right\}\times\nonumber\\
% &\sum\limits_{w = 1}^N\exp\bigg(i\frac{2\pi w}{N}\ell_u+i \frac{kR_tR_{r_q}}{r_q+(p-1)T_c v_q}\cos\frac{2\pi w}{N}\bigg),
% \end{align}
%
where $w=m-n$. For $r_{\bar{q}}\gg R_t, R_{r_{\bar{q}}}$ the $u$-th diagonal elements of $\mathbf{\bar{H}}^{\text{OAM}}_{\bar{q}}(p,l)$ can be approximately obtained as \cite{9139401}  
\setcounter{equation}{47}
\begin{align} \label{hiord}
\bar{h}^{\text{OAM}}_{\bar{q}}&(u,u)\approx  \frac{\beta}{2k_l \left[r_{\bar{q}}+(p-1)T_c v_{\bar{q}}\right]}\frac{N^2}{2^\tau}\frac{i^\tau}{\tau!}\cdot\nonumber\\
&\left\{\frac{k_lR_tR_{r_{\bar{q}}}}{r_{\bar{q}}+(p-1)T_c v_{\bar{q}}}\right\}^{\tau}\! e^{\left\{-ik_l\left[r_{\bar{q}}+(p-1)T_c v_{\bar{q}}\right]\right\}},
\end{align}
where $\tau={\min\left\{|\ell_u|, N-|\ell_u|\right\}}$. Given that $k_l$, $p$, $R_t$, $R_{r_{\bar{q}}}$, $T_c$, $N$ and $\ell_u$ are known to the communication Rx, the effective channel coefficient $\bar{h}_{\text{OAM}}(u,u)$ of each OAM mode is only a function of $r_{\bar{q}}$ and $v_{\bar{q}}$, which leads to very simple signal detection, i.e., $\mathbf{\Lambda}(p,l)=\textrm{diag}\{\lambda(k_l,p,1),\cdots,$ $  \label{}(k_l,p,U)\}$ and
\begin{align} \label{zeta}
\lambda&(k_l,p,u)= \frac{\beta}{2k_l \left[\hat{r}_{\bar{q}}+(p-1)T_c \hat{v}_{\bar{q}}\right]}\frac{N^2}{2^\tau}\frac{i^\tau}{\tau!}\cdot\nonumber\\
&\left\{\frac{k_lR_tR_{r_{\bar{q}}}}{\hat{r}_{\bar{q}}+(p-1)T_c \hat{v}_{\bar{q}}}\right\}^{\tau} e^{\left\{-ik_l\left[\hat{r}_{\bar{q}}+(p-1)T_c \hat{v}_{\bar{q}}\right]\right\}},
\end{align}
$u=1,2,\cdots,U.$ Finally, the detected downlink signal vector at the target $\bar{q}$ can be written as
\begin{align} \label{xp}
&\mathbf{x}_{{\bar{q}}}(p,l)=\nonumber\\
&\mathbf{\Lambda}_U^{-1}(p,l)\!\left(\mathbf{F}\!\odot\!\mathbf{B}(k)\!\right)\!\left(\mathbf{H}_{\bar{q}}(p,l) (\mathbf{F}^H_U\!\odot\!\mathbf{P}(k)\!)\mathbf{s}(p,l)\!+\!\mathbf{n}_\text{com}^{\bar{q}}(p,l)\!\right)\nonumber \\
&\approx\mathbf{s}(p,l)+\mathbf{\hat{n}}_\text{com}^{\bar{q}}(p,l),
\end{align}
where $\mathbf{\hat{n}}^{\bar{q}}_\text{com}(p,l) = \mathbf{\Lambda}_U^{-1}(p,l)\left(\mathbf{F}\odot\mathbf{B}(k)\right)\mathbf{n}_\text{com}^{\bar{q}}(p,l)$ is the corresponding noise vector.

Therefore, the precoding and post-processing matrices defined in \eqref{xTD} are
\begin{align}
    \begin{cases}\label{eq:beamforming_matrices}
    & \mathbf{D} = \mathbf{\Lambda}_U^{-1}(p,l)\!\left(\mathbf{F}^H\!\odot\!\mathbf{B}(k)\!\right) ,\\
    & \mathbf{T} = \mathbf{F}\!\odot\!\mathbf{P}(k). 
    \end{cases} 
\end{align}
% Explain the difference among $\equiv$ $\triangle$ and $=$.
% \textcolor{red}{Yuan: here we need to point out: $\mathbf{D} \equiv \mathbf{\Lambda}_U^{-1}(p,l)\!\left(\mathbf{F}\!\odot\!\mathbf{B}(k)\!\right)$ and $\mathbf{T} \equiv \mathbf{F}^H_U\!\odot\!\mathbf{P}(k)$  defined in \eqref{y}. }

% \subsection{Communication Spectrum Efficiency Design}
\subsection{{Sensing-Aided Communication Performance Analysis}}
Given the inevitable estimation errors in $(r_{\bar{q}},\vartheta_{\bar{q}},\varphi_{\bar{q}})$ and $v_{\bar{q}}$ in practice, we must assess the impact of these errors on the SE during the communication phase. Following \eqref{xp}, we have
\begin{align} \label{y2}
&x_{\bar{q}}(p, k_l,u)  = \nonumber\\
&\frac{1}{\lambda(k_l,p,u)}\sum\limits_{u' = 1}^U \bar{h}^{\text{OAM}}_{\bar{q}}(u,u')s_{\ell_u}(p,l) + \hat{n}_\text{com}^{\bar{q}}(p,k_l,u) \nonumber\\
& = s_{\ell_u}(p,l)+ 
\sum\limits_{u' \neq u} \bar{h}^{\text{OAM}}_{\bar{q}}(u,u')\frac{s_{\ell_u}(p,l)}{\lambda(k_l,p,u)} + \hat{n}_\text{com}^{\bar{q}}(p,k_l,u).
\end{align}
Therefore, the signal-to-interference-plus-noise ratio (SINR) of the $u$-th mode OAM can be formulated as
\begin{align} \label{SINR}
&\textrm{SINR}(p,k_l,u)= \nonumber\\ &\frac{\lambda^2(k_l,p,u)\mathbb{E}\left(\left|s_{\ell_u}(p,l)\right|^2\right)} {\sum\limits_{u'=1}^U\!|L(u,u')|^2 \mathbb{E}\left(\!\left|s_{\ell_u}(p,l)\right|^2\!\right) \!+\!\lambda^2(k_l,p,u)\mathbb{E}\!\left(\!\left|\hat{n}_\text{com}^{\bar{q}}(p,k_l,u)\right|^2\!\right)},
\end{align}
where
\begin{align}
L(u,u')=\left\{
\begin{matrix}
 \bar{h}^{\text{OAM}}_{\bar{q}}(u,u'), &u'\neq u;\\
\qquad0\qquad, &u'=u.
\end{matrix}\right.
\end{align}
{It is worth noting that the impact of CSI estimation errors is explicitly reflected in \eqref{y2} and \eqref{SINR}. The coefficient $\lambda(k,p,u)$ in \eqref{zeta} is constructed from the estimated sensing parameters $(\hat r_{\bar q},\hat\vartheta_{\bar q},\hat\varphi_{\bar q},\hat v_{\bar q})$. Therefore, any estimation error in these parameters leads to a mismatch between the reconstructed OAM channel and the true OAM channel. This mismatch distorts the desired gain $\lambda(k,p,u)$ in the numerator of \eqref{SINR}. Moreover, imperfect phase alignment prevents the equivalent OAM channel from being fully diagonalized, resulting in residual inter-mode leakage terms $L(u,u')$, $u'\neq u$, in the denominator of \eqref{SINR}. Consequently, the CSI estimation error affects the communication SINR through both the desired signal gain and the residual leakage, and the resulting SE is captured as}
% Thus, the SE during the communication phase described in \eqref{y2} can be written as
%
\begin{equation} \label{eq9}
{
C = \left(1-\frac{P_{\text{sen}}}{P}\right)\sum_{u=1}^U\sum_{p=1}^{P}\log_2\left(1 + \textrm{SINR}(p,k_l,u)\right).
}
\end{equation}
%
% where $P$ is the total number of symbols within a coherent block, and $P_{\text{sen}}$ is the number of symbols used for the sensing phase.
\begin{rem}
{
% The overall joint design of the sensing-aided communication framework is illustrated in Fig.~\ref{fig:operation_flow}.
In \eqref{eq9}, $\text{SINR}$ is a function of $P_\text{sen}$. An insufficient pilot length $P_{\text{sen}}$ can degrade CSI estimation and reduce the SINR during the communication phase, thereby suppressing the SE.
However, excessive pilot length $P_{\text{sen}}$ reduces the number of communication frames, thereby reducing SE.} 
Therefore, the frame allocation between the sensing and communication phases must be balanced.
{As proved in Appendix~\ref{App:proof_2}, the sensing-aided beamforming gain can offset the resource overhead only when the achieved SINR exceeds an overhead-dependent threshold, which provides a theoretical condition for selecting $P_{\rm sen}$.}
\end{rem}
% where sufficient $P_{\text{sen}}$ in the sensing phase improves the sensing performance at the cost of occupying more communication symbols. Meanwhile, the perfect CSI obtained from the sensing phase is essential for enabling OAM beamforming.

% \subsection{Joint design}
% %
% \begin{align} \label{eq9_}
% & C = \left(1-\frac{P_{\text{sen}}}{P}\right)\sum_{u=1}^U\sum_{p=P_{\text{sen}}+1}^{P}\log_2\left(1 + \textrm{SINR}(p,k_l,u)\right)+\nonumber\\
% &{\frac{P_{\text{sen}}}{P}\sum_{u=1}^U\log_2\left(1 + A(p,k_l,u)\right)},
% \end{align}
% %
% {WX: A is the SINR of the sensing + communication part of the communication Rx. The simulation of communication capacity versus SNR requires the estimation error of parameters under different SNR levels. We can further simulate the relationship with Psen to investigate whether a trade-off exists.}
\begin{table}[t!]
    \caption{Configurations used in simulation  }
    % \vspace{-0.2cm}
    \centering
    \begin{tabular}{lclclcl}
        \toprule
        Configurations &   Values & Descriptions  \\
        \midrule
        Carrier frequency $f_c$  &   $77$  [GHz] &  \\
        Wavelength $\lambda$ &  3.9 [mm]  & $\lambda = \frac{c}{f_c} $ \\  
        Bandwidth $B$  & $200$ [MHz] \\
        No. subcarriers $L$ &  128   \\
        Subcarrier $\Delta f$  & $1.5625$ [MHz]  & $\Delta f = \frac{B}{L}$ \\ 
        Subcarrier duration $T_c$ &  $6.67$ [$\mu$s]  & $T_c \geq T_\text{CP}+\frac{1}{\Delta f} $  \\
        No. symbols per CPI $P$ & $ 1024 $   &   \\
        CPI duration $T_\text{CPI}$  &    & $T_\text{CPI} = P_{\text{sen}} T_c$
        % , {$P_\text{sen} \leq P$} 
        \\
        {CP duration $T_\text{CP}$} & $0.635$~$\mu s$   &  $T_\text{CP} \geq \frac{2 R_\text{max}}{c}$ \\
        \midrule
        Unambiguous range $R_\text{max}$ &  $95.2$ [m] & $R_\text{max} =\frac{c}{2 \Delta f}$ \\ 
        Range resolution $\Delta_R$ &  $0.75$ [m] & $\Delta_R =\frac{c}{2B}$ \\ 
        Unambiguous velocity  $v_\text{max}$ &  $ 9.1315$ [m/s] & $v_\text{max} =\frac{\lambda}{4 T_c M}$ \\
        Velocity resolution $\Delta_v$ &  $0.285$-$2.283$ [m/s] & $\Delta_v =\frac{\lambda}{2 T_\text{CPI}}$ \\
        \midrule
        SNR & $0$-$20$ [dB] \\  
        % Simulation space [m$^2$] & [${00 \times 00}$] \\
        % NO. Tx $M$ & $1$  \\
        No. of Ants in UCAs & $ 16 $  \\
        Radius of UCAs & $ \frac{M\lambda}{4 \pi }  $  \\
        No. of modes   & $16$ \\
        \bottomrule
    \end{tabular}
    \label{tab:simu_ISAC_configs_1}
\end{table} 
\section{Simulation and discussion}\label{sec:simu_isac}
    The proposed framework focuses on the Doppler-induced angle-mode coupling and cross-mode interference in vortex wavefront sensing under the considered OAM-OFDM framework.
    Different from this spatial/mode-domain problem, OTFS improves Doppler resilience by exploiting the delay-Doppler domain. Therefore, the following numerical comparisons mainly evaluate the performance gains of the proposed design within the OAM-OFDM framework, rather than directly comparing with other Doppler-resilient ISAC waveforms.
    In this sense, the two designs are not conflicting, and the proposed framework can potentially be combined with Doppler-resilient waveforms such as OTFS.

The millimeter-wave OFDM signal is used for simulation and discussion of the proposed ISAC framework, as this is a potential frequency band considered in~\cite{8828030}. 
We consider LoS propagation among Tx, targets/UE, and Rx.  
Unless otherwise mentioned, the default simulation configurations are listed in Tab.~\ref{tab:simu_ISAC_configs_1}\footnote{{The signal-to-noise ratio (SNR) refers to Post-beamforming SNR}. }. 
{
There are $P$ symbols within one CPI, where $P_\text{sen}\leq P$ symbols are used for sensing. Therefore, $P_\text{sen}$ determines the effective sensing duration, the velocity resolution, and consequently the sensing performance. The velocity search interval is set according to the corresponding velocity resolution, i.e., $\Delta v=\frac{\lambda}{2P_\text{sen}T_c}$. The target velocities are not intentionally aligned with the velocity grid, reflecting practical off-grid cases.
}

% (In many OAM sensing works like \cite{10311524}, the simulation frequency is $9.5$ GHz.
% However, \textit{60-GHz IEEE 802.11ad standard wireless protocol} has been used with time division multiplexing of radar-only and radar communication frames . Hence, we use a $60$ GHz carrier.) 

% \subsection{Case Study of Azimuth-Doppler Ambiguity Elimination}
% % The same velocity, 
% % Change velocity to 5m/s and 50 m/s 
% \subsubsection{Multi-Targets Estimation Case}
% We set (let's say) 3 targets with velocities (0m/s, 10m/s, 20m/s). This is to say, in our simulation period, they have a moving trajectory, hence the beamforming in communication is also time varying. 
%%%%%%%%%%
\begin{figure*}[ht]
    \centering
        \begin{minipage}{0.34\textwidth}
        \centering
        \includegraphics[width=\textwidth]{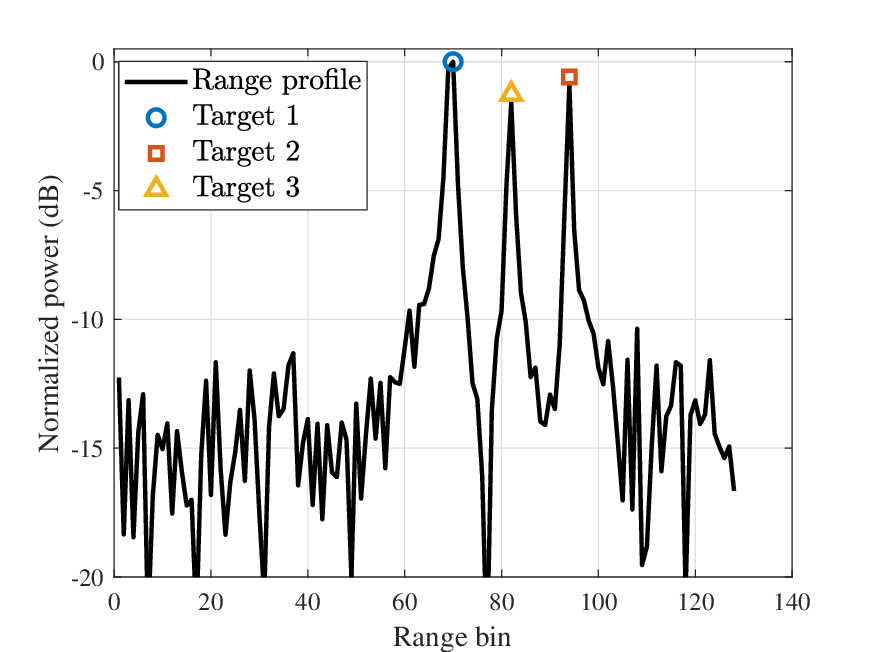}
        \subcaption{}
    \end{minipage}\hfill
    \begin{minipage}{0.31\textwidth}
        \centering
        \includegraphics[width=\textwidth]{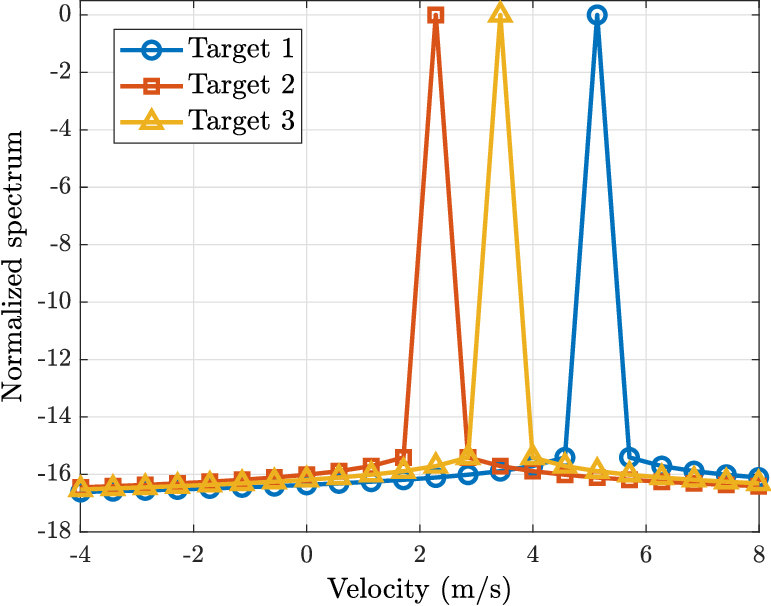}
        \subcaption{}
    \end{minipage}\hfill
        \begin{minipage}{0.31\textwidth}
        \centering
        \includegraphics[width=\textwidth]{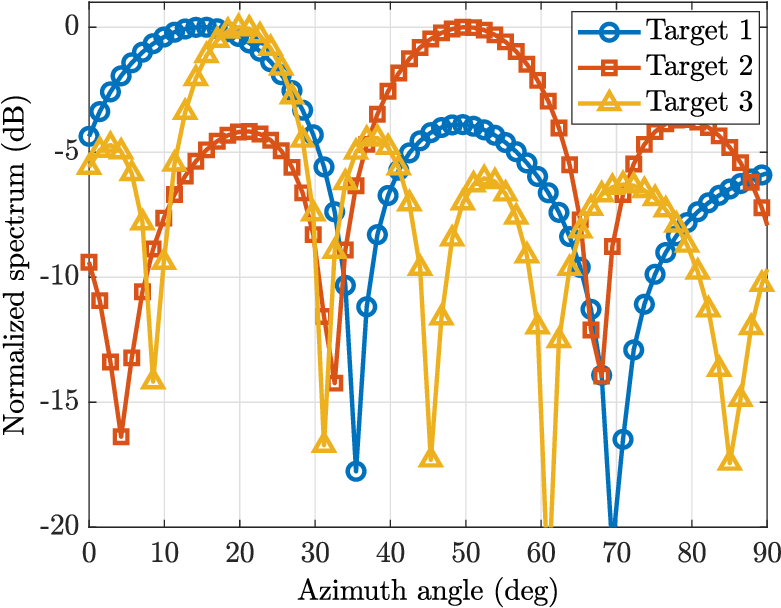}
        \subcaption{}
    \end{minipage}\hfill
        \begin{minipage}{0.31\textwidth}
        \centering
        \includegraphics[width=\textwidth]{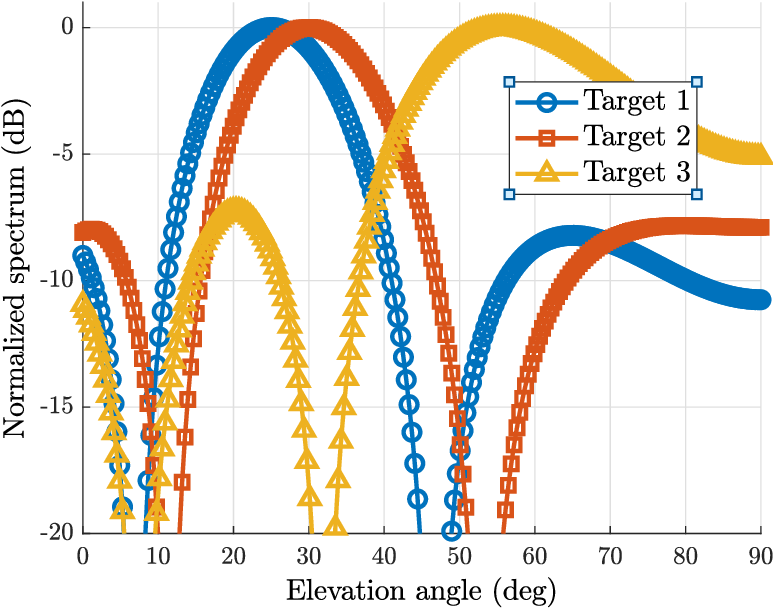}
        \subcaption{}
    \end{minipage}\hfill
    \begin{minipage}{0.36\textwidth}
        \centering
        \includegraphics[width=\textwidth]{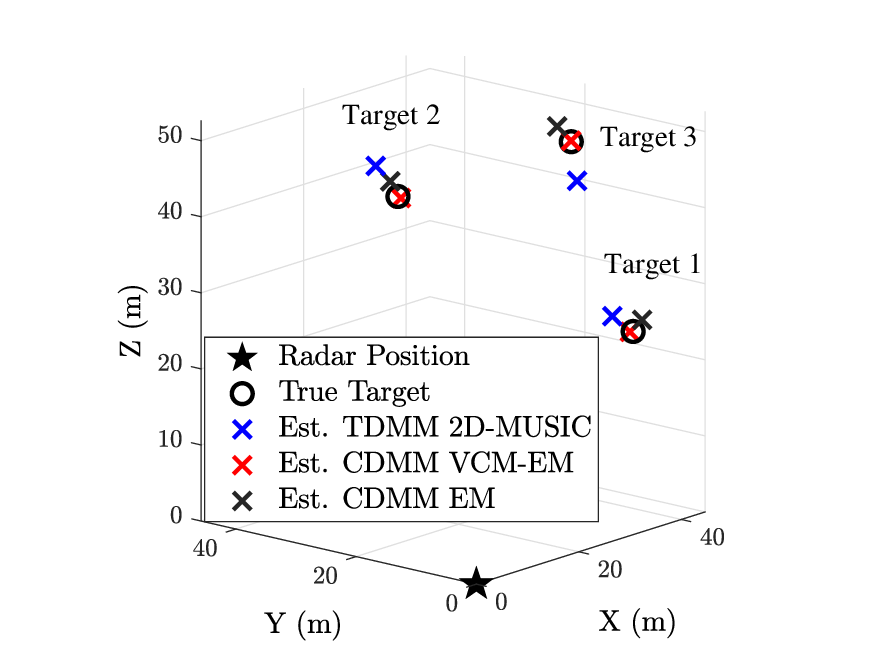}
        \subcaption{}
    \end{minipage}\hfill
        \begin{minipage}{0.33\textwidth}
        \centering
        \includegraphics[width=\textwidth]{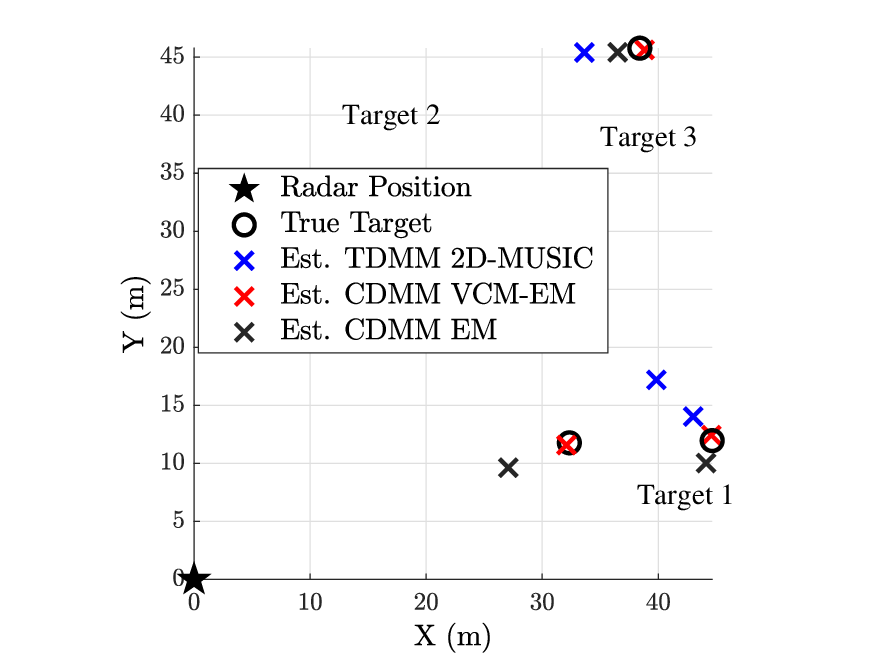}
        \subcaption{}
    \end{minipage}\hfill
        % \vspace{-0.5cm}
    \caption{(a) Range estimation of multiple targets; (b) Velocity estimation of multiple targets; (c) Azimuth estimation of multiple targets; (d) Elevation estimation of multiple targets; (e) {3-D localization comparison among the proposed CDMM-based modulation with VCM-EM, CDMM-based modulation with conventional EM, and TDMM-based modulation with FFT-assisted 2-D MUSIC}; (f) Top-view of localization comparison.   
    }\label{fig_multi_tar}
    % \vspace{-0.5cm}
\end{figure*}
%%%%%%%%%%
%%%%%%%%%%
\begin{figure}[t]
    \centering
    \includegraphics[width=0.45\textwidth]{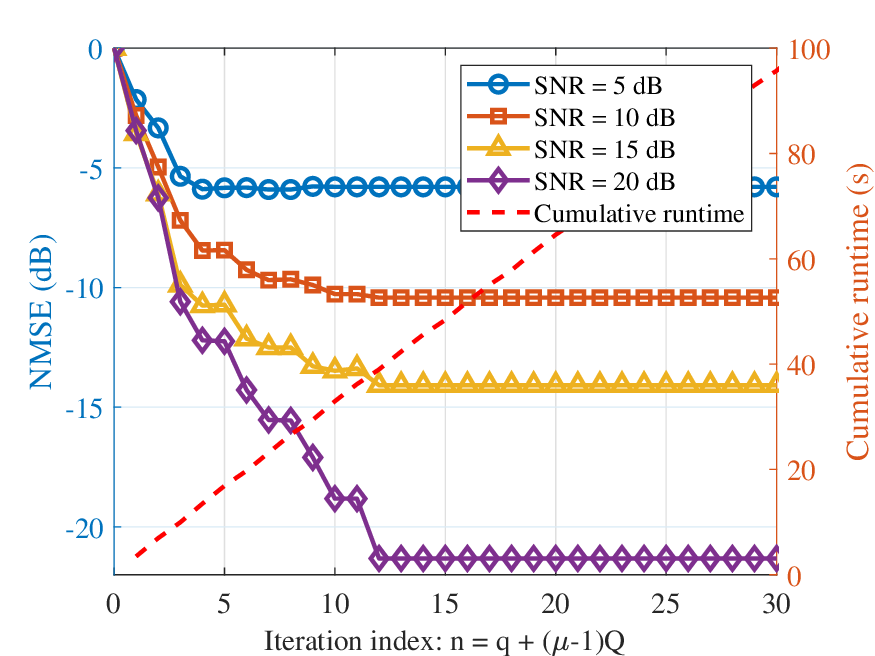}
          % \vspace{-0.3cm}
    \caption{
    {Convergence curves under different SNR levels and CPU times.}  
    }\label{fig:nmse}
     \vspace{-0.5cm}
\end{figure}
%%%%%%%%%%%%%%%
\subsection{Multi-Target Parameter Estimation and Localization}
\subsubsection{Scenario Settings}
For this experiment, three targets located at spherical coordinates $(51, 15, 25)$, $(69, 50, 30)$, and $(60, 20, 55)$, respectively, where the first coordinate denotes range in meters, and the last two coordinates denote azimuth and elevation angles in $\deg$. The radial velocities of the three targets are $5$ m$/$s, $2.2$ m$/$s, and $3.5$ m$/$s, respectively. The receiving SNR is $15$ dB. Codes used to generate the multi-target simulation results are available\footnote{\url{https://github.com/liuy2022/OAM_VortexSensing_CDMM}  }.
% 
% receiving SNR is assumed in this 
\subsubsection{Results Visualization}
% Figure~\ref{fig_multi_tar} illustrates the simulation result of a three-target scenario, where parameter sets of each target are sequentially estimated by the proposed framework.
% Specifically, Fig.~\ref{fig_multi_tar}(a) shows the range profile with three dominant peaks corresponding to the targets. Recalling the \textbf{VCM} step in \eqref{eq:estep_0_2} and \eqref{eq:estep_0_3}, the velocity is estimated from a bank of Doppler-matching spectra conditioned on the target's a-priori, here simply using the range bin from Fig.~\ref{fig_multi_tar}(a); Fig.~\ref{fig_multi_tar}(b) shows the resulting Doppler estimates, clearly separating the three velocities. Given the estimated range–Doppler, Fig.~\ref{fig_multi_tar}(d) shows the mode-domain spectrum for azimuth, while Fig.~\ref{fig_multi_tar}(e) shows the elevation spectrum obtained with the Bessel-based template; both display well-resolved peaks per target. Finally, Figs.~\ref{fig_multi_tar}(c) and \ref{fig_multi_tar}(f) assemble the estimates into 3-D positions and compare the proposed CDMM with a conventional TDMM baseline, with the top view included for clarity.
In Fig.~\ref{fig_multi_tar}, the estimation results of the three-target case are presented, where the proposed VCM-EM  algorithm sequentially estimates parameters of each target. 
In Fig.~\ref{fig_multi_tar}(a), the range profile shows three dominant peaks, one per target. 
{In the VCM~step in \eqref{eq:estep_0_2} and \eqref{eq:estep_0_3}, the velocity is estimated from a bank of Doppler-matched spectra conditioned on the target's a priori range information,} where the range bin is selected from Fig.~\ref{fig_multi_tar}(a). The resulting Doppler estimates in Fig.~\ref{fig_multi_tar}(b) clearly separate the three velocities. 
Given the range-Doppler estimates, Fig.~\ref{fig_multi_tar}(c) shows the mode-domain spectrum for azimuth, and Fig.~\ref{fig_multi_tar}(d) shows the elevation spectrum using the Bessel-based template. Both exhibit well-resolved peaks for each target. 

% Finally, Fig.~\ref{fig_multi_tar}(c) and (f) assemble the estimates into 3-D positions and compare the proposed CDMM with a conventional TDMM baseline; a top view is included for clarity.
{
The reconstructed 3-D target positions and the corresponding top view are further illustrated in Fig.~\ref{fig_multi_tar}(e) and Fig.~\ref{fig_multi_tar}(f). The proposed VCM-EM under CDMM-based OAM modulation is compared with two representative baselines. For the TDMM-based OAM modulation baseline, range-Doppler FFT is first applied to detect the three targets, followed by 2-D MUSIC angle estimation at the corresponding range bins. For the CDMM-based OAM modulation baseline, the conventional EM algorithm is employed to jointly estimate parameters. The comparison thus evaluates both the multiplexing strategies, namely TDMM and CDMM, and the parameter estimation methods, namely 2-D MUSIC~\cite{10620284,9968273,10304513}, conventional EM~\cite{feder1988parameter}, and the proposed VCM-EM.}
% Specifically, in Fig.~\ref{fig_multi_tar}(a), the range profile reveals three dominant peaks corresponding to the targets. 
% Recall the \textbf{VCM}-step \eqref{eq:estep_0_2} and \eqref{eq:estep_0_3}, the velocity estimation is based on a bunch of Doppler matching spectra, given by range bin of each target, where Fig.~\ref{fig_multi_tar}(b) shows the estimation of Doppler velocity, clearly separating the velocities of the three targets.
% Conditioned on estimated range-Doppler, Fig.~\ref{fig_multi_tar}(d) shows the mode-domain spectrum for azimuth estimation, while Fig.~\ref{fig_multi_tar}(e) shows the elevation spectrum obtained with the Bessel-based template; both exhibit well-resolved peaks per target. 
% Finally, Fig.~\ref{fig_multi_tar}(c) and (f) assemble the estimates into 3-D positions and provide a comparison between the proposed CDMM and a conventional TDMM baseline, together with the top view for visual clarity.
\subsubsection{Quantitative Comparison}
{Using the proposed VCM-EM for CDMM modulation, the estimated positions of the three targets are $(51, 15.59, 24.85)$, $(69, 49.6, 30.75)$, and $(60, 19.84, 55.36)$, respectively, with 3-D localization errors of $0.50$, $0.99$, and $0.39$ meters.
As benchmarks, the combination of  FFT and $2$-D MUSIC applied to} TDMM yields $(51, 18.08, 26.73)$, $(69, 54.22, 34.29)$, and $(60, 23.35, 43.68)$, respectively, with 3-D localization errors of $2.91$, $6.72$, and $12.05$ meters;
the conventional EM algorithm applied on CDMM yields  $(51, 15.06, 25.60)$, $(69, 50.46, 30.88)$, and $(60, 20.33, 56.86)$, respectively, with 3-D localization errors of $2.83$, $3.12$, and $6.68$ meters.
This demonstrates the benefits of both the proposed CDMM and Doppler-robust parameter estimation approach, particularly in scenarios involving multiple moving targets.

\subsubsection{Convergence}
The convergence behavior of the proposed VCM-EM algorithm\footnote{{As in many grid-based EM/SAGE and maximum-likelihood implementations, the discrete search grid introduces a quantization effect when the true parameter is located between two adjacent grid points. Reducing the grid spacing decreases this quantization loss at the cost of increased search complexity.
Similarly, increasing the number of OAM modes improves mode-domain diversity but increases the coding and VCM-dictionary dimensions. Therefore, the grid resolution and the number of modes should be selected to balance accuracy and complexity.
}} for the three-target case is shown in Fig.~\ref{fig:nmse}, which reports the normalized mean-square error (NMSE) in decibels.
Across all SNRs, the NMSE decreases with iteration and stabilizes at SNR-dependent levels, indicating numerically stable updates.
% Two additional observations are noteworthy.

Meanwhile, the central processing unit (CPU) runtime is evaluated in MATLAB~$2021$ on a laptop with an Intel Core i9-11950H CPU ($8$~cores, $2.60$~GHz), as shown in Fig.~\ref{fig:nmse}. The unoptimized implementation requires less than $3.2$~s per iteration, making it suitable for offline or quasi-real-time base-station processing, whereas real-time deployment would require further optimization.

% Meanwhile, the central processing unit (CPU) runtime is recorded and illustrated in Fig.~\ref{fig:nmse}.
% The runtime is measured in MATLAB 2021 environment on a laptop equipped with an $11$th Gen Intel Core $i9$-$11950H$ processor, with a base frequency of $2.60$~GHz, $8$~cores, $16$~logical processors, and a maximum clock speed of $4.9$~GHz.
% The average per-iteration runtime is less than $3.2$~seconds for this unoptimized MATLAB implementation.
% This indicates that the computational cost is manageable at the base station either in offline or quasi-real-time processing, while real-time deployment would require implementation-level optimization on dedicated hardware. }

\begin{figure*}[ht]
    \centering
    \begin{minipage}{0.45\textwidth}
        \centering
        \includegraphics[width=\textwidth]{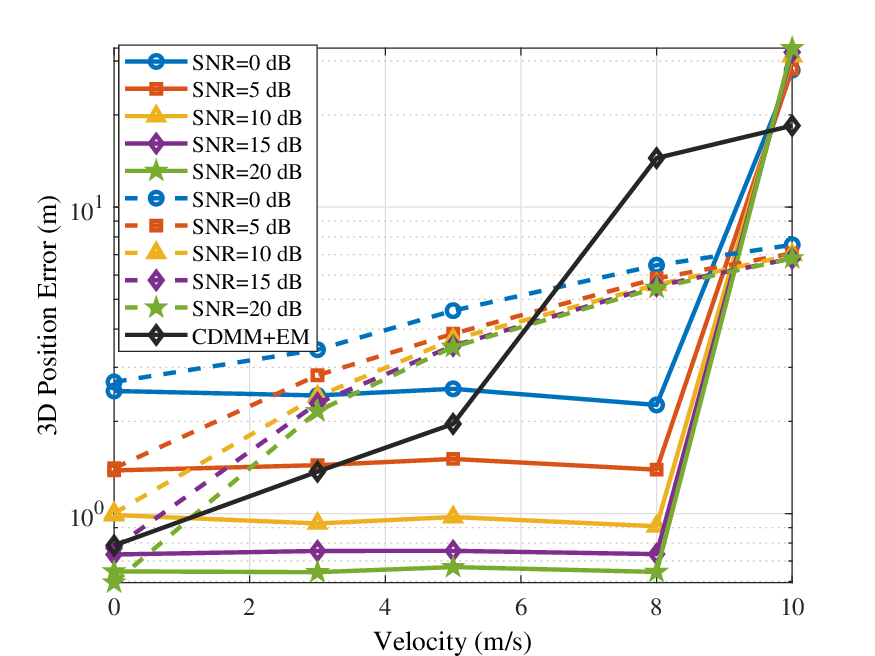}
        \subcaption{Comparison of 3-D localization errors}
    \end{minipage}\hfill
    \begin{minipage}{0.45\textwidth}
        \centering
        \includegraphics[width=\textwidth]{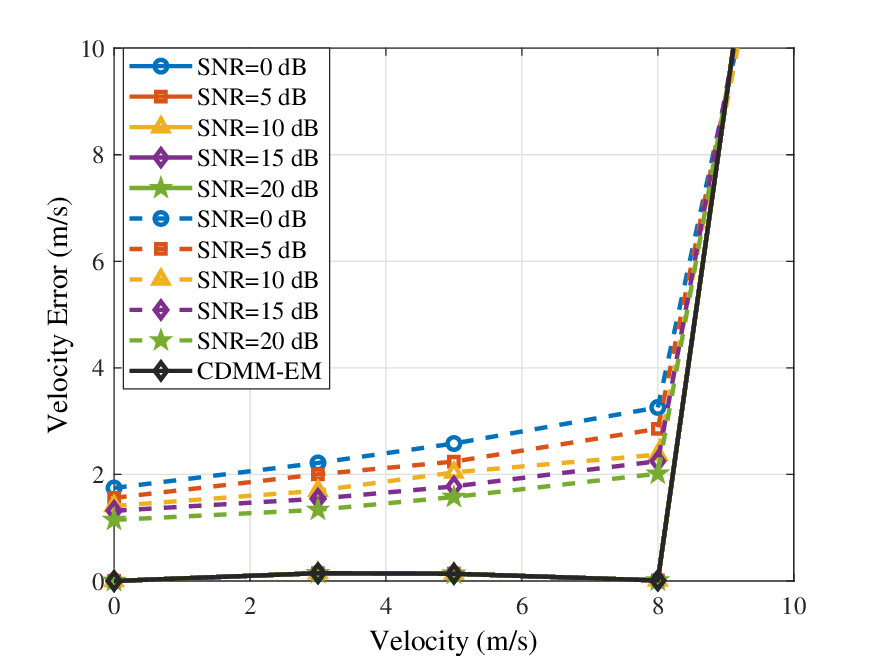}
        \subcaption{Comparison of velocity errors}
    \end{minipage}
        \begin{minipage}{0.45\textwidth}
        \centering
        \includegraphics[width=\textwidth]{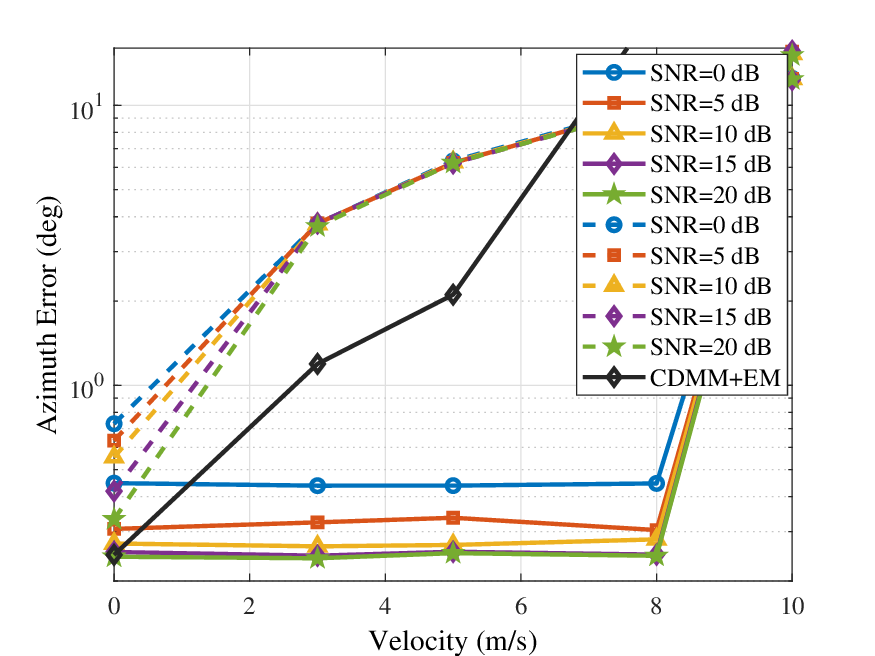}
        \subcaption{Comparison of azimuth angle errors}
    \end{minipage}\hfill
    \begin{minipage}{0.45\textwidth}
        \centering
        \includegraphics[width=\textwidth]{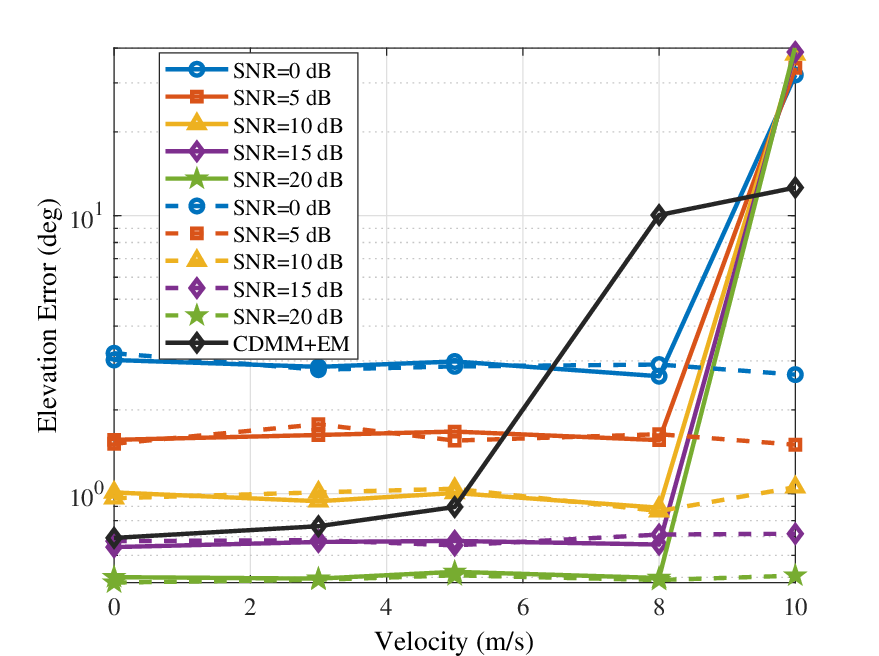}
        \subcaption{Comparison of elevation angle errors}
    \end{minipage}\hfill
    \caption{
    {Comparison of OAM processing methods in the dynamic multi-target scenario, including both multiplexing and high-resolution estimation baselines. The colored solid lines denote the proposed CDMM-based modulation with VCM-EM, while the dashed lines denote the TDMM-based modulation with range-Doppler FFT followed by 2-D MUSIC. The black solid line denotes the CDMM-based modulation with conventional EM at an SNR of $15$~dB.}}
    \label{fig:compare_CDMM_1}
    \vspace{-0.5cm}
\end{figure*}
%%%%%%%%%%%%%%%%%%%%%%%%%%%%%%%%%%%%%%%%%%%%%%%%%%%%%%%
\subsection{Monte Carlo Simulation of Sensing Performance}
The results of the performed Monte Carlo simulations are shown in Fig.~\ref{fig:compare_CDMM_1}, where each point on the curves in this section is averaged over $500$ independent runs for signal-target simulations. 
{For comparison, the widely used TDMM-based OAM radar processing methods are adopted as benchmarks, where range-Doppler FFT and 2-D MUSIC are combined for multi-dimensional parameter estimation. Specifically, range-Doppler FFT is first applied to identify the targets, and 2-D MUSIC is then performed at the detected range bins for high-resolution angle estimation.} 
In each case, simulations are conducted under different SNR levels, i.e., $0$~dB, $5$~dB, $10$~dB, $15$~dB, and $20$~dB, and for targets with different velocities, i.e., $0$~m/s, $3$~m/s, $5$~m/s, $8$~m/s, and $10$~m/s.
The target ranges are randomly set from $30$ to $60$ meters, while the azimuth and elevation angles are randomly set from $5$ to $80$ $\deg$. 
{
To further evaluate the proposed VCM-EM algorithm under Doppler-induced interference, CDMM-based modulation with conventional EM is included as an additional benchmark.  
}

\subsubsection{Localization and Velocity}
The averaged  position errors of 3-D estimation are shown in Fig.~\ref{fig:compare_CDMM_1}(a). In the static case, i.e., when the velocity is $0$~m/s, both the proposed method and the benchmark perform well and achieve comparable results, confirming the effectiveness of the benchmark in static scenarios. However, as the velocity increases, the estimation errors of the benchmark increase significantly, while the proposed method maintains a high level of accuracy. This demonstrates the Doppler-robust benefits of the proposed approach.
The mean error curves of the estimated velocities in Fig.~\ref{fig:compare_CDMM_1}(b) support this observation. The average estimated velocity error of the proposed method remains close to zero, while the benchmark method error increases with the target velocity.
It is also worth mentioning that, according to the configuration settings, the maximum unambiguous velocity \cite{9638345} $ \frac{\lambda}{4 T_c M} = 9.13$~m/s. Hence, when the target velocity exceeds this limit, all the estimation results deteriorate significantly. This indicates that the symbol duration $T_c$ must be carefully chosen.

\subsubsection{Azimuth and Elevation}
As the results of the angle estimation directly influence the beamforming in the communication phase, we also analyze these two parameters. 
The mean error curves of the estimated azimuth angle are shown in Fig.~\ref{fig:compare_CDMM_1}(c). As analyzed, the ambiguity increases with the target velocity. However, it is also observed that, with the proposed method, the angle estimation accuracy remains within $0.5^\circ$, which provides reliable prior information for communication beamforming.
Fig.~\ref{fig:compare_CDMM_1}(d) shows the elevation estimation results. First, the conventional TDMM-based methods perform at a similar level to the proposed method. This is because, as analyzed, the elevation estimation relies on amplitude searching over different mode Bessel functions and is therefore not influenced by the angle-Doppler ambiguity.
Moreover, the OAM-based method is known for its super-resolution capability, and this result serves as a validation. Since the amplitude of the Bessel function is highly dependent on the SNR, when the SNR is sufficiently high (e.g., $20$~dB), it achieves almost perfect estimation accuracy in both the proposed method and the benchmark.

\subsubsection{Comparison with conventional EM}
The black solid curves in Fig.~\ref{fig:compare_CDMM_1} correspond to CDMM-based modulation with conventional EM at an SNR of 15~dB and are compared with the purple solid curves of the proposed VCM-EM under the same SNR setting. At low target velocities, the two methods provide comparable estimation accuracy. As the velocity increases, conventional EM suffers from larger localization, velocity, azimuth, and elevation errors, while VCM-EM maintains more stable performance. This demonstrates the necessity of incorporating the Doppler-dependent vortex coding matrix into the iterative estimation for dynamic scenarios.

\subsection{{OFDM Configuration-Dependent Ambiguity, Trade-Off, and Algorithm Generality}}
% \textcolor{orange}{Comment from WX: Why is an additional subsection introduced here? Would it be more appropriate to directly rename Section C as “OFDM Configuration-Dependent Ambiguity, Trade-Off, and Algorithm Generality” and present the discussion without introducing another level of subsection?}
% 
%%%%%%%%%%
\begin{figure}[t]
    \centering
    \includegraphics[width=0.45\textwidth]{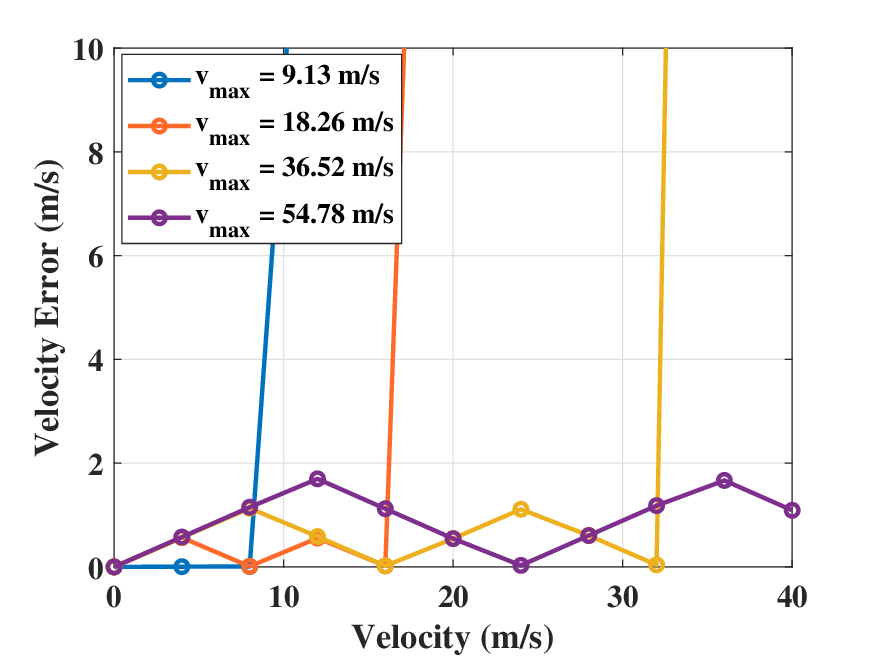}
          % \vspace{-0.3cm}
    \caption{
    {
    Estimated velocity errors as a function of target velocity under different $v_\text{max}$.}}
    \label{fig:velo_error_vmax}
     \vspace{-0.5cm}
\end{figure}
%%%%%%%%%%%%%%%
% \subsubsection{
% {OFDM Configuration-Dependent Sensing Performance Trade-off}}
{
Adopting OFDM for sensing, there is a configuration-dependent trade-off among maximum unambiguous range $R_\text{max}$, range resolution $\Delta R$, and maximum unambiguous velocity $v_\text{max}$, and velocity resolution $\Delta v$, whose expressions are given in Tab.~\ref{tab:simu_ISAC_configs_1}.  
Specifically, decreasing the subcarrier spacing $\Delta f$ enlarges $R_{\max}$, but for a fixed $L$, it reduces the bandwidth $B=L\Delta f$, thereby degrading $\Delta R$. Meanwhile, the increased symbol duration $T_c$ may reduce $v_\text{max}$, although it can improve $\Delta v$ through a longer CPI.
A detailed discussion on configuration-dependent trade-offs can be found in Fig.~5(a) of~\cite{11366124}.
Therefore, the adopted parameters should be selected according to the intended ISAC scenario.
}
 
{
The simulation settings in Tab.~\ref{tab:simu_ISAC_configs_1} are designed for short-range high-resolution OFDM-based
ISAC sensing.
Here, we reduce $T_c$ while keeping the other waveform configurations unchanged, in order to further examine the generality of the proposed algorithm in high-mobility cases. 
The modified setting increases $v_\text{max}$ and supports target velocities up to $54.78~\mathrm{m/s}$ ($197~\mathrm{km/h}$). 
However, this comes at the cost of a degraded velocity resolution, since the CPI becomes shorter when $P_{\mathrm{sen}}$ is fixed.
The simulation results are shown in Fig.~\ref{fig:velo_error_vmax}, where the estimated velocity errors are plotted as a function of the target velocity under different $v_{\max}$ settings, obtained by varying $T_c$. 
The results show that: 
(i) within the unambiguous velocity range determined by the OFDM configuration, the proposed OAM-CDMM coding and VCM-EM estimation exhibit strong robustness to target motion; 
(ii) as $T_c$ decreases, $v_\text{max}$ increases, but $\Delta v$ becomes coarser for a fixed $P_{\mathrm{sen}}$, leading to gradually increased quantization errors in the velocity estimates.
}
% Here we reduce the $T_c$, while keep the other parameters unchanged, hence exploring the the proposed algorithm generality in high-mobility cases. 
% The modified setting supports target velocities up to $54.78~\mathrm{m/s}$ (about $197~\mathrm{km/h}$), however, the corresponding velocity resolution degrades.  
% The simulation results are shown in Fig.~\ref{fig:velo_error_vmax}, where Estimated velocity errors as a function of target velocity under different $v_\text{max}$ ($T_c$ settings). 

% \subsubsection{Comparison with MIMO UCA Configurations}

\subsection{Joint Design of Sensing-Aided Communication Performance}
%%%%%%%%%%
\begin{figure*}[t]
    \centering
        \begin{minipage}{0.33\textwidth}
        \centering
        \includegraphics[width=\textwidth]{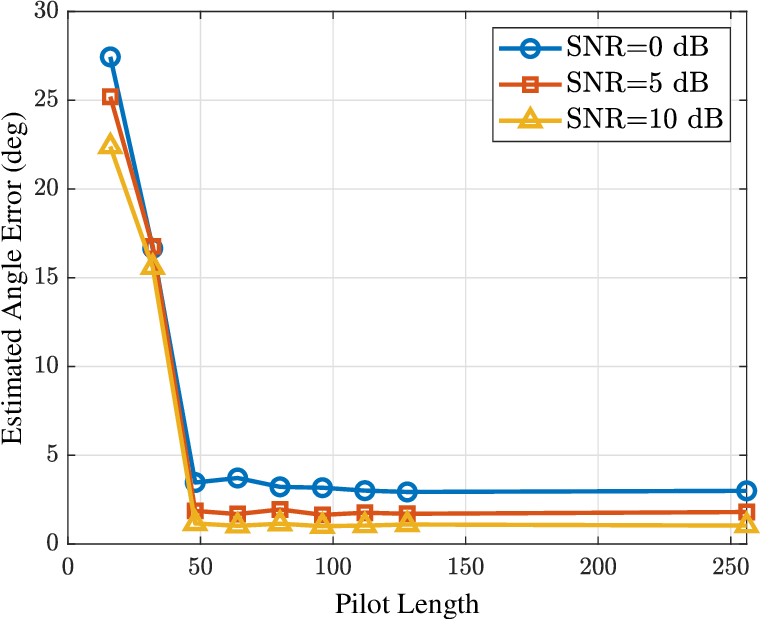}
        \subcaption{}
    \end{minipage}\hfill
    \begin{minipage}{0.33\textwidth}
        \centering
        \includegraphics[width=\textwidth]{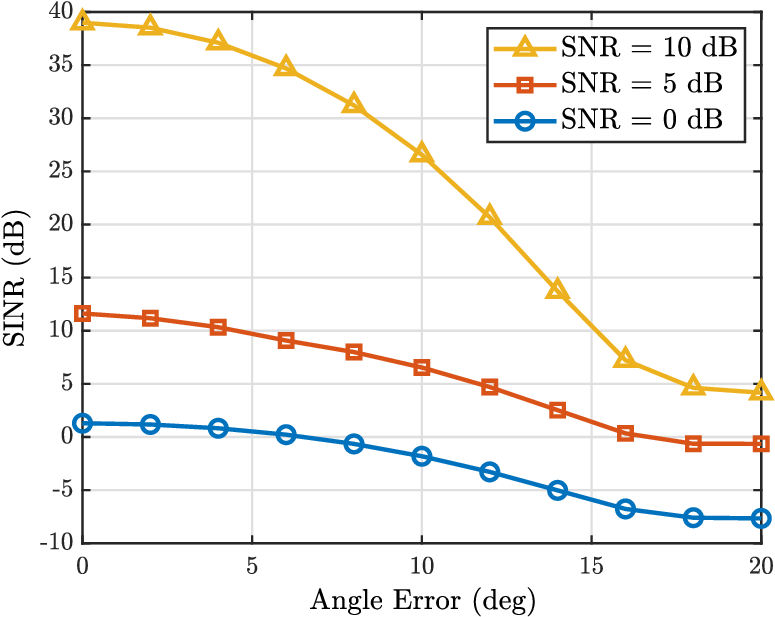}
        \subcaption{}
    \end{minipage}\hfill
    \begin{minipage}{0.33\textwidth}
        \centering
        \includegraphics[width=\textwidth]{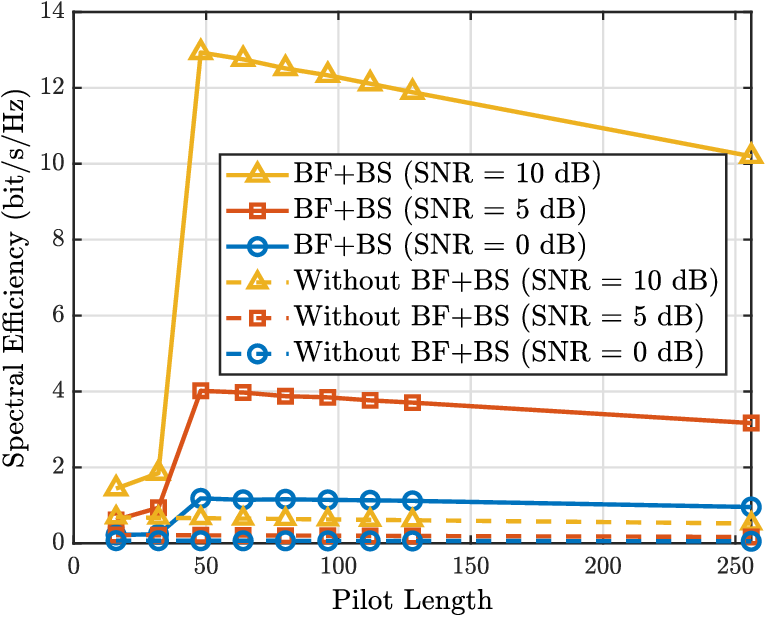}
        \subcaption{}
    \end{minipage}\hfill
    \caption{(a) Angle estimation error vs. pilot length at $U = 16$; (b) Received SINR vs. angle estimation error under different transmit SNRs; (c) Spectral efficiency vs. pilot length at $U=16$ under different transmit SNRs.  
    }\label{fig:ISAC_SE_1}
    \vspace{-0.5cm}
\end{figure*}
Fig.~\ref{fig:ISAC_SE_1} illustrates the impact of the pilot length, i.e., the number of symbols allocated for the sensing phase, on the communication performance under different transmit SNRs. As shown in Fig.~\ref{fig:ISAC_SE_1}(a), the angle estimation error is plotted as a function of the pilot length for various SNR levels. As expected, the estimation error gradually decreases with an increase in the pilot length. When $P_{\text{sen}}\geq 48$, the improvement in estimation accuracy becomes marginal, indicating that the DoA estimation has approached its performance limit. To further investigate the impact of the DoA estimation accuracy on the communication performance, Fig.~\ref{fig:ISAC_SE_1}(b) depicts the received SINR as a function of the angle estimation error. It can be observed that as the DoA estimation accuracy improves, the received SINR increases accordingly. However, when the estimation error becomes smaller than approximately $2^{\circ}$, the SINR enhancement tends to saturate. 

After that, Fig.~\ref{fig:ISAC_SE_1}(c) presents the SE of the integrated system when beamforming and beam steering are performed at the integrated Tx and the communication Rx, respectively, based on the estimated DoA, denoted as `BF+BS'. For comparison, the baseline SE without beamforming and beam steering at either terminal is also provided and denoted as `Without BF+BS'. By comparing Fig.~\ref{fig:ISAC_SE_1}(a) and Fig.~\ref{fig:ISAC_SE_1}(c), it is evident that when $P_{\text{sen}} < 48$, increasing the number of pilots leads to improved DoA estimation accuracy, which in turn enhances the system SE after applying beamforming and beam steering. However, when $P_{\text{sen}}\geq 48$, the estimation accuracy no longer improves significantly. In this regime, additional pilot symbols occupy the coherent interval that could otherwise be used for data transmission, thereby reducing the overall SE. This observation highlights an inherent trade-off between the pilot overhead for sensing and the effective data transmission duration for communication.

\section{Conclusion}\label{sec:conclude_isac}
This work studied a sensing-aided ISAC framework leveraging vortex wavefronts. 
Unlike previous OAM works, we designed a CDMM scheme to simultaneously transmit multi-mode sensing signals, where the vortex wavefront was exploited to address the Doppler-induced ambiguity commonly encountered in dynamic target sensing. 
In the sensing phase, a VCM-EM framework was developed to jointly perform target velocity estimation, sensing matrix decoding, and channel parameter estimation. 
In the communication phase, for mobile terminals, we proposed a joint beam alignment scheme implemented at both the Tx and Rx, and further analyzed the trade-off between the pilot length in the sensing and communication phases. 
Simulation results demonstrated that the proposed method effectively eliminated Doppler-induced interference and achieved Doppler-robust parameter estimation under the vortex-wavefront sensing paradigm. The resulting output provided reliable CSI for the communication Rx, leading to enhanced SE. 
The proposed framework thus offered valuable insights and design guidelines for ISAC in dynamic environments.

In this work, we also recognize that practical delays in sensing, control signaling, and scheduling may affect sensing-aided beamforming for high-mobility UEs, potentially leading to outdated CSI and beam misalignment. Latency-aware beamforming and predictive beam tracking are interesting to consider in future work.
Besides, practical work should consider receiver-side hardware constraints, including reduced-size, sparse, or non-UCA implementations for limited-mode OAM reception, as well as their impact on UE form factor and system performance.
Another interesting direction is to investigate the integration of the proposed spatial/mode-domain design with Doppler-resilient waveforms, such as OTFS, for jointly exploiting the spatial/mode and delay-Doppler domains in high-mobility ISAC scenarios.

% \newpage

\appendices

\section{}

\subsection{Calculation of the Disturbance Term of CDM Mode}\label{App:CDM_disturbance}
% \label{App:CDM_disturbance}
%
% \subsection{Calculation of items}\label{App:CDM_disturbance_A}
In this appendix, $i$, $j$, and $k = 1, 2, \cdots, U$ denote indices, which differ from their physical meanings in the main text.   
The element in the $i$th row and $j$th column $\widetilde{\mathbf{H}}_O$ can be expressed as 
% \textcolor{red}{(Here some notations are different. $U$ should be $M$, l should be ${\ell}_u$, ...)}
%
\begin{align} \label{eq:cdm_10}
 \widetilde{\mathbf{H}}_O(i,j) 
 & = \sum_{ k = 1 }^{ U } \dfrac{ \left[ \mathbf{W}^T\right]_{i,k}}{U} \left[  \mathbf{W}\odot \mathbf{\widetilde{\Omega}_q}  \right]_{k,j}   \nonumber \\
 & = \dfrac{1}{U} \sum_{k =1 }^{U} \mathbf{W}_{k,i} {\mathbf{W}}_{k,j} \left[\mathbf{\widetilde{\Omega}}_q\right]_{k,j}  .
\end{align}
\subsubsection{The diagonal, i.e., case of $i =j$}\label{App:CDM_disturbance_dia}
\begin{align} \label{eq:cdm_10_1}
 \tilde{\mathbf{H}}_O(i,i) 
 & =  \dfrac{1}{U} \sum_{k =1 }^{U} {\mathbf{W}}_{k,i} {\mathbf{W}}_{k,j} \left[\mathbf{\widetilde{\Omega}}_q\right]_{k,j} \nonumber \\
  & = \dfrac{1}{U} \sum_{k =1 }^{U} (1)^2 \tilde{\omega}_q^{k-1}   
 % \nonumber \\
 % % & = \sum_{l }^{U} (1)^2 \mathbf{\tilde{\Omega}}_{i,l} \nonumber \\
 % & 
 = \dfrac{1}{U} \frac{1 - \tilde{\omega}_q^{U} }{1 - \tilde{\omega}_q},
\end{align}
where the elements of the $k$th row in $\mathbf{\widetilde{\Omega}}$ are $\tilde{\omega}_q^{k-1}$.
\subsubsection{The off-diagonal, i.e., case of $i \neq j$}\label{App:CDM_disturbance_off}
\begin{align} \label{eq:cdm_10_2}
 \widetilde{\mathbf{H}}_O(i,j) 
  & = \dfrac{1}{U} \sum_{k =1 }^{U} {\mathbf{W} }_{k,i}   {\mathbf{W} }_{k,j} \left[\mathbf{\widetilde{\Omega}}_q\right]_{i,k}  
= \dfrac{1}{U} \sum_{k =1 }^{U} \{1, -1\}  \tilde{\omega}_q^{k -1}.
 % \nonumber \\
 % & = \mathcal{O}(\tilde{\omega}_q^{U}),
\end{align}
where the result is a weighted geometric series.

\subsection{Proof of \eqref{eq:cdm_11_0_1}}\label{App:proof_1}
\begin{proof}
For $\tilde{\omega}_q \neq 1$, the geometric-series identity gives
\begin{equation}
\frac{1-\tilde{\omega}_q^{U}}{1-\tilde{\omega}_q}
= \sum_{k'=0}^{U-1} \tilde{\omega}_q^{k'}.
\label{eq:geom-id}
\end{equation}
Dividing both sides of \eqref{eq:geom-id} by $U$ yields
\begin{equation}
\lim_{\tilde{\omega}_q \to 1}
\frac{1-\tilde{\omega}_q^{U}}{U(1-\tilde{\omega}_q)}
= \lim_{\tilde{\omega}_q \to 1}\frac{1}{U}\sum_{k'=0}^{U-1} \tilde{\omega}_q^{k}
% = \frac{1}{U}\sum_{k'=0}^{U-1} 1
= 1. \qedhere
\end{equation} 
\end{proof}

\subsection{Proof of the SE Gain Condition}\label{App:proof_2}
\begin{proof}
{
From \eqref{eq9}, consider the sensing-aided SE function 
\begin{equation}
\begin{aligned}
    C_{\rm Sen}(P_{\rm Sen}) & = \left(1-\frac{P_{\rm Sen}}{P}\right) \nonumber \\
                         &\sum_{u=1}^{U} \sum_{p=1}^{P} \log_2 \left(1+{\rm SINR}_{\rm Sen}(p,k,u;P_{\rm Sen})\right),  \nonumber
\end{aligned}
\end{equation}
with $\quad 0\leq P_{\rm Sen}\leq P $. \\
For $P_{\rm Sen}=0$, ${\rm SINR}_{\rm Sen}(p,k,u;P_{\rm Sen})$ is small, $C_{\rm Sen}(P_{\rm Sen}) =0$;
For $P_{\rm Sen}=P$, $C_{\rm Sen}(P_{\rm Sen}) =0$. \\
${\rm SINR}_{\rm Sen}(p,k,u;P_{\rm Sen})$ is non-negative and continuous with respect to $P_{\rm Sen}$. Then $C_{\rm Sen}(P_{\rm Sen})$ is continuous on the compact interval $[0,P]$. \\
By Rolle's  theorem, $C_{\rm Sen}(P_{\rm Sen})$ attains a maximum.  
Therefore, there exists at least one interior maximizer
\begin{equation}
P_{\rm Sen}^{\star}\in(0,P). \nonumber
\end{equation}}
\end{proof}

\section*{Acknowledgment}

The authors would like to thank Prof. Michele Morelli of the University of Pisa for his valuable support and helpful discussions regarding the ICI-related analysis during the preparation of this manuscript.

\bibliographystyle{IEEEtran}
\bibliography{IEEEabrv,refs_2.bib}

% \begin{IEEEbiography}
% \section*{Biographies}
% {YUAN LIU} (Member, IEEE) is a Postdoctoral Researcher with the Interdisciplinary Centre for Security, Reliability and Trust (SnT),
% University of Luxembourg, Luxembourg. 
% His research interests include channel modeling and signal processing for sensing systems. 
% % \end{IEEEbiography}

% {Wen-Xuan Long} (Member, IEEE)

% {BHAVANI SHANKAR M. R.} (Senior Member, IEEE) is an Assistant Professor with SnT, University of Luxembourg. 
% His research interests include radar signal processing and wireless communications.

\end{document}